\documentclass[11pt,a4paper]{article}
\usepackage[left=2cm, right=2cm, top=2cm, bottom=1.8cm]{geometry}
\usepackage{authblk}
\usepackage{float}
\usepackage[T1]{fontenc}
\usepackage[utf8]{inputenc}
\usepackage{amsmath,amsthm,amsfonts,amssymb}
\usepackage{nccmath}
\usepackage{comment}
\usepackage{graphicx}
\usepackage{enumerate}
\usepackage{cite}
\usepackage{caption}
\usepackage{longtable}
\usepackage[labelformat=parens]{subcaption}
\usepackage{bm}
\usepackage[round]{natbib}

\title{Bifurcation of frozen orbits in a gravity field with zonal harmonics}\date{}
\author[1]{Irene Cavallari,}
\author[2]{Giuseppe Pucacco}
\affil[1]{Dipartimento di Matematica, Universit\`a di Pisa}
\affil[2]{Dipartimento di Fisica and INFN -- Sezione di Roma II, Universit\`a di Roma ``Tor Vergata''}

\newtheorem{remark}{\bf Remark}

\def\J2{J_2}

\def\K{\mathcal{K}}
\def\ZMAX{\mathcal{E}}
\def\jjC{j_C}
\def\jCL{\tilde{j}_C}
\def\jj4{j_4}
\def\rhoUJ2{\rho_{E_+}}
\def\rhoLJ2{\rho_{E_-}}
\def\rhoSolUJ4{{\bar{\rho}}_{E_+}}
\def\rhoSolUJ4{{\bar{\rho}}_{E_-}}

\newcommand\beq[1]{ \begin{equation}\label{#1} }
\newcommand{\eeq}{ \end{equation} }
\newcommand\beqa[1]{ \begin{eqnarray} \label{#1}}
\newcommand{\eeqa}{ \end{eqnarray} }
\newcommand{\bv}{\boldsymbol}

\begin{document}
\maketitle

\begin{abstract}
	
	We propose a methodology to study the bifurcation sequences of frozen orbits when the 2nd-order fundamental model of the satellite problem is augmented with the contribution of octupolar terms and relativistic corrections. The method is based on the analysis of twice-reduced closed normal forms expressed in terms of suitable combinations of the invariants of the Kepler problem, able to provide a clear geometric view of the problem. 
		
\end{abstract}

\section{Introduction}
\label{intro}
Among the manifold versions of the perturbed Kepler problem, the investigation of the gravity field expanded 
in multipole terms has traditionally received great attention for its relevance in applications. Therefore, several 
analytical tools have been developed to highlight the most important phenomena. Perturbation theory 
with the construction of normal forms is the standard method since the first pioneering studies \citep{Brouwer1959,Kozai1962b}.
The case in which only zonal terms are included in one of the settings in which we
can obtain explicit approximations of the regular dynamics since the normal form is integrable. 
However, the presence of several parameters, both dynamical (or ``distinguished'' in the language of the
theory of integrable systems) and physical like the multipole coefficients, hinders a global description 
of the dynamics. More efficient geometric and group-theoretic tools have been exploited to study 
the bifurcation of invariant objects when these parameters are varied \citep{Cushman1983,Coffey1986,
	Coffey1994,Palacian2007}. 

Here we study the bifurcation sequences of frozen orbits when the 2nd-order 
fundamental model of the satellite problem is augmented with further features of a typical planetary gravity 
field. We consider the contribution of the octupolar term \citep{Vinti1963,Coffey1994} and the relativistic correction 
due to the quadrupolar term \citep{Heimberger1990}. We implement a twice-reduced normal form \citep{Cushman1988,Marchesiello2014, P19} 
which allows us to obtain in an efficient way the conditions for relative equilibria corresponding to 
the family of periodic orbits with fixed eccentricity and inclination. The method is tested in the 2nd-order $J_2$-problem 
in which known results are reproduced \citep{Palacian2007} and then applied to the above-mentioned 
perturbations. For the  $J_4$-problem, interesting features around the parameter values of the ``Vinti problem''
are highlighted with an additional family of stable frozen orbits. For the relativistic $J_2$-correction, the treatment 
extends and completes several results obtained by \cite{Jupp1991}.

The plan of the paper is as follows: in Section \ref{sec:NF} we recall the model problem based on the normal from 
obtained after averaging with respect to the mean anomaly; in Section \ref{GR} we review the reduction methods 
adapted to the symmetries of the present model, discuss the version adopted here to cope with the structure of 
the Brouwer class of hamiltonians and show how it works in locating relative equilibria; in Section \ref{AppSect} 
we illustrate the results in concrete cases; in Section
\ref{EndSect} we conclude with some hint for possible developments and future works.

\section{The model in closed normal form}
\label{sec:NF}
We are discussing some aspects of the general problem described by a Hamiltonian of the form
\beq{NHZ}
{\cal H} (L,H,G,\ell,g,h) = \sum_{j=0}^{\infty} \epsilon^j {\cal H}_j (L,H,G,\ell,g,h),
\eeq
where
${\cal H}_0$ is the Kepler Hamiltonian and the canonical Delaunay variables have the 
following expression in terms of the standard Keplerian elements $(a,e,i,\ell,\omega,\Omega)$
\begin{align}
	&L =\sqrt{\mu a}, & &  G = \sqrt{\mu a}  \sqrt{1 - e^2}, & & H = \sqrt{\mu a}  \sqrt{1 - e^2} \cos i, \label{DAz}\\
	& \ell = M , & &  \quad g=\omega, & & h=\Omega. \label{DAn}
\end{align}
In the above equation, $\epsilon$ is a formal parameter, called \sl book-keeping \rm parameter, suitably chosen to order the hierarchy of perturbing terms \citep[see][]{Christos}. Therefore, we have a \sl perturbed Kepler problem. \rm

Specifically, in the even zonal artificial satellite problem, we assume to start with the ``original Hamiltonian''
\beq{Hrela}
{\cal H} ({\bv q}, {\bv p}) = \frac12 p^2 + {\cal V}_{CGF} - \frac1{c^2} \left( \frac{p^4}8 - \frac{{\cal V}_{CGF}^2}2 - \frac32{\cal V}_{CGF} p^2 \right)\eeq
in standard Cartesian form, where $\bv q =\{x,y,z\}$, $\bv p =\{\dot{x},\dot{y},\dot{z}\}$, $p=\vert\bv p\vert$, ${\cal V}_{CGF}$ is the classical gravity field and $c$ is the speed of light. 
We include the classical gravity field ${\cal V}_{CGF}$ expanded in terms of the zonal harmonics of even degree\footnote{In this work, we focus on the even zonal problem. Thus, only the even zonal harmonics are considered in the expansion of the gravitational potential. The complete expansion, including also tesseral terms, can be found in \citep{Kaula1966}.}
\beq{GPZ}
{\cal V}_{CGF} = - \frac{\mu}r \left[1- \sum_{k=1}^{\infty} J_{2k} \frac{R_{P}^{2k}}{r^{2k}} P_{2k} (\sin \theta) \right],
\eeq

where $\mu = {\cal G} M_P$ is the product of Newton constant and the mass of the ``planet'', $R_P$ is its radius
and the $P_k$ are the Legendre polynomials with
$$ \sin \theta = \frac{z}r, \quad r = \sqrt{x^2 + y^2 + z^2}.$$
We also add the first-order relativistic corrections following e.g. \cite{Weinberg1972}.

To simplify the structure of the Hamiltonian, 
we then perform a \sl closed-form normalisation \rm like in \citep{Coffey1994} and \citep{Heimberger1990}. 
This method, inspired by works of \citet{Deprit1981,Deprit1982}, has the advantage of avoiding 
expansions in the eccentricity and inclination \citep{Palacian2002,IreneChristos2022}. 
The model in \eqref{Hrela} is rich enough to convey several interesting dynamical features 
keeping the closed form structure at the lowest level of complexity. In fact, after the Delaunay reduction and the elimination of the ascending node, we deal with a secular
Hamiltonian in closed form which depends on only one degree of freedom, corresponding to the pair $G$ and $g$ (the argument of the perigee): 
\begin{equation}
{\cal K} (L,H,G,g) = \sum_j \epsilon^j {\cal K}_j (L,H,G,g),
\label{NFZ}
\end{equation}
with $L$ and $H$ formal integrals of the motion. 
The zero-order term is clearly
\begin{equation}\label{KP}
{\cal K}_0 = {\cal H}_0 = -\frac{\mu^2}{2 L^2}.
\end{equation}
The first-order term is
\begin{equation}
	{\cal K}_1= \frac{\mu^4  J_2 R_{P}^2 (G^2 - 3H^2)}{4 G^5 L^3 } - \frac{\mu^4}{c^2 L^4} \left[3 \frac{L}{G} - \frac{15}{8} \right].
	\label{NFZ1}
\end{equation}
The second-order term ${\cal K}_2$ consists of two contributions: 
$$
{\cal K}_2 = {\cal T}_2 + \langle \mathcal{H}_2 \rangle.
$$
The first is related to the propagation at second order of the $J_2$ term in 
the normalising transformation \citep{Deprit1969,Christos},
\begin{equation}
	\begin{split}
	 {\cal T}_2 & = \frac{3 \mu^6 J_2^2 R_{P}^4}{128 L^5 G^{11} }\bigg[ -5 G^6 - 4 G^5 L + 24 G^3 H^2 L - 36 G H^4 L - 35 H^4 L^2  + G^4 (18 H^2 \\ & \quad + 5 L^2) - 5 G^2 (H^4 + 2 H^2 L^2) +  2 (G^2 - 15 H^2) (G^2 - L^2) (G^2 - H^2) \cos 2 g \bigg] \\ & \quad-\frac{3\mu^6 J_2 R_{P}^2}{4 c^2 L^5 G^7} \left[ (G^2 - 3 H^2) (4 G^2 - 3 GL - 5 L^2) +  (L^2 - G^2) (G^2 - H^2)\cos 2 g \right].
	\end{split}
	\label{HZ21}
\end{equation}
The second is associated directly with the average of the $\mathcal{H}_2$ term:
\begin{equation}
	\begin{split}
	\langle \mathcal{H}_2 \rangle  & = \frac1{2\pi} \int_0^{2\pi} \mathcal{H}_2 d \ell = \frac{\mu^6 J_2 R_{P}^2}{8 c^2 L^5 G^7} \big[ (G^2 - 3 H^2) (6 L^2 - 5 G^2) \\ & \quad - 3  (L^2 - G^2) (G^2 - H^2) \cos 2 g \big] +  \frac{3\mu^6 J_4 R_{P}^4}{128 L^5 G^{11}} \big[ (3 G^4 - 30 G^2 H^2 \\[5pt] & \quad
	+ 35 H^4) (5 L^2 - 3 G^2) - 10 (G^2 - 7 H^2) (L^2 - G^2) (G^2 - H^2)  \cos 2 g\big]
	\end{split}
	\label{NFZ2}
\end{equation}

In this work, we do not consider terms of order higher than $j=2$. Hamiltonians of this type are generally denoted as ``Brouwer's'' ones \citep{Brouwer1959,Cushman1983}. They are characterised by the independence on the mean anomaly $\ell$ and the longitude of the node $h$ (with corresponding conservation of the actions $L$ and $H$), whereas the argument of perigee appears only with the harmonic $\cos 2 g$. These symmetries will all be exploited in the geometric approach described in the following.

The two relativistic terms proportional to $J_2/c^2$ appearing in \eqref{HZ21} and \eqref{NFZ2} have the same structure. However, in the literature \citep{Heimberger1990,Soffel2018}, they are usually kept separate and are respectively referred to as the \sl indirect \rm and \sl direct term \rm 
related to the non-trivial relativistic contribution of the quadrupole of the gravity field of the central body. The ordering of the perturbing terms is performed by assuming (with a certain degree of arbitrariness) the $J_2$ and $c^{-2}$ terms to be of order $\epsilon$ and the $J_4$ term of order $\epsilon^2$, like the $J_2^2$ and $J_2 \times c^{-2}$ terms. 

We remark that, with a slight abuse of notation, we have denoted with the same symbols the Delaunay variables appearing in \eqref{NHZ} and \eqref{NFZ}. 
We have to recall that actually they are respectively the \sl original \rm and the \sl new \rm variables related by the normalising transformation. In the
present work, we are not interested in the explicit construction of particular solutions. Therefore, we will not detail the back-transformation from the new to the 
original coordinates. Moreover, we are not going to investigate any issue connected with the convergence of the expansions. We rely on the asymptotic properties 
of these series and their ability to provide reliable approximations, especially in the cases of Earth-like gravity fields.

For sake of completeness, the different parts of the normalised Hamiltonian $\mathcal{K}=\mathcal{K}_0+\mathcal{K}_1\epsilon+\left(\mathcal{T}_2+\langle \mathcal{H}_2 \rangle\right)\epsilon^2$, expressed in terms of the orbital elements $(a,e,i,\omega)$, are given by
\begingroup
\allowdisplaybreaks
\begin{align*}
\mathcal{K}_0 & = -\frac{\mu}{2a}, \\[5pt]
\mathcal{K}_1& =   \frac{1}{4}\, \frac{\mu J_2R_P^2}{a^3\eta^3}\left(1-3\cos^2 i\right)-\frac{3}{8}\,\frac{\mu^2}{c^2 a^2}\left(\frac{1}{\eta}-5 \right), \\[5pt]
{\cal T}_2 & = \frac{3\mu J_2^2 R_{P}^4}{128\,a^5\eta^7}\, \Big[ -\left(5\,{\eta}^{2}+36\,\eta+35 \right) \sin^{4}i+ 8\left( -\eta^{2}+6\,\eta+10 \right)\sin^{2}i\\
& \quad +8\left({\eta}^{2}-2\,
\eta-5\right) 2\sin^2i\left(1-\eta^2\right)\left(1-15\cos^2i\right)\cos 2\omega
\Big]\\
& \quad-\frac{3\mu^2J_2R_P^2}{4c^2a^4\eta^5}\Big[\left(4\eta^4-3\eta-5\right)(1-3\cos^2i) +\sin^2i(1-\eta^2)\cos2\omega\Big], \\[5pt]
\langle \mathcal{H}_2 \rangle & = \frac{\mu^2 J_2R_P^2}{8c^2a^4\eta^5}\left[(6-5\eta^2)(1-3\cos^2i)-3\sin^2i(1-\eta^2)\cos 2\omega\right]+\\
& \quad\frac{3\mu J_4 R_P^4}{128 a^5\eta^7}\big[(5-3\eta^2)(35\sin^4i-40\sin^2i+8) \\ & \quad
-10\sin^2i(1-\eta^2)(1-7\cos^2i)\cos2\omega\big],
\end{align*}
\endgroup
with $\eta=\sqrt{1-e^2}$.

\section{Geometric reduction}
\label{GR}
The secular Hamiltonian in closed form in \eqref{NFZ}, while computed with an ingenious combination of tools based on the Lie transform method \citep{Deprit1969,Christos} and the elimination of the parallax \citep{Deprit1981}, 
is nonetheless standard in being essentially an average with respect to the mean anomaly \citep{Deprit1982,Palacian2002}. However, it is liable to be treated with a group theoretically 
approach. It can be interpreted as a suitable combination of the invariants generating the $SO(3)$ symmetry of the Kepler problem. In fact, the dynamics ensues from 
the {\sl reduction} of the Hamiltonian defined on the space of the trajectories having, for the unperturbed Kepler problem with negative energy, the structure of the direct product of two spheres. 
The additional symmetries of the closed form of the perturbed problem are exploited to identify a regular reduced phase space with the topology of the 2-sphere. 
In practice, we will use a further transformation leading to a singular reduction on a surface with equivalent topology, which produces a clearer geometric view of the bifurcation sequence of frozen orbits. Here, we provide a quick reminder of the 
invariant theory of the Kepler problem and then apply the reduction process to perturbed Kepler problems described by Brouwer's Hamiltonians.

\subsection{Invariants of the Kepler problem}

Let us call $\bm{G}$ the angular momentum and $\bm{A}$ the  Laplace-Runge-Lenz vector, given by
\[
\bm{G}=G\left[\begin{array}{c}
\sin i\sin h \\- \sin i \cos h \\ \cos i
\end{array}\right], \qquad
\bm{A}=	\sqrt{1-\frac{G^2}{L^2}}\left[\begin{array}{c}
\cos g\cos h-\sin g\sin h \cos i\\ \cos g\sin h+\sin g\cos h \cos i\\ \sin g\sin i
\end{array}\right],
\]
with $i = \arccos\left({H/G}\right)$ the orbital inclination. By defining 
\beq{xGA}
\bm{x} = \bm{G} + L \bm{A}, \quad \bm{y} = \bm{G} - L \bm{A},
\eeq
we get the Poisson structure of the generators of $SO(3)$
\[
\begin{split}
\{x_1,x_3\}=&x_2, \quad \{x_3,x_2\}=x_1, \quad \{x_2,x_1\}=x_3,\\
\{y_1,y_3\}=&y_2, \quad \{y_3,y_2\}=y_1, \quad \{y_2,y_1\}=y_3,
\end{split}
\]
and phase-space defined by the direct product of the two 2-spheres
\beq{KPS}
x_1^2 + x_2^2 + x_3^2 = L^2, \quad y_1^2 + y_2^2 + y_3^2 = L^2.\eeq
It can therefore be imagined as the invariant space of the states characterised 
by given eccentricity, inclination, and arguments of perigee and node, but 
nonetheless equivalent for what pertains to the mean anomaly. In the unperturbed 
problem, the state is a given still point of the invariant space. 
The state point is kept moving on it by the action of the perturbation.

 \subsection{Reduction of the axial symmetry}

Perturbed Kepler problems described by Hamiltonians of the form \eqref{NFZ} are characterised by axial symmetry with $H$ as formal third integral. 
In \cite{Cushman1983} and \cite{Coffey1986} it is shown that, if $0<\vert H\vert<L$, the two-dimensional phase space of such problems is still diffeomorphic to a sphere. Two different sets of variables, both functions of the Keplerian invariants $x_k,y_k \; (k=1,2,3)$ and suitable to analyse the dynamics, are proposed. The variables  $(\pi_1,\pi_2,\pi_3)$ are defined as 
\[
\begin{split}
\pi_1=& \frac12 (x_3 - y_3) = L(\bm{A}\cdot \bm{k}), \\ 
\pi_2=& x_1 y_2 - x_2 y_1 = 2L(\bm{A}\times \bm{G})\cdot\bm{k}, \\ 
\pi_3 =& x_1 y_1 + x_2 y_2 =  \vert \bm{G}\times\bm{k}\vert^2-L^2\vert \bm{A}\times\bm{k}\vert^2,
\end{split}
\]
where $\bm{k}=(0,0,1)^T$ \citep[see][]{Cushman1983}. The phase-space is then
\begin{equation}
\mathcal{P}=\left\{\left(\pi_1,\pi_2,\pi_3\right)\in \mathbb{R}^3 :\, \pi_2^2+\pi_3^2=((L+\pi_1)^2-H^2)((L-\pi_1)^2-H^2) \right\}.
\end{equation}
Instead, in \cite{Coffey1986}, the variables $(\xi_1,\xi_2,\xi_3)$ are introduced, defined as
\[
\xi_1 = {L}(\bm{G}\times \bm{A})\cdot \bm{k},\qquad \xi_2 = {L}\vert\bm{G}\vert (\bm{A}\cdot\bm{k}), \qquad \xi_3=\frac{1}{2}\left(\vert\bm{G}\times\bm{k}\vert^2-{L^2}\vert \bm{A}\vert^2\right),
\]
or, in terms of Delaunay variables,  
\begin{equation}
\begin{split}
\xi_1 = &\sqrt{(G^2-H^2)(L^2-G^2)}\cos g,\\
\xi_2 = &\sqrt{(G^2-H^2)(L^2-G^2)}\sin g,\\
\xi_3 = &G^2-\frac{L^2+H^2}{2}.
\end{split}
\label{xideldef}
\end{equation}
In this case, the phase-space is the sphere of radius $(L^2-H^2)/2$:
\beq{RKPS}
\mathcal{S}=\left\{(\xi_1,\xi_2,\xi_3)\in \mathbb{R}: \xi_1^2+\xi_2^2+\xi_3^2=\frac{(L^2-H^2)^2}{4}\right\}.
\eeq
The relation between the $\pi_k$ and the $\xi_k$ is
\[
\begin{split}
\pi_1= & \frac{\sqrt{2}\xi_2}{\sqrt{2\xi_3+L^2+H^2}},  \\
\pi_2 = & -2\xi_1,\\
\pi_3 = & 2\xi_3+\frac{2\xi_2^2}{{2\xi_3+L^2+H^2}}.
\end{split}
\]
The advantage of both these sets of variables with respect to the Delaunay variables is well explained in \cite{Coffey1986} with an imaginative metaphor. In simpler words, we can say that the Kepler reduction allows us to translate the closed form dynamics in terms of the invariants of the unperturbed problem (formal conservation of $L$) and the further reduction generated by the invariants $\xi_k$ is readily apt to account for the axial symmetry associated with the formal conservation of $H$. Recalling the 
description of the states of the space defined in \eqref{KPS}, we now have that the states of \eqref{RKPS}, given a value of $H$, are characterised by the eccentricity and the perigee but are nonetheless equivalent for what concerns $h$. The dynamical evolution of the system is then determined by the intersections of the reduced phase-space $\mathcal{S}$ with the Hamiltonian expressed in terms of the invariants, e.g. $\mathcal{K} (\xi_1,\xi_2,\xi_3)$.

Whenever one uses the $(G,g)$ chart to analyse the dynamics of the closed form for given values of $L$ and $H$, one excludes circular and equatorial orbits. Indeed, when either the orbital eccentricity or the orbital inclination is zero, the argument of the perigee $g$ is not defined, thus the Delaunay variables result unsuitable to evaluate the stability of such orbits, if they are periodic as typically happens in the artificial satellite problem. Following \cite{Cushman1983}, in \cite{Inarrea2004} it is shown that when ${\cal K}$ possesses independent symmetries of the type 
\[
\begin{split}
\mathcal{R}_1:&(\pi_1,\pi_2,\pi_3)\rightarrow(-\pi_1,\pi_2,\pi_3),\\
\mathcal{R}_2:&(\pi_1,\pi_2,\pi_3)\rightarrow(\pi_1,-\pi_2,\pi_3),\\
\mathcal{R}_3:&(\pi_1,\pi_2,\pi_3)\rightarrow(-\pi_1,-\pi_2,\pi_3),\\
\end{split}
\]
the phase-space can be further reduced, and the variables $\sigma_1,\sigma_2$, defined as
\[
\sigma_1 = (L-\vert H\vert)^2-\pi_1^2, \qquad \sigma_2 = \frac{\sqrt{L^2+H^2-\pi_1^2+\pi_3}}{\sqrt{2}},
\]
are introduced, where $\sigma_2=G$. We propose here to exploit a further set of variables, which is particularly suitable when the normalised Hamiltonian possesses symmetries of the type
\begin{equation}
\begin{split}
{R}_1:&(\xi_1,\xi_2,\xi_3)\rightarrow(-\xi_1,\xi_2,\xi_3),\\
{R}_2:&(\xi_1,\xi_2,\xi_3)\rightarrow(\xi_1,-\xi_2,\xi_3),\\
{R}_3:&(\xi_1,\xi_2,\xi_3)\rightarrow(-\xi_1,-\xi_2,\xi_3).
\label{symm}
\end{split}
\end{equation}
We introduce the variables $(X,Y,Z)$ defined as
\begin{equation}
\begin{split}
X = &\xi_1^2-\xi_2^2,\\
Y = &2\xi_1\xi_2,\\
Z = &\xi_3,
\end{split}
\label{XYZdef}
\end{equation}
which turn the spherical phase space $\mathcal{S}$ into a \textit{lemon} space:
\[
\mathcal{L}=\left\{(X,Y,Z)\in \mathbb{R}: X^2+Y^2=\left(-Z^2+\ZMAX^2\right)^2\right\}, \qquad \ZMAX=\frac{L^2-H^2}{2}.
\]
This kind of reduction was proposed for the first time by \cite{Hanssmann2001}. It is an example of \sl singular \rm reduction \citep{CB} as opposed to the regular setting generated by the invariants $\xi_k$. This occurs here due to the appearance of \sl cusps \rm in the reduced phase-space $\mathcal{L}$ contrary to the smoothness of the 2-sphere $\mathcal{S}$. However, as it will appear clear in the following, this fact does not pose any practical issue in the induction process implemented hereafter.

Even though the phase-space is still three-dimensional, we see that, in the case in which symmetries \eqref{symm} are fulfilled (such as in the problem of the geo-potential when only even zonal harmonics are retained), the transformed closed form does not depend on the variable $Y$: $\mathcal{K}=\mathcal{K}(X,Z)$. In particular, in the case of the Brouwer's Hamiltonian \eqref{NFZ}, $\mathcal{K}$ depends linearly on $X$, i.e. it is of the form
\begin{equation}
\mathcal{K}(X,Z;\bm{a}) = g(Z;\bm{a})+f(Z;\bm{a})X,
\label{normalformstructure}
\end{equation}
where $\bm{a}$ is the set of parameters characterising the problem, including the ``distinguished parameter'' $\ZMAX$. For such a problem, the analysis of the intersection of the reduced phase-space $\mathcal{L}$ with the function \eqref{normalformstructure} is simplified by the extra symmetry of the Brouwer's Hamiltonian since, rather than working in the full 3D-space, all significant information can be obtained by projection on the $(Z,X)$ plane. 
As a matter of fact, when expressed in Delaunay variables, $(X,Y,Z)$ are equal to 
\begin{equation}
\begin{split}
X = &(G^2-H^2)(L^2-G^2)\cos 2g,\\
Y = &(G^2-H^2)(L^2-G^2)\sin 2g,\\
Z = &G^2-\frac{L^2+H^2}{2},
\end{split}
\label{XYZdeldef}
\end{equation}
and, considering the structure of the normalised Hamiltonian presented in the previous section, the possibility of using the general form \eqref{normalformstructure} 
appears immediately justified.

\subsection{Equilibrium points}
Relative equilibria of the reduced systems correspond to periodic orbits of the 
original closed form in \eqref{NFZ}, which in turn are approximations of the periodic 
orbits of the model problem in \eqref{NHZ}. Our main concern refers to frozen orbits 
which play a major role in shaping the phase-space structure of the system. They can be 
identified by locating ``contacts'' between the surfaces defined by the 
Hamiltonian function \eqref{normalformstructure} 
and the lemon space $\mathcal{L}$  \citep{Marchesiello2014} or in some peculiar case 
we will encounter in what follows if the Hamiltonian possesses a 1-dimensional level set 
whose intersection with the phase-space produces additional (unstable) critical points. In the 
present subsection we describe the general procedure to locate equilibria, postponing 
to the next section the details of each case.

Considering $G$ as a function of $Z$, $G=\sqrt{Z+(L^2+H^2)/2}$, the Poisson structure of the 
$(X,Y,Z)$ variables is 
\[
\begin{split}
\{X,Y\}= &8GZ\sqrt{X^2+Y^2},\\
\{X,Z\}= &-4GY,\\
\{Y,Z\}= & 4GY.
\end{split}
\]
Henceforth, given a Hamiltonian of the form $\K$ in \eqref{normalformstructure}, 
the equations of motion are 
\[
\begin{split}
\frac{d X}{d t}= & \{X,\K\}=-4GY\frac{\partial \K}{\partial Z}, \\
\frac{d Y}{d t}= & \{Y,\K\}=4G\left(-2Z\sqrt{X^2+Y^2}\frac{\partial \K}{\partial X}+X\frac{\partial\K}{\partial Z}\right),\\
\frac{d Z}{d t}= & \{Z,\K\}=4GY\frac{\partial \K}{\partial X}. \\
\end{split}
\]
Since we are typically interested in elliptic trajectories, which implies $G\neq 0$, there exist equilibrium points whenever
\begin{equation}
\left\{\begin{array}{l}
Y=0,\\
X\left(2Z\frac{\partial \K}{\partial X}\mathrm{sign}(X)+\frac{\partial \K}{\partial Z}\right)=0,
\end{array} \right.
\label{eqp_yeqzero}
\end{equation}
or
\begin{equation}
\frac{\partial \K}{\partial Z}=\frac{\partial \K}{\partial X}=0.
\label{eqp_yneqzero}
\end{equation}
The variables $X,Y,Z$ are particularly useful in the first case when conditions \eqref{eqp_yeqzero} are fulfilled. 
On the $(X,Z)$ plane, the contour of the lemon space $\mathcal{L}$ is $\mathcal{C}=\mathcal{C}_+\bigcup\mathcal{C}_-$, with
\[
\mathcal{C}_{\pm}= \left\{(X,Z)\in\mathbb{R}^2:\, \vert Z\vert\le\ZMAX,\, X=\pm\hat{X}(Z;\ZMAX)\right\},
\]
where
\begin{equation}
\hat{X}(Z;\ZMAX)=-Z^2+\ZMAX^2.
\label{Xlemonspace}
\end{equation}
For any values of the parameters $\bm{a}$, condition \eqref{eqp_yeqzero} is fulfilled if $X=0$. 
Thus, the normalised Hamiltonian $\K$ always possesses the equilibrium points
\[
E_1 = \left(0,0,-\ZMAX\right), \qquad E_2 = \left(0,0,\ZMAX\right).
\]
From \eqref{XYZdeldef}, $Z=-\ZMAX$ implies $G=H$; thus, the equilibrium point $E_1$ represents the family of equatorial orbits. Instead, $Z=\ZMAX$ implies $G=L$: the equilibrium point $E_2$ represents the family of circular orbits. Condition \eqref{eqp_yeqzero} is also fulfilled whenever a level curve $\tilde{X}(Z;\bm{a},k)$ is tangent to the contour $\mathcal{C}$, with $k$ a given level set of the Hamiltonian $\mathcal{K}(X,Z;\bm{a})$.
We can therefore have an equilibrium point of coordinates $(\hat{X}(Z_+;\ZMAX),0,Z_{+})$ if there exists $Z=Z_{+}$ such that 
\begin{equation}
\left\{\begin{array}{l}
\frac{d \tilde{X}}{d Z}\left(Z_+;\bm{a},k\right)= \frac{d \hat{X}}{d Z}\left(Z_+;\ZMAX\right), \\
\tilde{X}\left(Z;\bm{a},k\right)=\hat{X}\left(Z_+;\ZMAX\right),
\end{array}\right.
\label{gen_case1}
\end{equation}
where
\begin{equation}
\tilde{X}(Z;\bm{a},k)=\frac{k-g(Z;\bm{a})}{f(Z;\bm{a})}
\label{Xtilde}
\end{equation}
is defined by recalling \eqref{normalformstructure}.
From \eqref{gen_case1} and \eqref{Xtilde}, we obtain that $Z_{+}$ is a zero of the function $s_+(Z;\bm{a})$ equal to
\begin{equation}
s_+(Z;\bm{a})=-\frac{1}{f(Z;\bm{a})}\frac{\partial \mathcal{K}}{\partial Z}\left\vert_{X=\hat{X}(Z;\ZMAX)}\right.+2Z.
\label{gen_case1_fun}
\end{equation}
Function $s_+(Z;\bm{p})$ can have multiple zeros corresponding to acceptable equilibrium solutions. In the following, we will refer to them as equilibrium points of type $E_{+}$. 
On the other hand, we can have an equilibrium point of coordinates $(-\hat{X}(Z_-;\ZMAX),0,Z_{-})$, if there exists $Z=Z_{-}$ such that
\begin{equation}
\left\{\begin{array}{l}
\frac{d \tilde{X}}{d Z}\left(Z_-;\bm{a},k\right)= -\frac{d \hat{X}}{d Z}\left(Z_-;\ZMAX\right), \\
\tilde{X}\left(Z_-;\bm{a},k\right)=-\hat{X}\left(Z_-;\ZMAX\right).
\end{array}\right.
\label{gen_case2}
\end{equation}
In this case, $Z_-$ results to be a zero of the function $s_-(Z;\bm{a})$ given by
\begin{equation}
s_-(Z;\bm{a})=-\frac{1}{f(Z;\bm{a})}\frac{\partial \mathcal{K}}{\partial Z}\left\vert_{X=-\hat{X}(Z;\bm{a})}\right.-2Z.
\label{gen_case2_fun}
\end{equation}
Similarly as before, equation $s_-(Z;\bm{a})=0$ can have multiple acceptable solutions. In this case, we are going to talk about equilibrium points of type $E_-$.  
From the first of \eqref{XYZdeldef}, we have that equilibrium points of type $E_{+}$ correspond to the families of periodic orbits with $g=0,\pi$, while those of type $E_{-}$ correspond to the families of periodic orbits with $g=\pm \frac{\pi}{2}$.

In the second case of \eqref{eqp_yneqzero}, if there exist $\bar{X}\in\mathbb{R}$ and $\bar{Z}\in\mathbb{R}$ fulfilling these conditions, one must verify whether the two resulting equilibrium points $\bar{E}_1=(\bar{X},\bar{Y}_1,\bar{Z})$ and $\bar{E}_2=(\bar{X},\bar{Y}_2,\bar{Z})$, with
\[
\bar{Y}_1= \sqrt{\left(-\bar{Z}^2+\ZMAX^2\right)^2-\bar{X}^2}, \qquad \bar{Y}_2=-\bar{Y}_1,
\]
belong to $\mathcal{L}$, i.e. whether $\bar{Y}_1,\bar{Y}_2\in \mathbb{R}$. It is interesting to notice that for every $Y$ the level curves of the Hamiltonian, $\left\{\mathcal{K}=k\right\}$, given by \eqref{Xtilde}, have a singularity at $Z=\bar{Z}$ as $\frac{\partial \K}{\partial X}(\bar{Z})=f(\bar{Z};\bm{a})=0$. The value $Z=\bar{Z}$ gives a vertical {\it asymptote} that is a vertical plane in the 3D space $X,Y,Z$. The condition 
\[
\frac{\partial \K}{\partial Z}= g'(Z) + f'(Z) X = 0,
\qquad
\mbox{namely} \qquad
X = - \frac{g'(Z)}{f'(Z)},
\]
gives an oblique {\it asymptote}, a tilted surface in the 3D space $X,Y,Z$. The two surfaces cross in a straight line, 
orthogonal to the $(Z,X)$ plane, which ``pierces''
the lemon in the symmetric fixed points $\bar{E}_1,\bar{E}_2$.

\begin{remark}
	Each equilibrium point $E_{1,2}$ in $X,Y,Z$, corresponds to one equilibrium in the variables $(\xi_1,\xi_2,\xi_3)$, respectively equal to 
	\[
	\bm{\xi}^{E_1}=(0, 0, -\ZMAX), \qquad \bm{\xi}^{E_2}=(0, 0, \ZMAX).
	\]
	Instead, each equilibrium point of type $E_{\pm}$ and of type $\bar{E}_{1,2}$ correspond to two equilibria. 
	We have the following list of correspondences:
	\[
	\bm{\xi}^{E_+}_{1,2}=(\pm\sqrt{\hat{X}(Z_+;\ZMAX)},0,Z_+); \qquad \bm{\xi}^{E_-}_{1,2}=(0,\pm\sqrt{\hat{X}(Z_-;\ZMAX)},Z_-);
	\]
	\[
	\bm{\xi}^{\bar{E}_{1}}_{1,2} = \left(\frac{\bar{Y_1}}{\bar{\xi}_2}, \pm\bar{\xi}_2,\bar{Z}\right);  \qquad \bm{\xi}^{\bar{E}_{2}}_{1,2} = \left(\frac{\bar{Y_2}}{\bar{\xi}_2}, \pm\bar{\xi}_2,\bar{Z}\right);\qquad \bar{\xi}_2=\sqrt{\frac{-\bar{X}+\sqrt{\bar{X}^2+\bar{Y}_1^2}}{2}}.
	\]
	\label{remark:eqpoint_number}
\end{remark}

\subsection{Stability of the equilibria}
\label{sec:SEP}
To study the stability of the equilibrium points, it is more convenient to come back to the variables $\xi_1,\xi_2, \xi_3$ \citep{Coffey1994}. The transformed closed form is 
\[
\mathsf{K} = g(\xi_3;\bm{a})+ f(\xi_3;\bm{a})\left(\xi_1^2-\xi_2^2\right).
\]
Let us set $\bm{\xi} = (\xi_1, \xi_2, \xi_3)^T$. We have
\[
\dot{\bm{\xi}} = \bm{F}(\bm{\xi}), \qquad \bm{F}(\bm{\xi}) = 2G\left(\frac{\partial \mathsf{K}}{\partial \bm{\xi}}\times \bm{\xi}\right).
\]
We recall that $G=G(\xi_3)= \sqrt{\xi_3+\frac{L^2+H^2}{2}}$. Let us call $\bm{\xi}_E$ an equilibrium point and $\delta \bm{\xi} =  \bm{\xi} - \bm{\xi}_E$ a small displacement from it. The linearised system around the equilibrium is
\[
\delta \dot{\bm{\xi}} = D\bm{F}\vert_{\bm{\xi = \bm{\xi}_E}}\delta\bm{\xi},
\] 
where 
\[
D\bm{F}(\bm{\xi}) = 2G\left[\begin{array}{c c c}
-2\frac{\partial f}{\partial \xi_3}\xi_1\xi_2 & -\left(\frac{\partial \mathsf{K}}{\partial \xi_3}+2f\xi_3-2\frac{\partial f}{\partial \xi_3}\xi_2^2\right) & -\xi_2\left(\frac{\partial^2 \mathsf{K}}{\partial \xi_3^2}+2 f +2\frac{\partial f}{\partial \xi_3}\xi_3\right)\\
\left(\frac{\partial \mathsf{K}}{\partial \xi_3}-2 f\xi_3+2\frac{\partial f}{\partial \xi_3}\xi_1^2\right) & -2\frac{\partial f}{\partial \xi_3}\xi_1\xi_2 & 
\xi_1\left(\frac{\partial^2 \mathsf{K}}{\partial \xi_3^2}-2 f -2\frac{\partial f}{\partial \xi_3}\xi_3\right)\\
4 f \xi_2 & 4 f \xi_1 & 4 \frac{\partial f}{\partial \xi_3} \xi_1\xi_2
\end{array}\right].
\]
Since the $\bm{\xi} \in \mathcal{S}$, the solution of the previous differential system must identically satisfy the constraint
\[
\xi_1\delta\xi_1+\xi_2\delta\xi_2+\xi_3\delta\xi_3=0.
\]
Thus, we obtain the reduced system 
\[
\left[\begin{array}{c}
\delta \dot{\xi}_1 \\
\delta \dot{\xi}_2 
\end{array}\right] = DF_R(\bm{\xi = \bm{\xi}_E})\left[\begin{array}{c}
\delta {\xi}_1 \\
\delta {\xi}_2 
\end{array}\right],
\]
with
\[
DF_R(\bm{\xi})=2G\left[\begin{array}{c c}
\frac{\xi_1\xi_2}{\xi_3}\left(\frac{\partial^2 \mathsf{K}}{\partial \xi_3^2}+2 f\right) &  \frac{\xi_2^2}{\xi_3}\left(\frac{\partial^2 \mathsf{K}}{\partial \xi_3^2}+2 f\right)-\left(\frac{\partial \mathsf{K}}{\partial \xi_3}+2f\xi_3-4\frac{\partial f}{\partial \xi_3}\xi_2^2\right)\\

\left(\frac{\partial \mathsf{K}}{\partial \xi_3}-2 f\xi_3+4\frac{\partial f}{\partial \xi_3}\xi_1^2\right)-\frac{\xi_1^2}{\xi_3}\left(\frac{\partial^2 \mathsf{K}}{\partial \xi_3^2}-2 f\right)  &-\frac{\xi_1\xi_2}{\xi_3}\left(\frac{\partial^2 \mathsf{K}}{\partial \xi_3^2}-2 f\right)\\
\end{array}\right].
\]
To evaluate the stability of the equilibrium point we have to compute the eigenvalues $\alpha_{1,2}$ of $DF_R(\bm{\bm{\xi}_E})$, by solving the characteristic equation
\[
\alpha^2-\mathsf{Tr}\, DF_R(\bm{\bm{\xi}_E})\alpha + \mathsf{det}DF_R(\bm{\bm{\xi}_E})=0,
\]
with $\mathsf{Tr}\, DF_R(\bm{\bm{\xi}_E})$ and $\mathsf{det}DF_R(\bm{\bm{\xi}_E})$ the trace and the determinant of $DF_R(\bm{\bm{\xi}_E})$. By using transformation \eqref{XYZdef}, we obtain that the characteristic equation for the equilibrium point $E_1$ is 
\begin{equation}
\alpha^2+4H^2f^2\left(-\ZMAX;\bm{a}\right)s_-(-\ZMAX;\bm{a})s_+(-\ZMAX;\bm{a})=0,
\label{careq_E1}
\end{equation}
while the one for $E_2$ is
\begin{equation}
\alpha^2+4L^2f^2\left(\ZMAX;\bm{a}\right) s_-(\ZMAX;\bm{a})s_+(\ZMAX;\bm{a})=0,
\label{careq_E2}
\end{equation}
with $s_+(Z;\bm{a})$ and $s_-(Z;\bm{a})$ given in \eqref{gen_case1_fun} and \eqref{gen_case2_fun}. Note that whenever the parameters $\bm{a}$ are such that an equilibrium point of  either type $E_+$ or $E_-$ coincides with $E_1$ (i.e. $Z=-\ZMAX$ is a zero of either $s_+(Z;\bm{a})$ or $s_-(Z;\bm{a})$) $E_1$ becomes degenerate. The same holds true for $E_2$. 
For an equilibrium point of type $E_{+}$ of coordinates $(\hat{X}(Z_+;\ZMAX),0,Z_+)$, it can be proved that the characteristic equation is
\begin{equation}
\alpha^2+16G^2f^2(Z_+;\bm{a})\hat{X}(Z_+;\ZMAX) \left(\frac{d^2\tilde{X}}{dZ^2}(Z_+;\bm{a},k_+)-\frac{d^2\hat{X}}{dZ^2}(Z_+;\ZMAX)\right)=0,
\label{careq_E3}
\end{equation}
with $k_+$ the value of the Hamiltonian such that $\tilde{X}(Z_+;\bm{a},k_+)=\hat{X}(Z_+;\ZMAX)$. Similarly, for an equilibrium point of type $E_-$ of coordinates $(-\hat{X}(Z_-;\ZMAX),0,Z_-)$ we have
\begin{equation}
\alpha^2-16G^2f(Z_-;\bm{a})^2\hat{X}(Z_-;\ZMAX) \left(\frac{d^2\tilde{X}}{dZ^2}(Z_-;\bm{a},k_-)+\frac{d^2\hat{X}}{dZ^2}(Z_-;\ZMAX)\right)=0,
\label{careq_E4}
\end{equation} 
with $k_-$ such that $\tilde{X}(Z_-;\bm{a},k_-)=-\hat{X}(Z_-;\ZMAX)$. Since for $Z\neq\pm\ZMAX$, $\hat{X}(Z;\ZMAX)>0$, the stability of the equilibrium points of type $E_+$ and $E_-$ can be determined by comparing the concavities of the level curve $\tilde{X}(Z;\bm{a},k)$ and of the contour $\mathcal{C}$ of $\mathcal{L}$ at their point of tangency. Finally, the characteristic equations for  $\bar{E}_1$ and $\bar{E}_2$ are
\[
\alpha^2  +4\bar{G}^2\frac{\bar{Y}^2_{1,2}}{\bar{Z}^2}\left(2\bar{\mathcal{K}}_{ZZ}^2-16\bar{f}^2_{Z}\bar{Z}^2\right)=0,
\]
with
\[
\bar{\mathcal{K}}_{ZZ}=\frac{\partial^2\K}{\partial Z^2}\left(\bar{X},\bar{Z};\bm{p}\right),\qquad \bar{f}_{Z}=\frac{\partial f}{\partial Z}(\bar{Z};\bm{p}), \qquad \bar{G}=\sqrt{\bar{Z}^2+\frac{L^2+H^2}{2}}.
\]
As $\bar{Y}^2_{1}=\bar{Y}^2_{2}$, the two characteristic equations coincide: $\bar{E}_1$ and $\bar{E}_2$ have the same stability. When $\bar{E}_1$ and $\bar{E}_2$ coincide since $\bar{Y}_1=\bar{Y}_2=0$, the resulting equilibrium point is degenerate. 

\begin{remark}
	To evaluate the stability we can also exploit  the Poincar\'e-Hopf index theorem:
	
	\textit{Let $M$ be a compact manifold and $w$ a smooth vector field on $M$ with isolated zeros. 
		The sum $\sum \iota$ of the indices of the zeros of $w$ is equal to the Euler characteristic of $M$} \citep{Milnor1965}.
	
	As the phase space is a sphere in the coordinates $(\xi_1,\xi_2,\xi_3)$, its Euler characteristic is equal to $2$. Whenever $E_1$ and $E_2$ are Lyapunov stable, in the linearised reduced system they are centres; thus, their indexes $\iota$ are both equal to $+1$. Instead, when one of them is Lyapunov unstable, it corresponds to a saddle with $\iota = -1$. Each equilibrium point of type $E_{\pm}$ corresponds to two equilibrium points in $(\xi_1,\xi_2,\xi_3)$ (see Remark \ref{remark:eqpoint_number}), both either Lyapunov stable or unstable. The bifurcation of a first stable pair, implies a stability/instability transition of one of the cusps so that the indexes are $(+1-1+1+1)$. The bifurcation of a second unstable pair implies that the cusp regains stability and the indexes are $(+1+1+1+1-1-1)$. Due generalisation applies in the case of the points of type $\bar{E}_{1,2}.$
\end{remark}

\section{Applications}
\label{AppSect}

\subsection{The $J_2$-problem}
\label{sec:J2}
We are going to apply the variables $X,Y,Z$ to analyse a classical and well-known problem in the framework of the artificial satellite theory: the study of the secular Hamiltonian in which only the second zonal harmonic of the gravitational potential is retained, i.e. the $J_2$ terms. From \eqref{KP}, \eqref{NFZ1}, \eqref{HZ21} and \eqref{NFZ2}, the resulting closed form is
\begin{equation*}
\begin{split}
\mathcal{K}_{J_2}& =-\frac{\mu^2}{2 L^2}+\frac{\mu^4  J_2 R_{P}^2 (G^2 - 3H^2)}{4 G^5 L^3 }+\frac{3 \mu^6 J_2^2 R_{P}^4}{128 L^5 G^{11} }  \bigg[ -5 G^6 - 4 G^5 L \\
& \hspace{5mm}+ 24 G^3 H^2 L - 36 G H^4 L - 35 H^4 L^2 +  G^4 (18 H^2 + 5 L^2) \\
& \hspace{5mm}- 5 G^2 (H^4 + 2 H^2 L^2) +  2 (G^2 - 15 H^2) (G^2 - L^2) (G^2 - H^2) \cos 2 g \bigg].
\end{split}
\end{equation*}
To simplify its analysis,
we make the system dimensionless by performing the following choice of units: we take the orbital semi-major axis $a$ as unit of length and the unit of time such that $\mu=1$. Let us call $\rho=H/L$. In the adimensional system the Delaunay actions $L$ and $H$ become
\[
L = 1, \qquad H = \rho,
\]
where $\rho=G\cos i$, with $i$ the orbital inclination. Moreover, the action $G$ coincides with $\eta=\sqrt{1-e^2}$, being $e$ the orbital eccentricity. Since in the adimensional system  the planet's radius $R_{P}<1$, let us set 
\begin{equation}
\lambda=J_2R_{P}^2;
\label{lambdadef}
\end{equation}
$\lambda$ plays here the role of small parameter, of the same order as the book-keeping parameter $\epsilon$. We drop the constant Keplerian term and we perform a transformation of the time variable $t\mapsto \tau$, defined as 
\begin{equation}
\frac{\partial \tau}{\partial t} = \lambda.
\label{timetrans}
\end{equation}
Thus, the secular Hamiltonian in closed form becomes
\[
\begin{split}
\mathcal{K}_{J_2}& =\frac{(G^2 - 3\rho^2)}{4 G^5 }+\frac{3\lambda}{128 G^{11} }  \bigg[ -5 G^6 - 4 G^5 + 24 G^3 \rho^2  - 36 G \rho^4  - 35 \rho^4 +  G^4 (18 \rho^2 + 5 ) \\
& \hspace{5mm}- 5 G^2 (\rho^4 + 2 \rho^2) +  2 (G^2 - 15 \rho^2) (G^2 -1) (G^2 - \rho^2) \cos 2 g \bigg].
\end{split}
\]
In these units, the $X,Y,Z$ variables are
\begin{equation}
X = (G^2-\rho^2)(1-G^2)\cos 2g,\quad
Y = (G^2-\rho^2)(1-G^2)\sin 2g,\quad
Z = G^2-\frac{1+\rho^2}{2},
\label{XYZdeldef_nondimsys}
\end{equation}
and we also have $\ZMAX=(1-\rho^2)/2$.
The introduction of the variables $X,Y,Z$ leads to a closed form with the same structure of $\K$ in \eqref{normalformstructure}, with
\[
\begin{split}
g(Z,\bm{a}) &=  \frac {-5{{\rho}}^{2}+2Z+1}{\sqrt{2} \left( {{\rho}}^{2}+2Z+1
	\right) ^{\frac{5}{2}}} -\frac{3\lambda}{16\sqrt{2}\left(\rho^{2}+2Z+1\right)^{\frac{11}{2}}}\Big( 40{Z}^{3}+ \left( -84{{\rho}}^{2}+20 \right) {Z}^{2} \\ & \quad + \left( -
74{{\rho}}^{4}-44{{\rho}}^{2}-10 \right) Z-11{{\rho}}^{
	6}+273{{\rho}}^{4}-{{\rho}}^{2}-5 \\ & \quad +4\sqrt {2\,{{\rho}}^{2}+4Z+2} \left(-5{{\rho}}^{2}+2Z+1\right) ^{2}\Big),
\\
f(Z,\bm{a}) & =  -\frac{3}{2}\,\lambda {\frac { \left( -29\,{{\rho}}^{2}+2\,Z+1 \right) }{ \sqrt{2}\left( {{\rho}}^{2}+2\,Z+1 \right) ^{\frac{11}{2}}}},
\end{split}
\]
and $\bm{a}=(\rho; \lambda)$.
Note that if the terms proportional to $\lambda$ are neglected, the problem has one equilibrium solution for
\begin{equation}
Z = \frac{9\rho^2-1}{2},\quad \forall\, X,Y,
\label{j2prob_zeroordersol}
\end{equation}
which implies
\begin{equation}
G = G_{c} = \sqrt{5}\rho.
\label{critG}
\end{equation}
Since $\rho=G\cos i$, the orbit has then a stationary pericentre at the so-called critical inclination:
\begin{equation*}
i_c = \arccos\frac{1}{\sqrt{5}}.
\label{criticalinclination}
\end{equation*}
In the following, we study the $\J2$-problem for $\vert\rho\vert\in(0,1)$ and $\lambda\in(0,1)$: we discuss the existence and the stability of frozen orbits by analysing the corresponding properties of the equilibrium points of the reduced system. 

First of all, we show that the equilibrium point $E_1$, representative of the family of equatorial orbits, is always stable.
Then, we analyse the stability of the equilibrium point $E_2$, representative of the family of circular orbits. In particular, we determine the values $\rho_+$ and $\rho_-$ of $\vert\rho\vert$ at which pitchfork bifurcations occur: for $\vert\rho\vert$ between $\rho_-$ and $\rho_+$ $E_2$ is unstable, otherwise it is stable; moreover, there exist a stable equilibrium point of type $E_+$ for $\vert\rho\vert<\rho_+$ and an unstable equilibrium point of type $E_-$ for $\vert\rho\vert<\rho_-$. At last, we show that the equilibrium points $\bar{E}_1$ and $\bar{E}_2$ do not exist for any $\rho$ and $\lambda$. 

For the $J_2$ problem and the other problems analysed in the following, all the equilibrium points of type $E_+$ are indicated with an odd integer number larger than $1$ as a subscript; similarly, the subscript of the equilibrium points of type $E_-$ is an even integer number larger than $2$.

We recall that $\rho=G\cos i$ and $G=\sqrt{1-e^2}$. In the procedure we follow, we select a planet and we fix the value of the semi-major axis, on which $\lambda$ depends through the dimensionless $R_P$. Suppose to select a value of $\rho$ such that an equilibrium point of type $E_+$ exists and to compute the value of the action $G$ of such equilibrium point; from the selected $\rho$ and the value of $G$ we can obtain the orbital eccentricity and inclination of the family of orbits represented by the equilibrium point itself. The same holds for all the equilibrium points. In the same way, since the eccentricity of the orbits represented by $E_2$ is equal to zero, from $\rho_+$ and $\rho_-$ we can compute the values of the orbital inclination at which the stability of the circular orbits changes. 

\subsubsection{Stability of $E_1$}
From \eqref{careq_E1} it results that the stability of the equilibrium point $E_1$ depends on the sign of $s_+(-\ZMAX;\bm{a})$ and   $s_-(-\ZMAX;\bm{a})$. We have
\begin{equation*}
s_+(-\ZMAX;\bm{a}) = -\frac{2}{7}\,{\frac {8{{\rho}}^{4}+\lambda\left(-7{{\rho}}^{2}+12
		\vert\rho\vert+31\right)}{\lambda}},
\label{sME3_j2}
\qquad
s_-(-\ZMAX,\bm{a}) = -\frac{8}{7}\, \frac {2{{\rho}}^{4}+3\lambda \vert\rho\vert+6\lambda}{\lambda}.
\end{equation*}
It is straightforward that $s_+(-\ZMAX;\bm{a})<0$ and $s_-(-\ZMAX,\bm{a})<0$ $\forall\lambda\in(0,1)$ and $\forall\vert\rho\vert\in(0,1)$. Thus, the equilibrium point $E_1$ is stable and  it does never coincide with equilibrium points of either type $E_+$ or $E_-$.

\subsubsection{Stability of $E_2$}
\label{sec:j2prob_E2}
In analogy to $E_1$, from \eqref{careq_E2} we obtain that the stability of $E_2$ depends on the sign of $s_+(\ZMAX;\bm{a})$ and   $s_-(\ZMAX;\bm{a})$. We have 
\begin{equation*}
s_+(\ZMAX;\bm{a}) = -\frac{1}{2}\,
\frac {425\lambda{{\rho}}^{4}-146\lambda{{\rho}}^{2}+80\,{{\rho}}^{2}+9\lambda-16}{\lambda \left( 15{{\rho}}^{2}-1\right)},
\label{sPE3_j2}
\end{equation*}
and 
\begin{equation*}
s_-(\ZMAX;\bm{a}) = -\frac{1}{2}\,\frac {\left( 365\lambda{{
			\rho}}^{4}-82\lambda{{\rho}}^{2}+80\,{{\rho}}^{2}+5\lambda-16 \right)}{\lambda\left( 15\rho^2-1 \right) }.
\label{sPE4_j2}
\end{equation*}
The function $s_+(\ZMAX;\bm{a})$ has two real zeros $\rho=\pm\rho_+$, where
\begin{equation}
\rho_+=\sqrt{\frac{1}{5}+\frac{ -4(3\lambda+10)+4\sqrt {85\lambda^2+(3\lambda+10)^2}}{425\lambda}}. 
\label{rhopiu_j2}
\end{equation}
Similarly, $s_-(\ZMAX;\bm{a})$ possesses two real zeros $\rho=\pm\rho_-$, with
\begin{equation}
\rho_-= \sqrt{\frac{1}{5}+ \frac {-4(8\lambda+10)+4\sqrt {-73 \lambda^2+(8\lambda+10)^2}}{365\lambda}}.
\label{rhomeno_j2}
\end{equation}
It holds that $\rho_+>\rho_-$, $\forall \lambda\in(0,1)$. Thus,
\begin{itemize}
	\item for $\rho_+<\vert\rho\vert<1$ and for $0<\vert\rho\vert<\rho_-$, $E_2$ is stable;
	\item for $\rho_-<\vert\rho\vert<\rho_+$, $E_2$ is unstable.
\end{itemize}
Moreover, at $\vert\rho\vert=\rho_+$ the degenerate $E_2$ coincides with an equilibrium point of type $E_+$, while at $\vert\rho\vert=\rho_-$ it coincides with an equilibrium point of type $E_-$. 

If we approximate $\rho_+$ and $\rho_-$ as series in $\lambda$, we obtain
\begin{align}
\rho_+ \sim & \frac{1}{\sqrt{5}}\Big(1+\frac{1}{10}\lambda-\frac{7}{200}\lambda^2-\frac{7}{800}\lambda^3+\mathcal{O}(\lambda^4)\Big), \label{rhoPseries}\\
\rho_- \sim &  \frac{1}{\sqrt{5}}\,\Big(1-\frac{1}{10}\lambda+\frac{3}{40}\lambda^2-\frac{299}{4000}\lambda^3+\mathcal{O}(\lambda^4)\Big). \label{rhoMseries}
\end{align}
Thus, the bifurcations occur nearby the zero-order solution \eqref{j2prob_zeroordersol} when $Z=\ZMAX$. 

\subsubsection{Existence and stability of the equilibrium points of type $E_+$ and $E_-$}
\label{sec:j2prob_EpEm}
The equilibrium points of type $E_+$ correspond to the zeros of $s_+(Z;\bm{a})$. 
Performing the change of variable $Z\mapsto G$ using \eqref{XYZdeldef_nondimsys}, we obtain
\begin{equation*}
s_+\left(G^2-\frac{1+\rho^2}{2};\bm{a}\right) = \frac{S_+(G;\bm{a})}{16\lambda G(28\rho^2+2G)},
\end{equation*}
with
\[
\begin{split}
S_+(G;\bm{a})= &32G^8+(-160\rho^2-15\lambda)G^6-24G^5\lambda+(-98\lambda\rho^2+21\lambda)G^4+192G^3\lambda\rho^2\\
& +(225\lambda\rho^4+198\lambda\rho^2)G^2-360G\lambda\rho^4-715\lambda\rho^4.
\end{split}
\]
The zeros of $s_+(G^2-\frac{1+\rho^2}{2};\bm{a})$ are the zeros of $S_+(G;\bm{a})$. This is a polynomial function of degree $8$ in $G$. Thus, finding its zeros is not straightforward. However, some pieces of information can be inferred by inverting the roles of $G$ and $\rho$: we consider $G$ as a parameter and $\rho$ becomes the independent variable of the problem. We obtain $S_+(G;\bm{a})=0$ for
$\rho^2={\rhoUJ2^2}_{1,2}$, with
\begin{equation}
{\rhoUJ2^2}_{1,2}= \frac{G^2}{5\lambda}\frac{A_+\pm 4\sqrt{B_+}}{C_+},
\label{rhoE312_j2}
\end{equation}
\begin{align*}
A_+ = & \left( 49{G}^{2}-96\,G-99 \right) {\lambda}+80\,{G}^{4}, & 
B_+ = & \frac{1}{16}\,(A_+^2-5\lambda C_+D_+), \\
C_+ = & 45G^2-72G-143, &
D_+ = & 32G^4-15G^2\lambda-24G\lambda+21\lambda.
\end{align*}
The solutions are admissible if $0<{\rhoUJ2^2}_{1,2}<G^2$. Since $G\in(0,1]$ and $\lambda\in(0,1)$, it is easy to verify that $C_+<0$ and \[
D_+>(32G^4-15G^2-24G+21)\lambda >0, 
\]
which implies $B_+>A_+^2/16$.
Thus, ${\rhoUJ2^2}_{1}<0$ and is not an admissible solution. Instead, ${\rhoUJ2^2}_{2}>0$. Since it also holds $A_+-5\lambda C_+>0$ and
\[
16B_+-(A_+-5\lambda C_+)^2 = 80\lambda C_+(8G^4-7G^2\lambda+12G\lambda+31\lambda)<0,
\]
we have ${\rhoUJ2^2}_{2}<G^2$. It follows that ${\rhoUJ2^2}_{2}$ is admissible $\forall \lambda\in(0,1)$ and $\forall G\in(0,1]$. For $G=1$ we obtain ${\rhoUJ2^2}_{2}=\rho_+$; furthermore, it can be proved that 
\begin{equation}
\frac{d {\rhoUJ2^2}_{2}}{dG}>0 \quad \forall G\in(0,1], \quad \forall\lambda\in(0,1), 
\label{j2prob_derivativeP}
\end{equation}
see the Appendix. It follows that for each $\vert\rho\vert\le\rho_+$ there exists only one value of $G$ solving $S_+(G;\bm{a})=0$. Thus,  for each $\vert\rho\vert\le\rho_+$ there exists only one equilibrium point of type $E_+$, which we call $E_3$ and which coincides with $E_2$ for $\vert\rho\vert=\rho_+$. 

The value of the only meaningful solution of $S_+(G;\bm{a})=0$ can be approximated with a perturbation method. We observe that the solution of the ``unperturbed'' problem with $\lambda=0$ is just the critical value \eqref{critG}. Then, we can look for a solution of the form
\[
G_+ = \sqrt{5}\rho + \sum_{k \ge 1} a_k \lambda^k.
\]
At third order in $\lambda$, we find
\begin{align}
G_+ & =  \sqrt{5}\rho +  \frac{-1 + 4 \rho^2}{10 \sqrt{5} \rho^3} \lambda - 
\frac{14 + 6 \sqrt{5} \rho - 81 \rho^2 - 24 \sqrt{5} \rho^3 + 100 \rho^4}{5000 \sqrt{5} \rho^7} \lambda^2 \nonumber\\
& \quad + \frac{1}{5000000 \sqrt{5} \rho^{11}}\Bigg(353 - 336 \sqrt{5} \rho - 4077 \rho^2 + 1944 \sqrt{5} \rho^3 \\
& \quad
+ 12310 \rho^4 - 
2400 \sqrt{5} \rho^5 - 6600 \rho^6\Bigg)\lambda^3.\label{Gplus}
\end{align}

We apply the same technique to verify the existence of equilibrium points of type $E_-$. They correspond to the zeros of a function $S_-(G;\bm{a})$ equal to
\[
\begin{split}
S_-(G;\bm{a})= & 32{G}^{8}+ \left( -160{{\rho}}^{2}-35\,{\lambda} \right) {G}
^{6}-24{G}^{5}{\lambda}+ \left( 350\,{\lambda}\,{{\rho}}^{2
}+49\,{\lambda} \right) {G}^{4}\\
&+192\,{G}^{3}{\lambda}\,{{\rho
}}^{2}+ \left( -315\,{\lambda}\,{{\rho}}^{4}-378\,{\lambda}\,
{{\rho}}^{2} \right) {G}^{2}-360\,G\,{\lambda}\,{{\rho}}^{4}-
55\,{\lambda}\,{{\rho}}^{4}.
\end{split}
\]
We have 
$S_-(G;\bm{a})=0$ for $\rho^2={\rhoLJ2^2}_{1,2}$, with
\begin{equation}
{\rhoLJ2^2}_{1,2} = -\frac{G^2}{5\lambda}\frac{A_-\pm4\sqrt{B_-}}{C_-}, 
\label{rhoE412_j2}
\end{equation}
\begin{align*}
A_- = &  \left( -175\,{G}^{2}-96\,G+189 \right) {\lambda}+80\,{G}^{4}, & 
B_- = & \frac{1}{16}\, (A_-^2+5\lambda C_-D_-),\\
C_- = & 63G^2+72G+11, & 
D_- = & (32G^4-35G^2\lambda-24G\lambda+49\lambda.
\end{align*}
$\forall G\in(0,1]$ and $\forall\lambda\in(0,1)$ we have $C_->0$  and
\[
D_->(32G^4-35G^2-24G+49)\lambda >0.
\]
Thus, ${\rhoLJ2^2}_{1}<0$ and is not an admissible solution. Instead, ${\rhoLJ2^2}_{2}>0$; as $5\lambda C_-+A_->0$ and 
\[
16B_--(5\lambda C_-+A_-)^2=-320\lambda C_-(2G^4+3G\lambda+6\lambda)<0,
\]
we have ${\rhoLJ2^2}_{2}<G^2$:  ${\rhoLJ2^2}_{2}$ is admissible $\forall G\in(0,1]$ and $\forall\lambda\in(0,1)$. For $G=1$, ${\rhoLJ2^2}_{2}=\rho_-^2$; it can also be proved that 
\begin{equation}
\frac{d {\rhoLJ2^2}_{2}}{dG}>0, \quad \forall\lambda\in(0,1),
\label{j2prob_derivativeM}
\end{equation}
see the Appendix. As a consequence, also in this case we obtain that there exists one equilibrium point of type $E_-$ for any $\vert\rho\vert\le\rho_-$. We call it $E_4$. For $\vert\rho\vert=\rho_-$, it coincides with $E_2$. The solution of $S_-(G;\bm{a})=0$ can be approximated in analogy with what seen above. At third order in $\lambda$, we find
\begin{align}
G_- & =  \sqrt{5}\rho +  \frac{9 - 35 \rho^2}{100 \sqrt{5} \rho^3} \lambda - 
\frac{1305 -108 \sqrt{5} \rho - 7910 \rho^2 +420 \sqrt{5} \rho^3 + 11025 \rho^4}{100000 \sqrt{5} \rho^7} \lambda^2 \nonumber\\
& \quad + \frac{1}{100000000 \sqrt{5} \rho^{11}}\Bigg(267309 - 31320 \sqrt{5} \rho - 2226905 \rho^2 + 189840 \sqrt{5} \rho^3 \\
& \quad + 5775175 \rho^4 - 
264600 \sqrt{5} \rho^5 - 4501875 \rho^6\Bigg)\lambda^3.\label{Gminus}
\end{align}

For $\rho_-<\vert\rho\vert<\rho_+$, the equilibrium $E_3$ is stable as a consequence of the Poincar\'e-Hopf theorem. By applying this last theorem, we also obtain that for $0<\vert\rho\vert<\rho_-$, one equilibrium points between $E_3$ and $E_4$ is stable, while the other is unstable. Since $E_3$ does not undergo any bifurcation at $\vert\rho\vert=\rho_-$, it is stable, while $E_4$ is unstable. 

\subsubsection{About the existence of $\bar{E}_1$ and $\bar{E}_2$}
The coordinates $\bar{X}$ and $\bar{Z}$ of the equilibrium points $\bar{E}_1$ and $\bar{E}_2$ are 
\[
\bar{X}=-\frac{1}{3}{\rho}^{2} \left( -144\sqrt {15}\vert \rho \vert-2835{\rho}^{2}+307\right) -18000\frac {{\rho}^{6}}{{\lambda}}, \qquad \bar{Z} = \frac{29\rho^2-1}{2}.
\]
In order to have $\bar{Z}\in[-\ZMAX,\ZMAX]$ it is necessary that $\vert\rho\vert<1/\sqrt{15}$. Let us call $\mathcal{Y}$ the square of $Y$ coordinates of the equilibrium points, $\bar{Y}_{1,2}$. It holds
\begin{equation*}
\begin{split}
\mathcal{Y}= & -\frac{\rho^4}{27\lambda^2}\Big( -10800{{\rho}}^{4}\sqrt{15}+441{\lambda}{{\rho}}^{2}\sqrt {15}-53{\lambda}\sqrt {15}+432{
	\lambda}\vert{\rho}\vert  \Big)  \Big( -54000{{\rho}}^{4}\sqrt {15}\\
& +3465{\lambda}{{\rho}}^{2}\sqrt {15}-349{\lambda}\sqrt{15}+2160{\lambda}\vert{\rho}\vert\Big).
\end{split}
\label{Eq15}
\end{equation*}
We have to verify whether there exist values of $\vert\rho\vert<1/\sqrt{15}$ such that $\mathcal{Y}\ge0$. Since $\lambda\in(0,1)$ we have
\[
\begin{split}
&-10800{{\rho}}^{4}\sqrt{15}+441{\lambda}{{\rho}}^{2}\sqrt {15}-53{\lambda}\sqrt {15}+432{\lambda}\vert{\rho}\vert<\lambda\big(-10800{{\rho}}^{4}\sqrt{15}+441{{\rho}}^{2}\sqrt {15}\\ & -53\sqrt {15}+432\vert{\rho}\vert\big)<0,
\end{split}
\]
and 
\[
\begin{split}
& -54000{{\rho}}^{4}\sqrt {15}+3465{\lambda}{{\rho}}^{2}\sqrt {15}-349{\lambda}\sqrt{15}+2160{\lambda}\vert{\rho}\vert <\lambda\big(-54000{{\rho}}^{4}\sqrt {15}+3465{{\rho}}^{2}\sqrt {15}\\
&-349\sqrt{15}+2160\vert{\rho}\vert \big)<0.
\end{split}
\]
It follows that $\mathcal{Y}<0$, $\forall \rho\in(0,1/\sqrt{15})$ and $\forall \lambda \in (0,1)$. Thus, the equilibrium points $\bar{E}_1$ and $\bar{E}_2$ never exist for the $J_2$-problem.

\subsubsection{Summary and comparison with previous works}
Here we summarise the results for the $J_2$-problem and we compare them with those  previously obtained by \citet{Coffey1986} and \citet{Palacian2007}.

\begin{figure}
	\centering
	\begin{subfigure}{0.4\textwidth}
		\includegraphics[width=\textwidth]{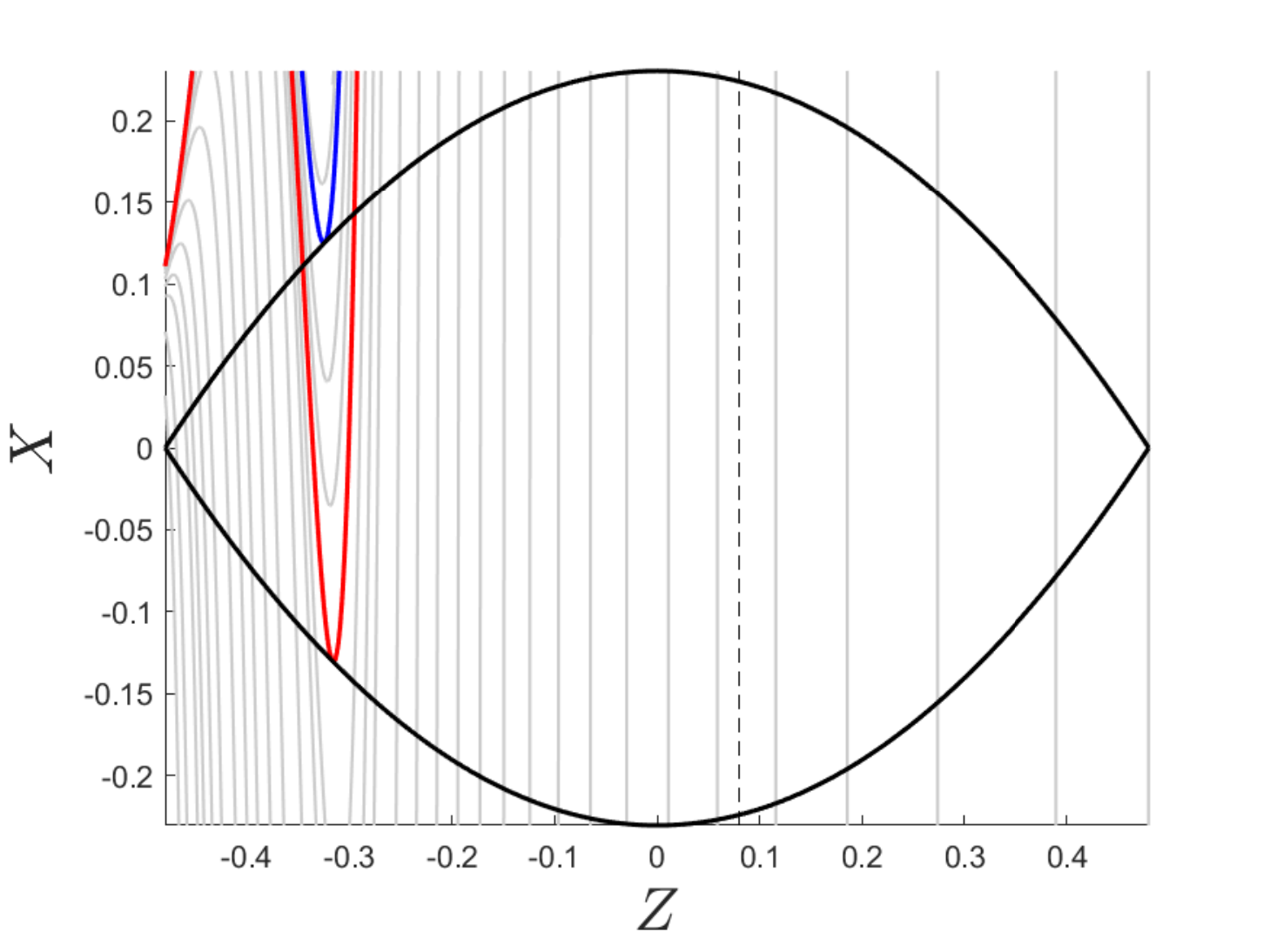}
	\end{subfigure}
	\begin{subfigure}{0.4\textwidth}
		\includegraphics[width=\textwidth]{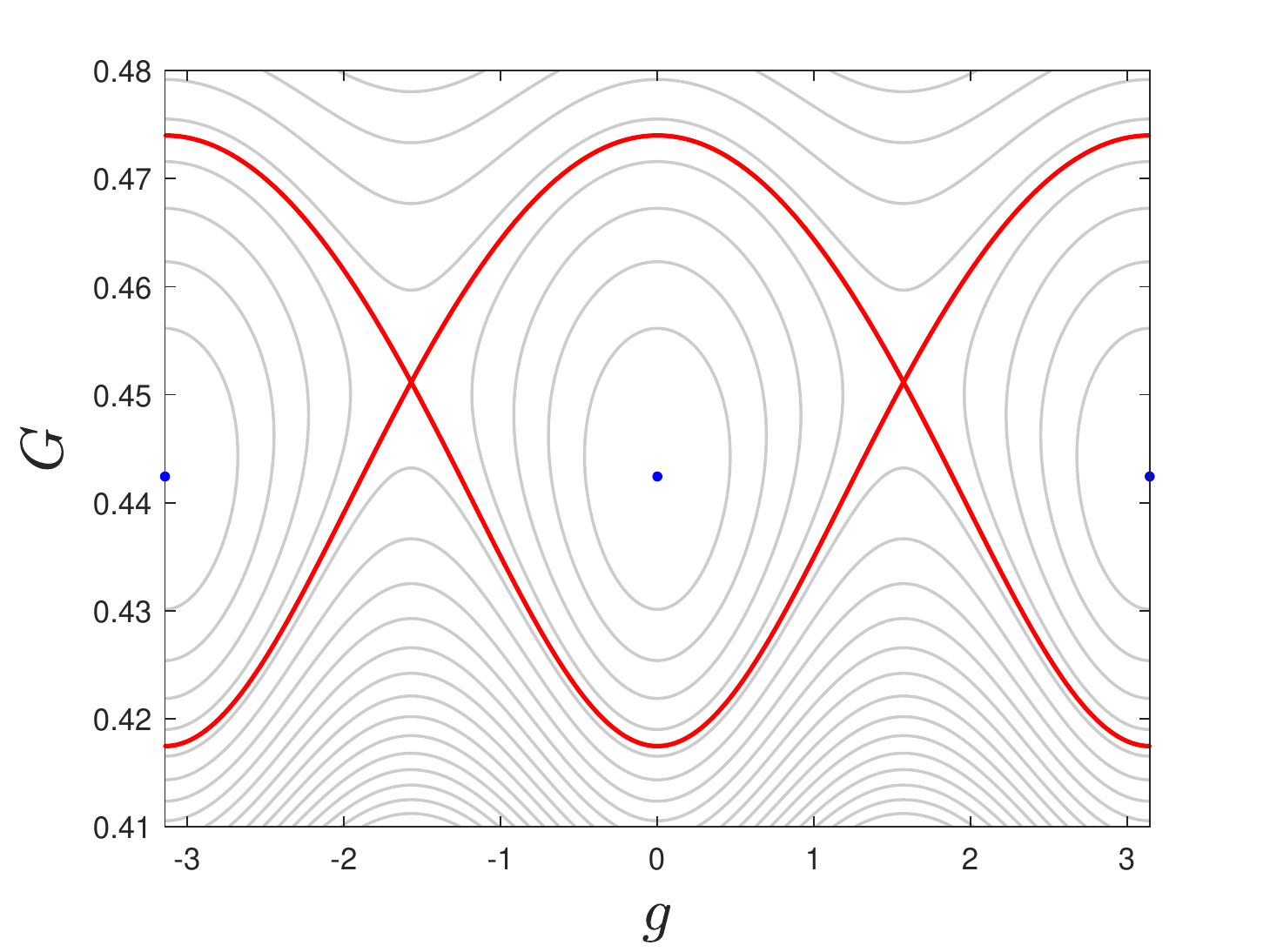}
	\end{subfigure}
	\caption{Level curves for the $\J2$-problem. Here, $\rho=0.2$ and $\lambda=0.001$. On the left, they are shown in the $(Z,X)$ plane. The black line represents the contour $\mathcal{C}$ of the \textit{lemon} space. The blue line represents the level curve tangent to $\mathcal{C}$ at the stable equilibrium point $E_3$, while the red one represents the level curve tangent to $\mathcal{C}$  at the unstable $E_4$. The dashed black line corresponds to $Z=\bar{Z}$, for which the level curves have a singularity. On the right, the level curves are shown in an enlargement of the $(g,G)$ plane surrounding the equilibrium points. The blue dots are the stable equilibrium points corresponding to $E_3$; the red curve is the separatrix of the equilibrium points at $g=\pm \frac{\pi}{2}$ corresponding to $E_4$.}
	\label{fig:j2ex}
\end{figure}

At $\vert\rho\vert=\rho_+$ and $\vert\rho\vert=\rho_-$, with $\rho_+$ and $\rho_-$ defined in \eqref{rhopiu_j2} and \eqref{rhomeno_j2}, there are two pitchfork bifurcations. In particular, we have that
\begin{itemize}
	\item for $\rho_+<\vert\rho\vert<1$ there exist only the equilibrium points $E_1$ and $E_2$ and they are stable;
	\item at $\vert\rho\vert=\rho_+$ there is a bifurcation: $E_2$ is degenerate and coincides with $E_3$, while $E_1$ is still stable; $E_3$ is an equilibrium point of type $E_+$;  
	\item for $\rho_-\le\vert\rho\vert<\rho_+$ there exist the equilibrium points $E_1$, $E_3$, which are stable, and $E_2$ which is unstable; 
	\item at $\vert\rho\vert=\rho_-$ there is a bifurcation: $E_2$ is degenerate and coincides with $E_4$ of type $E_-$, while $E_1$ and $E_3$ are still stable;   
	\item for $\vert\rho\vert<\rho_-$ there exist the equilibrium points $E_1$, $E_2$ and $E_3$, which are stable, and $E_4$, which is unstable. 
\end{itemize}
In Fig.\ref{fig:j2ex} (left panel), we show the level curves of the closed form in the $(Z,X)$ plane when $\vert\rho\vert<\rho_-$. The blue and red lines are tangent to the contour $\mathcal{C}$ of the lemon space respectively at the equilibrium points $E_3$ and $E_4$. Note that the blue line has a concavity larger than that of $\mathcal{C}$ at their tangency point as $E_3$ is stable (see \ref{careq_E3}). Also the concavity of the red line is larger than that of  $\mathcal{C}$ at their tangency point, which in this case implies that $E_4$ is unstable as follows from \eqref{careq_E4}. In the right panel of Fig.\ref{fig:j2ex} we show the level curves in an enlargement of the $(g,G)$ plane containing the equilibrium points. Here, $E_3$ corresponds to the stable equilibrium points at $g=0,\pi$ (blue dots). Instead, $E_4$ corresponds to the two unstable equilibrium points at $g=\pm\pi/2$: the separatrix is in red. By using \eqref{Gplus} and \eqref{Gminus}, we are able to compute the approximated values of $G_{\pm}$ of the equilibrium points: $G_+ = 0.4424$ and $G_- = 0.4512$. 

We remind that $\rho=H/L$. Through a transformation of variables and units we can determine the values of $\vert H\vert$ at which the bifurcations occur in the original dimensional system. Let us call them $H_+$ and $H_-$. From \eqref{rhoPseries} and \eqref{rhoMseries}, we obtain  
\begin{align*}
H_+ \sim & \frac{L}{\sqrt{5}}\left(1+\frac{1}{10} \frac{\J2 \mu^2 R_{P}^2}{L^4}-\frac{7}{200}\frac{\J2^2\mu^4R_{P}^4}{L^8}-\frac{7}{800}\frac{\J2^3\mu^6R_{P}^6}{L^{12}}+\mathcal{O}\left(\frac{\J2^4 \mu^8 R_{P}^8}{L^{16}}\right)\right), \\
H_-  \sim &  \frac{L}{\sqrt{5}}\,\left(1-\frac{1}{10}\frac{\J2 \mu^2 R_{P}^2}{L^4}+\frac{3}{40}\frac{\J2^2\mu^4R_{P}^4}{L^8}-\frac{299}{4000}\frac{\J2^3\mu^6R_{P}^6}{L^{12}}+\mathcal{O}\left(\frac{\J2^4 \mu^8 R_{P}^8}{L^{16}}\right)\right).
\end{align*}
With the same transformation, by using \eqref{Gplus} and \eqref{Gminus}, the values $G_{\pm}(L,H)$ for the two bifurcated families can be expressed as series in $J_2$.
In conclusion, by exploiting the $(X,Y,Z)$ variable and the geometrical approach we have recovered the results found in \citep{Coffey1986} and in \citep{Palacian2007}.

\subsection{The $J_4$-problem}
\label{sec:probj4}
We study now the zonal problem containing both the $J_2$ and the $J_4$ terms. From \eqref{KP}, \eqref{NFZ1}, \eqref{HZ21} and \eqref{NFZ2}, the closed form is
\begin{equation*}
\begin{split}
\mathcal{K}_{J_4}& =-\frac{\mu^2}{2 L^2}+\frac{\mu^4  J_2 R_{P}^2 (G^2 - 3H^2)}{4 G^5 L^3 }+\frac{3 \mu^6 J_2^2 R_{P}^4}{128 L^5 G^{11} }  \big[ -5 G^6 - 4 G^5 L \\
& \hspace{5mm}+ 24 G^3 H^2 L - 36 G H^4 L - 35 H^4 L^2 +  G^4 (18 H^2 + 5 L^2) \\
& \hspace{5mm}- 5 G^2 (H^4 + 2 H^2 L^2) +  2 (G^2 - 15 H^2) (G^2 - L^2) (G^2 - H^2) \cos 2 g \big]+\\
& \hspace{5mm} \frac{3\mu^6 J_4 R_{P}^4}{128 L^5 G^{11}} \big[ (3 G^4 - 30 G^2 H^2 + 35 H^4) (5 L^2 - 3 G^2) 
\\
& \hspace{5mm} - 10 (G^2 - 7 H^2) (L^2 - G^2) (G^2 - H^2)  \cos 2 g\big].
\end{split}
\end{equation*}
Let us set
\[
\jj4 = -\frac{J_4}{J_2^2}.
\]
After introducing it in the Hamiltonian, we adopt the same adimensional system and perform the same transformations described in Sect.\ref{sec:J2}. Also for this problem, we obtain a secular Hamiltonian in closed form with the structure of $\mathcal{K}$ in \eqref{normalformstructure}, with
\[ \begin{split} g(Z,\bm{a}) &=  \frac {-5{{\rho}}^{2}+2Z+1}{\sqrt{2} \left( {{\rho}}^{2}+2Z+1\right) ^{\frac{5}{2}}} -\frac{3\lambda}{16\sqrt{2}\left(\rho^{2}+2Z+1\right)^{\frac{11}{2}}}\Big( 40{Z}^{3}+ \left( -84{{\rho}}^{2}+20 \right) {Z}^{2} \\ & \quad + \left( -74{{\rho}}^{4}-44{{\rho}}^{2}-10 \right) Z-11{{\rho}}^{6}+273{{\rho}}^{4}-{{\rho}}^{2}-5 \\ & \quad +4\sqrt {2\,{{\rho}}^{2}+4Z+2} \left(-5{{\rho}}^{2}+2Z+1\right) ^{2}\Big) -  \frac{3\lambda j_4}{16\sqrt{2}\left(\rho^{2}+2Z+1\right)^{\frac{11}{2}}}\Big(-72{Z}^{3}\\ & \quad +12 \left(51{{\rho}}^{2}+1\right) {Z}^{2} + 3\left( -58 {{\rho}}^{4}-156{{\rho}}^{2}+22\right) Z-249{{\rho}}^{ 6}+743{{\rho}}^{4}-387{{\rho}}^{2}+21 \Big), \\
 f(Z,\bm{a}) & =  -\frac{3}{2}\,\lambda {\frac { \left( -29\,{{\rho}}^{2}+2\,Z+1 \right) }{ \sqrt{2}\left( {{\rho}}^{2}+2\,Z+1 \right) ^{\frac{11}{2}}}}+\frac{15}{2}\lambda j_4{\frac { \left( -13{{\rho}}^{2}+2Z+1 \right) }{ \sqrt{2}\left( {	{\rho}}^{2}+2Z+1 \right) ^{\frac{11}{2}}}},  \end{split}\]
and $\bm{a}=(\rho;\lambda,\jj4)$.

In the following, we discuss the dynamical behaviour of the problem for $\vert\rho\vert\in(0,1)$ and $j_4\in[-6,6]$. This range of $\jj4$ is coherent with the book-keeping scheme used for the computation of the normalised Hamiltonian in Sect.\ref{sec:NF} and its extent allows us to include Earth and Mars. Our results are both the outcomes of analytical considerations and numerical studies. In this case, to simplify the analysis, we fix the value of $\lambda$, taking $\lambda=0.001$.  We expect the main features of the dynamics to be qualitatively similar also for other values of $\lambda$ sufficiently small. 

First, we analyse the stability of the equilibrium points $E_1$ and $E_2$. Then, we discuss the existence of the equilibrium points of type $E_+$ and $E_-$ and the existence of $\bar{E}_1$ and $\bar{E}_2$. Finally, we discuss their stability and we trace a bifurcation diagram. We find out that the stability of $E_1$ depends on both $j_4$ and $\rho$. In particular, there are ranges of $j_4$ in which pitchfork bifurcations occur: they affect the stability of $E_1$ and can give rise to either a stable equilibrium point of type $E_-$ or an unstable equilibrium point of type $E_+$. For each $j_4\in[-6,6]$, we also have pitchfork bifurcations affecting the stability of $E_2$. We determine the values $\rho_+$ and $\rho_-$ of $\vert\rho\vert$ at which they occur. As in the $J_2$ problem, $E_2$ is unstable if the value of $\vert\rho\vert$ lies between $\rho_-$ and $\rho_+$, otherwise it is stable. Following the pitchfork bifurcation occurring at  $\vert\rho\vert=\rho_+$, an equilibrium point of type $E_+$ is generated; similarly, for $\vert\rho\vert<\rho_-$ there exists an equilibrium point of type $E_-$. The stability of these points depends on both $j_4$ and $\rho$. An interesting result is that there are ranges of $j_4$ where their stability changes as a consequence of further pitchfork bifurcations, which affect the existence of the equilibrium points $\bar{E}_1$ and $\bar{E}_2$. We show that, when existing, these last ones are always unstable. Finally, we find out that for some $j_4$ saddle-node bifurcations also occur. They can give rise to either a pair of equilibrium points of type $E_+$ or a pair of equilibrium points of type $E_-$. Independently of the type, one of the point of the pair is stable, while the other is unstable. 

We recall once again that, for the selected planet and the fixed value of the semi-major axis (i.e. for the given $j_4$ and $\lambda$), the values of $\rho$ and $G$ of one considered equilibrium point allow us to determine the eccentricity and the inclination of the family of orbits represented by the equilibrium point itself. In the following, we perform a general analysis not taking into account some physical limitations, for example, the fact that the orbits corresponding to a given equilibrium point may be collisional.

\subsubsection{Stability of $E_1$}
\label{sec:probj4_e1}
The stability of $E_1$ depends on the sign of the product $s_+(-\ZMAX,\bm{a})s_-(-\ZMAX,\bm{a})$. For the $J_4$-problem, we have
\[
s_+(-\ZMAX,\bm{a}) = 2\,\frac{8\rho^4 + \lambda \left(31 + 12\vert\rho\vert -7\rho^2 -5\jj4(3\rho^2 - 7)\right)}{\lambda(15\jj4-7)},
\]
and 
\[
s_-(-\ZMAX,\bm{a}) = 4\,\frac{4\rho^4 + \lambda \left(6(2+\vert\rho\vert)-5\jj4(3\rho^2 - 5)\right)}{\lambda(15\jj4-7)}.
\]
For $\jj4\ge-31/35$, function $s_+(-\ZMAX,\bm{a})$ has no real zeros; instead, for $\jj4<-31/35$ there exists a real value of $\vert\rho\vert$, $\vert\rho\vert={\rho}_{\vartriangle}$, solving equation $s_+(-\ZMAX,\bm{a})=0$. 
Similarly, if $\jj4<-12/25$ there exists one real value of $\vert\rho\vert$, $\vert\rho\vert={\rho}_{\triangledown}$, which is a zero of $s_-(-\ZMAX,\bm{a})$. Thus, if $\jj4\ge-12/25$ $E_1$ is always stable. Instead, if $-31/35\le\jj4<-12/25$, 
\begin{itemize}
	\item for ${\rho}_{\triangledown}<\vert\rho\vert<1$, $E_1$ is stable; 
	\item for $\vert\rho\vert<{\rho}_{\triangledown}$, $E_1$ is unstable.
\end{itemize}
Finally, if $\jj4<-31/35$, it holds ${\rho}_{\triangledown}>{\rho}_{\vartriangle}$ so that
\begin{itemize}
	\item for ${\rho}_{\triangledown}<\vert\rho\vert<1$ and $0<\vert\rho\vert<{\rho}_{\vartriangle}$, $E_1$ is stable;
	\item for ${\rho}_{\vartriangle}<\vert\rho\vert<\rho_{\triangledown}$, $E_1$ is unstable.
\end{itemize}
At $\vert\rho\vert={\rho}_{\vartriangle}$, the degenerate $E_1$ coincides with an equilibrium point of type $E_+$. At $\vert\rho\vert={\rho}_{\triangledown}$, it coincides with an equilibrium point of type $E_-$. In the following, we call $j_{4_{\rm bif 6}}= -12/25$ and $j_{4_{\rm bif 9}}= -31/35$.

\subsubsection{Stability of $E_2$}
The stability of $E_2$ depends on the solutions of equation \eqref{careq_E2}. We have
\[
s_+(\ZMAX,\bm{a}) = -\frac{1}{2}\,\frac{16 - 80 \rho^2 - \lambda 
	\left((420\jj4+425)\rho^4-(280\jj4+146)\rho^2+20\jj4+9 \right)}{\lambda\left((35\jj4-15)\rho^2-5\jj4+1\right)},
\]
and
\[
s_-(\ZMAX,\bm{a}) = -\frac{1}{2}\,\frac{16 - 80 \rho^2 - \lambda 
	\left((560\jj4+365)\rho^4-(440\jj4+82)\rho^2+40\jj4+5 \right)}{\lambda\left((35\jj4-15)\rho^2-5\jj4+1\right)}.
\]
For $\lambda$ sufficiently small, $\forall \jj4\in[-6,6]$ $s_+(\ZMAX,\bm{a})$ possesses two real zeros at $\rho=\pm \rho_+$ with 
\begin{equation}
\rho_+ = \sqrt{ 
	\frac{\lambda(140\jj4+73)-40+4\sqrt{2}\sqrt{50 + (30 - 140 \jj4)\lambda + (47 + 255 \jj4 +350 \jj4^2) \lambda^2}}{5(84\jj4+85)\lambda}},
\label{rhoP_j4}
\end{equation}
and $s_-(\ZMAX,\bm{a})$ possesses two real zeros at $\rho=\pm \rho_-$ with 
\begin{equation}
\rho_- = \sqrt{\frac{\lambda(220\jj4+41)-40+4\sqrt{100 + (160 - 540 \jj4)\lambda - (9 - 40 \jj4 -1625 \jj4^2) \lambda^2}}
	{5(112\jj4+73)\lambda}}.
\label{rhoM_j4}
\end{equation}
More manageable expressions are given by the series expansions
\[
\rho_+ = \frac{1}{\sqrt{5}}\, \left(1 + \frac{1+6\jj4}{10}  \lambda - \frac{7+20\jj4-132\jj4^2}{200}  \lambda^2  \right) 
+\mathcal{O}\left(\lambda^3\right),\]
and
\[
\rho_- = \frac{1}{\sqrt{5}}\, \left(1 - \frac{1-8\jj4}{10}  \lambda + \frac{15-166\jj4+368\jj4^2}{200}  \lambda^2  \right) 
+\mathcal{O}\left(\lambda^3\right).\]
At  first order they coincide with those found by \citet{Coffey1994}. For $\jj4=j_{4_{\rm bif 1}}$, with 
\begin{equation}
j_{4_{\rm bif 1}}= 1 - \frac{14}5 \lambda + \frac{1239}{50} \lambda^2 +\mathcal{O}\left(\lambda^3\right) , 
\label{j4V}
\end{equation}
it holds $\rho_+=\rho_-$; if $\jj4>j_{4_{\rm bif 1}}$, $\rho_->\rho_+$, while for $\jj4<j_{4_{\rm bif 1}}$, $\rho_-<\rho_+$. As in the $\J2$-problem, when the value of $\vert\rho\vert$ is between $\rho_-$ and $\rho_+$ $E_2$ is unstable; at either $\vert\rho\vert=\rho_+$ or $\vert\rho\vert=\rho_-$, it is degenerate and coincides respectively with an equilibrium point of type $E_+$ and $E_-$. For all the other values of $\vert\rho\vert$ $E_2$ is stable. 

\subsubsection{Existence of the equilibrium points of type $E_+$ and $E_-$}
\label{sec:probj4_EPEM}
We use here the same strategy applied for the $\J2$-problem. After the change of variables $Z\mapsto G$, we obtain
\[
s_+\left(G^2-\frac{1+\rho^2}{2};\bm{a}\right)=\frac{\hat{S}_+(G;\bm{a})}{4G^2\lambda(5G^2\jj4-35\jj4\rho^2-G^2+15\rho^2)},
\]
with
\[
\begin{split}
\hat{S}_+ = &  (315G^2\jj4+225G^2-360G-1155\jj4-715)\lambda\rho^4-2G^2\big(80G^4+\lambda(35G^2\jj4\\
& +49G^2-96G-315\jj4-99)\big)\rho^2+G^4\big(32G^4+\lambda(-5G^2\jj4-15G^2
-24G-35\jj4+21)\big).
\end{split}
\]
We have that $\hat{S}_+=0$ for $\rho^2= \hat{\rho}^2_{E_{+_{1,2}}}$, where
\[
\hat{\rho}^2_{E_{+_{1,2}}} = \frac{G^2}{5\lambda}\,\frac{\hat{A}_+\pm\sqrt{\hat{B}_+}}{\hat{C}_+},
\]
\begin{align*}
\hat{A}_+ = &80G^4+ \lambda\left((35G^2-315)\jj4+49G^2-96G-99\right), \\
\hat{B}_+ = &  \frac{\hat{A}_+^2-5\lambda\hat{C}_+\hat{D}_+}{16},\\
\hat{C}_+ = & (63G^2-231)\jj4+45G^2-72G-143, \\ \hat{D}_+ = & {32}G^4+  \lambda\left((-5G^2- 35)\jj4-15G^2-24G+21\right).
\end{align*}
For $\lambda$ sufficiently small, it turns out that $0<\hat{\rho}^2_{E_{+_{2}}}<G^2$, $\forall G\in(0,1]$ and $\forall \jj4\in[-6,6]$. For $G=1$ it holds $\hat{\rho}^2_{E_{+_{2}}}=\rho_+$. Moreover, we numerically verified that $\hat{\rho}^2_{E_{+_{2}}}$ is increasing with respect to $G$. 
Consequently, for each $\vert\rho\vert\in(0,\rho_+)$ there exists one equilibrium point, $E_3$, which coincides with $E_2$ for $\vert\rho\vert=\rho_+$.  By applying the same perturbation method used in Section \ref{sec:j2prob_EpEm}, we can determine the value of $G$ corresponding to $E_3$. At third order in $\lambda$ we obtain
\begin{align*}
G_+ & =  \sqrt{5}\rho +  \frac{-5 -7\jj4+ 20 \rho^2+5\jj4\rho^2}{50 \sqrt{5} \rho^3} \lambda + 
\frac{1}{25000 \sqrt{5} \rho^7}\big(70 -384\jj4 -392\jj4^2 -30\sqrt{5}\rho\\
& \quad -42\sqrt{5}\jj4\rho +405 \rho^2 +1215\jj4\rho^2+70\jj4^2\rho^2+ 120 \sqrt{5} \rho^3+30\sqrt{5}\jj4\rho^3 - 500 \rho^4\\
& \quad +475\jj4\rho^4+150\jj4^2\rho^4\big)
\lambda^2 +\frac{1}{25000000 \sqrt{5} \rho^{11}}\big(1765 - 16569 \jj4 - 70021 \jj4^2 - 60711 \jj4^3\\
& \quad - 1680 \sqrt{5} \rho - 9072 \sqrt{5} \jj4 \rho - 9408 \sqrt{5} \jj4^2 \rho - 20385 \rho^2 + 
86215 \jj4 \rho^2 + 189665 \jj4^2 \rho^2 \\
& \quad - 20335 \jj4^3 \rho^2 + 
9720 \sqrt{5} \rho^3 + 29160 \sqrt{5} \jj4 \rho^3 + 1680 \sqrt{5} \jj4^2 \rho^3 + 
61550 \rho^4 - 19900 \jj4 \rho^4 \\
& \quad + 265125 \jj4^2 \rho^4 + 53375 \jj4^3 \rho^4 - 
12000 \sqrt{5} \rho^5 + 11400 \sqrt{5} \jj4 \rho^5 + 3600 \sqrt{5} \jj4^2 \rho^5 	- 
33000 \rho^6 	\\ & \quad
- 316750 \jj4 \rho^6 - 99625 \jj4^2 \rho^6 - 5625 \jj4^3 \rho^6\big)\lambda^3.
\end{align*}
The other solution  $\hat{\rho}^2_{E_{+_{1}}}$ is only admissible for some values of $\jj4$. The analysis of the 
$\hat{\rho}^2_{E_{+_{1}}}$ is complex and we are forced to fix the value of $\lambda$ at $0.001$. Anyway, we expect similar outcomes for all values of $\lambda$ sufficiently small.
For $\jj4>j_{4_{\rm bif 2}}$, where $j_{4_{\rm bif 2}}\sim 0.5695$, there exists a range of values of $G$ such that $0< \hat{\rho}^2_{E_{+_{1}}}<G^2$. The function $\hat{\rho}^2_{E_{+_{1}}}$ is not monotone with respect to $G$. Let us set 
\begin{equation}
\rho_{\blacktriangle} = \sqrt{\max_{G} \hat{\rho}^2_{E_{+_{1}}}(G;\jj4)}.
\label{rhoSaddleNodeP}
\end{equation}
For $\vert\rho\vert=\rho_{\blacktriangle}$ there exists one equilibrium point $E_5$ of type $E_+$. Instead, for  $\vert\rho\vert<\rho_{\blacktriangle}$ there are multiple equilibrium points of type $E_+$; they are typically two and we call them $E_7$ and $E_9$. Also for $\jj4<j_{4_{\rm bif 9}}$, it holds $\hat{\rho}^2_{E_{+_{1}}}> 0$; through a numerical study, we observed that the function is increasing with $G$ and that $\hat{\rho}^2_{E_{+_{1}}}\le G^2$ up to a certain value of $G$ lower than $1$, for which it holds $\hat{\rho}^2_{E_{+_{1}}}=\rho_{\vartriangle}^2$. Thus, for $\jj4<j_{4_{\rm bif 9}}$ and $\vert\rho\vert\in(0,\rho_{\vartriangle})$ there exists an equilibrium point $E_{11}$, which coincides with $E_1$ for $\vert\rho\vert = \rho_{\vartriangle}$. 

Concerning the equilibrium points of type $E_-$, we have 
\[
s_-(G^2-\frac{1+\rho^2}{2};\bm{a})=\frac{\hat{S}_-(G;\bm{a})}{4G^2\lambda(5G^2\jj4-35\jj4\rho^2-G^2+15\rho^2)},
\]
with
\[
\begin{split}
\hat{S}_- =&(1575G^2\jj4-315G^2-360G-2695\jj4-55)\lambda\rho^4-2G^2(80G^4+\lambda(595G^2\jj4\\
&-175G^2-96G-1035\jj4+189))\rho^2+G^4(32G^4+\lambda(95G^2\jj4-35G^2-24G-175\jj4+49)).
\end{split}
\]
$\hat{S}_-=0$ for $\rho^2= \hat{\rho}^2_{E_{-_{1,2}}}$, where
\[
\hat{\rho}^2_{E_{-_{1,2}}} = \frac{G^2}{5\lambda}\,\frac{\hat{A}_-\pm4\sqrt{\hat{B}_-}}{\hat{C}_-},
\]
\begin{align*}
\hat{A}_- = & 80G^4 + \lambda\left((595G^2-1035)j_4 -175G^2-96G+189\right), \\
\hat{B}_- = & \frac{\hat{A}_-^2-5\lambda\hat{C}_-\hat{D}_-}{16},\\
\hat{C}_- = & (315G^2-539)j_4-63G^2-72G-11, & \\
\hat{D}_- = &  {32}G^4 + \lambda\left((95G^2-175)j_4-35G^2-24G+49\right).
\end{align*}
For $\lambda$ sufficiently small, solution $\hat{\rho}^2_{E_{-_{2}}}$ is admissible $\forall G\in(0,1]$ and $\forall j_4\in[-6,6]$. For $G=1$ we have $\hat{\rho}^2_{E_{-_{2}}}=\rho_-$. Moreover, we numerically verified that ${\partial \hat{\rho}^2_{E_{-_{2}}}}/{\partial G}> 0$.
Thus, for each $\vert\rho\vert\in(0,\rho_-)$ there exists the equilibrium point $E_4$ which coincides with $E_2$ for $\vert\rho\vert=\rho_-$ and whose value is
\begin{align*}
G_- & = 	\sqrt{5} \rho + \frac{125 {\jj4} \rho ^2-41 {\jj4}-35 \rho ^2+9}{100 \sqrt{5} \rho ^3}\lambda +
\frac{1}{100000 \sqrt{5} \rho ^7}\Big(-103125 \jj4^2 \rho ^4+90450 \jj4^2 \rho ^2\\
& \quad -18573 {\jj4}^2+68250
\jj4 \rho ^4+1500 \sqrt{5} \jj4 \rho ^3-54320 \jj4 \rho ^2-492
\sqrt{5} \jj4 \rho +10022 \jj4\\
& \quad -11025 \rho ^4-420 \sqrt{5} \rho ^3+7910
\rho ^2+108 \sqrt{5} \rho -1305\Big)\lambda^2 \\
& \quad
-\frac{1}{100000000 \sqrt{5} \rho ^{11}}\Big(-106171875 \jj4^3 \rho ^6 +113728125 \jj4^2 \rho ^6+2475000 \sqrt{5}
\jj4^2 \rho ^5\\
& \quad+168643125 \jj4^3 \rho ^4-168503125 \jj4^4-2170800 \sqrt{5} \jj4^2 \rho ^3 -80521425 \jj4^3 \rho ^2+75097235
\jj4^2 \rho ^2 \\
& \quad +445752 \sqrt{5} \jj4^2 \rho +12014271
\jj4^3-10434331 \jj4^2-39598125 \jj4 \rho ^6 -1638000 \sqrt{5}
\jj4 \rho ^5 \\
& \quad +54647375 \jj4 \rho ^4+1303680 \sqrt{5} \jj4\rho
^3-22678075 \jj4 \rho ^2 -240528 \sqrt{5} \jj4 \rho +2929289
\jj4  \\
& \quad +4501875 \rho ^6+264600 \sqrt{5} \rho ^5-5775175 \rho ^4-189840 \sqrt{5}
\rho ^3+2226905 \rho ^2 +31320 \sqrt{5} \rho \\
& \quad -267309\Big)\lambda^3.
\end{align*}

Concerning the other solution $\hat{\rho}^2_{E_{-_{1}}}$, its admissibility depends on $\jj4$. {Here too, we set $\lambda = 0.001$}. Through an analysis similar to the one done for $\hat{\rho}^2_{E_{+_{1}}}$, we reach the following conclusions:
\begin{itemize}
	\item	for $\jj4>j_{4_{\rm bif 5}}$, with $j_{4_{\rm bif 5}}\sim 0.2755$, at $\vert\rho\vert=\rho_{\blacktriangledown}$ there exists one equilibrium solution $E_6$ of type $E_-$, while for  $\vert\rho\vert<\rho_{\blacktriangledown}$ there exist typically two equilibrium solutions of type $E_-$ which we call $E_{8}$ and $E_{10}$; here,
	\begin{equation}
	\rho_{\blacktriangledown} = \sqrt{\max_{G} \hat{\rho}^2_{E_{-_{1}}}(G;\jj4)};
	\label{rhoSaddleNodeM}
	\end{equation}
	\item for $\jj4<j_{4_{\rm bif 6}}$ and for $\vert\rho\vert<\rho_{\triangledown}$ there exists an equilibrium point $E_{12}$, which coincides with $E_1$ for $\vert\rho\vert=\rho_{\triangledown}$.
\end{itemize}



\subsubsection{About the existence of $\bar{E}_1$ and $\bar{E}_2$}
\label{sec:probj4_E1bE2b}
If existing, the equilibrium points $\bar{E}_1$ and $\bar{E}_2$ have coordinates $(\bar{X},\bar{Y}_{1,2},\bar{Z})$, $\bar{Y}_1^2=\bar{Y}_2^2=\mathcal{Y}$, where 
\[  \begin{split}  \bar{X} = &  -\frac{\rho^2}{\lambda\left(4375j_4^5-5375 j_4^4+2550 j_4^3-590  j_4^2+67 j_4-3\right)}\Big(-2000\rho^4(j_4-1)(7j_4-3)^3\\  & +\lambda\big( (5 j_4-1)(55125 j_4^4\rho^2-28700 j_4^3\rho^2-14875 j_4^4-23130 j_4^2\rho^2 +3350 j_4^3\\  & +17460j_4\rho^2+7000 j_4^2-2835\rho^2-2950 j_4+307)\\  & +48\sqrt{\frac{7 j_4-3}{5j_4-1}}\sqrt{5}\vert \rho\vert (125 j_4^4-250 j_4^3+160 j_4^2-38j_4+3)\big)  \Big),  \end{split} \]
\begin{equation*}
\bar{Z} = \frac{65j_4\rho^2-29\rho^2-5j_4+1}{5j_4-1}, \qquad \mathcal{Y} = (-\bar{Z}^2+\ZMAX^2)^2-\bar{X}^2.
\label{Zbarj4}
\end{equation*}
$\bar{E}_1$ and $\bar{E}_2$ exist if 
\begin{align}
& \bar{Z}\in[-\ZMAX,\ZMAX], \label{cond1}\\
& \mathcal{Y}\ge0. \label{cond2}
\end{align}
Let us remark that when $\bar{Y}_1=\bar{Y}_2=0$ and $\bar{X}=\hat{X}(\bar{Z};\ZMAX)$, $\bar{E}_1$ and $\bar{E}_2$ coincide with an equilibrium point of type $E_+$, i.e. they correspond to zeros of $s_+(Z;\bm{a})$ defined in \eqref{gen_case1_fun}. We call $\rho_{\diamond}$, $\rho_{\diamond,\rm bis}$  the values of $\vert\rho\vert$ for which this occurs.  Similarly, when $\bar{Y}_1=\bar{Y}_2=0$ and $\bar{X}=-\hat{X}(\bar{Z};\ZMAX)$ $\bar{E}_1$ and $\bar{E}_2$ coincide with an equilibrium point of type $E_-$.
In this case, we call $\rho_{\square,}$, $\rho_{\square, \rm bis}$ the corresponding values of $\vert\rho\vert$.

Let us set $\lambda = 0.001$. If either $\jj4\in [j_{4_{\rm bif 1}},j_{4_{\rm bif 4}})$, with $j_{4_{\rm bif 4}}\sim 0.546$ or $\jj4\le j_{4_{\rm bif 7}}$, with $j_{4_{\rm bif 7}}\sim-0.4840$, there exists an interval of values of $\vert\rho\vert<\frac{5\jj4-1}{7\jj4-3}$ for which both conditions \eqref{cond1} and \eqref{cond2} are fulfilled. In particular, through a numerical study we obtain that 

\begin{itemize}
	\item in the range $j_{4_{\rm bif 3}}<\jj4\le j_{4_{\rm bif 1}}$, with $j_{4_{\rm bif 3}}\sim 0.552$, $\bar{E}_1$ and $\bar{E}_2$ exist for  $\vert\rho\vert\in[\rho_{\diamond},\rho_{\square}]$; 
	\item for  $j_{4_{\rm bif 4}}<\jj4\le j_{4_{\rm bif 2}}$ $\bar{E}_1$ and $\bar{E}_2$ exist for  $\vert\rho\vert\in(0,\rho_{\square}]$;
	\item in the range $j_{4_{\rm bif 8}}<\jj4\le j_{4_{\rm bif 7}}$, with $j_{4_{\rm bif 8}}\sim-0.4886$, $\bar{E}_1$ and $\bar{E}_2$ exist for $\vert\rho\vert\in[\rho_{\square,\rm bis},\rho_{\square}]$; 
	\item for $j_{4_{\rm bif 10}}<\jj4\le j_{4_{\rm bif 7}}$, with $ j_{4_{\rm bif 10}}\sim-1.3454$,  $\bar{E}_1$ and $\bar{E}_2$  exist for $\vert\rho\vert\in(0,\rho_{\square}]$; 
	\item in the range $j_{4_{\rm bif 11}}<\jj4\le j_{4_{\rm bif 10}}$, with $j_{4_{\rm bif 11}}\sim-1.3533$, $\bar{E}_1$ and $\bar{E}_2$ exist for $\vert\rho\vert\in[\rho_{\diamond},\rho_{\square}]$ and for $\vert\rho\vert\in(0,\rho_{\diamond, \rm bis}]$; 
	\item for $\jj4\le j_{4_{\rm bif 11}}$ $\bar{E}_1$ and $\bar{E}_2$ exist for $\vert\rho\vert\in[\rho_{\diamond},\rho_{\square}]$.
\end{itemize} 
We numerically verified that the equilibrium point of type $E_+$ coinciding with $\bar{E}_1$ and $\bar{E}_2$ at $\vert\rho\vert=\rho_{
	\diamond}$ and $\vert\rho\vert=\rho_{
	\diamond, \rm bis}$ is $E_3$. Similarly, we also verified that 
at $\vert\rho\vert=\rho_{
	\square}$ and $\vert\rho\vert=\rho_{
	\square, \rm bis}$ $\bar{E}_1$ and $\bar{E}_2$ coincide with $E_4$. Thus, $\rho_{\diamond}<\rho_+$ and $\rho_{\square}<\rho_-$.

\subsubsection{Stability analysis and bifurcation diagram}
In the following we discuss the evolution of the dynamics. We set $\lambda=0.001$, but we expect similar outcomes for all values of $\lambda$ sufficiently small. 
\begin{figure}
	\centering
	\begin{subfigure}{0.55\textwidth}
		\includegraphics[width=1\textwidth]{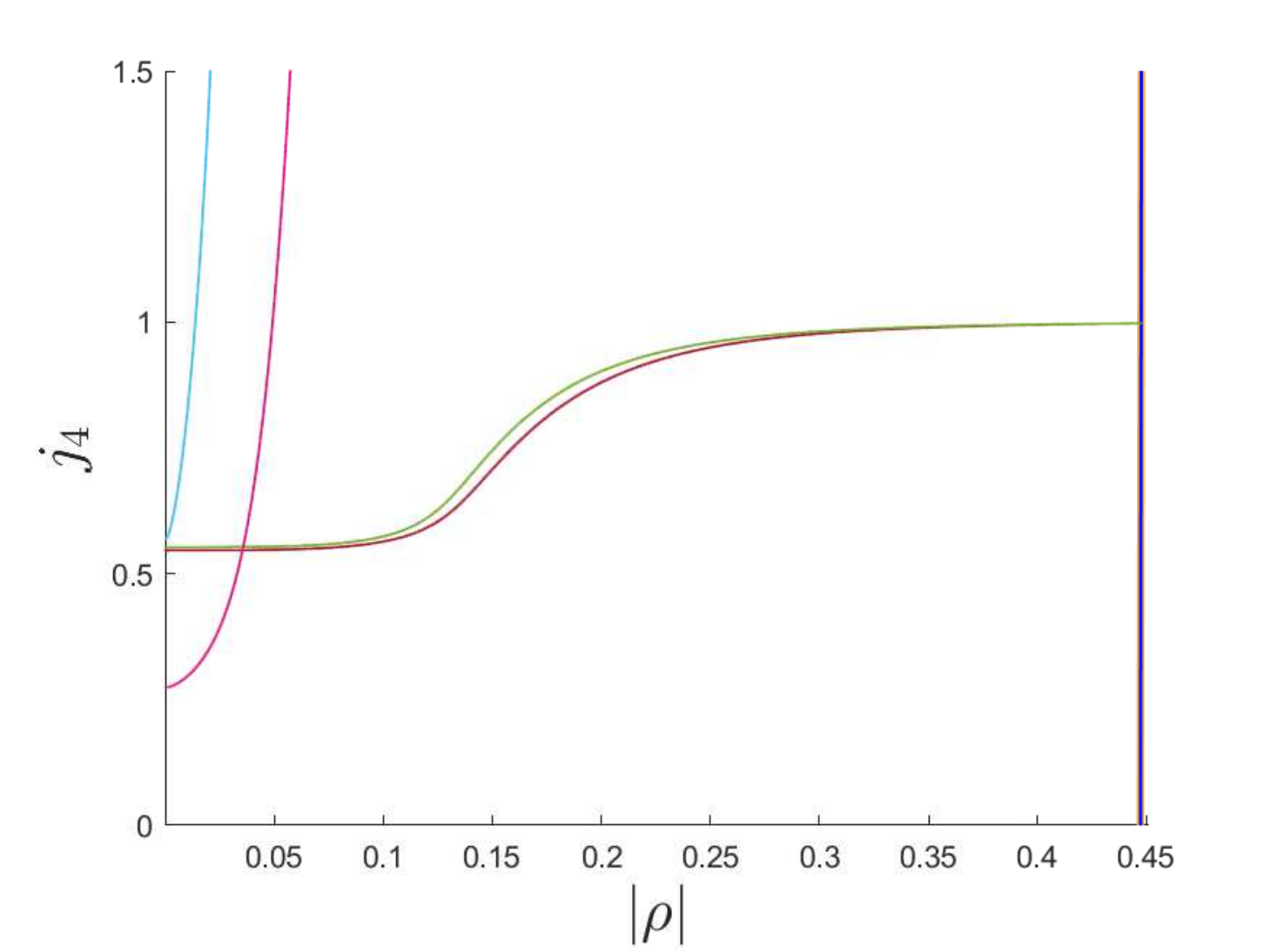}
	\end{subfigure}
	\begin{subfigure}{0.55\textwidth}
		\includegraphics[width=1\textwidth]{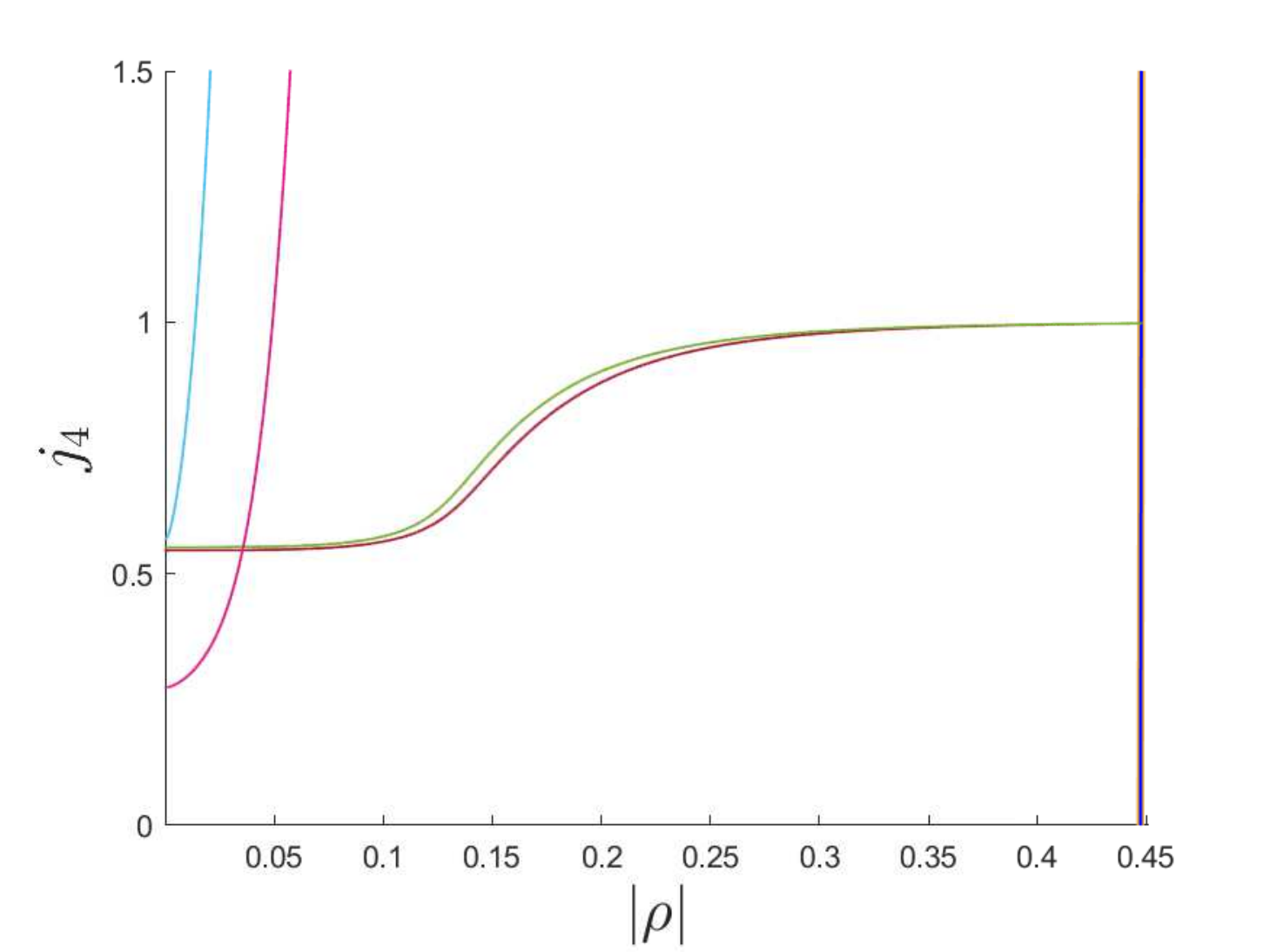}
	\end{subfigure}
	\begin{subfigure}{0.55\textwidth}
		\includegraphics[width=\textwidth]{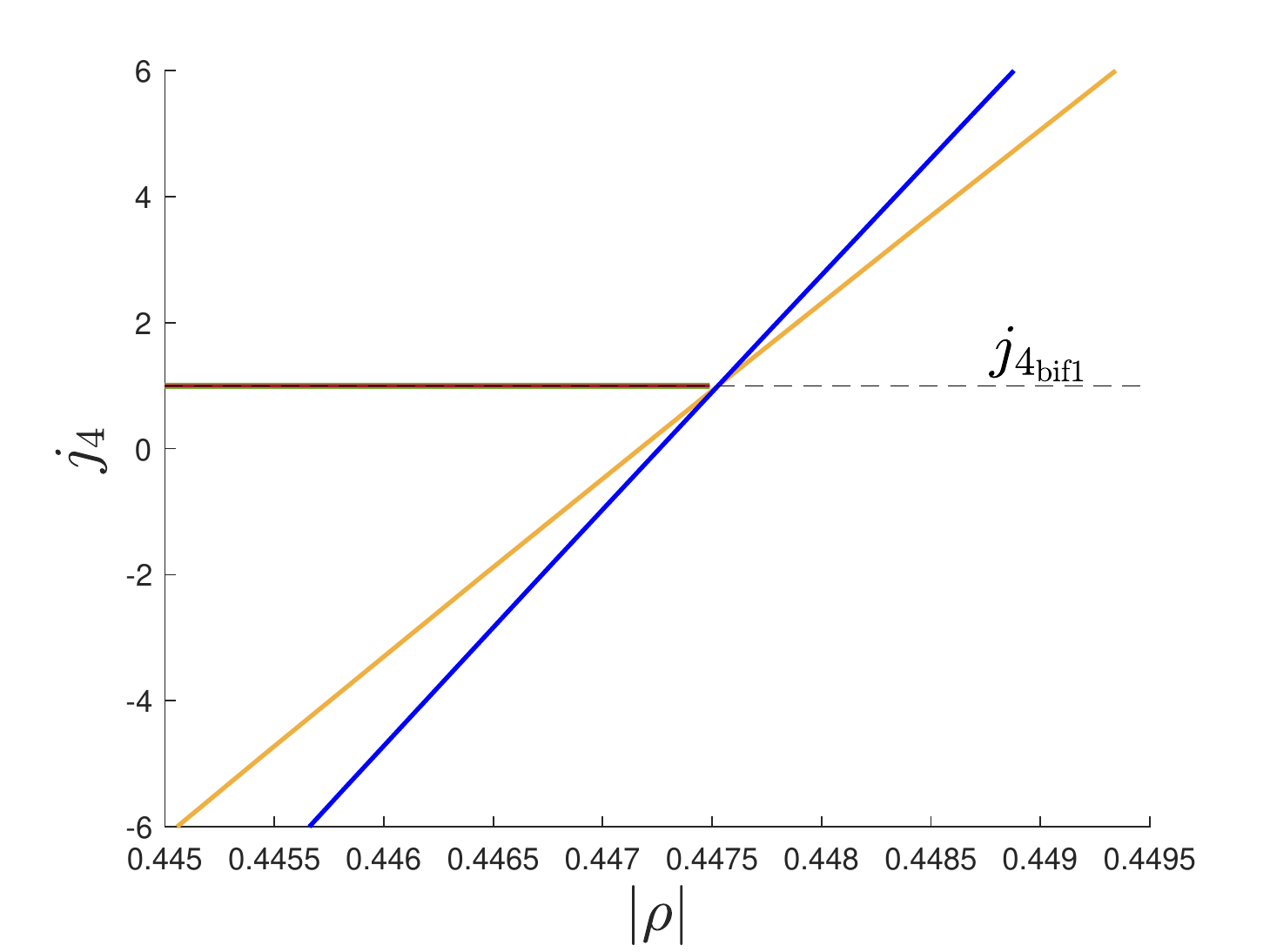}
	\end{subfigure}
	\caption{In the upper panel we show the bifurcation diagram for the $J_4$-problem with $\lambda=0.001$. The lower panels show two enlargements of the diagram. Further enlargements of interesting regions are shown in Fig.\ref{fig: bifdiagramlambda0001_en}. The blue and the orange lines represent respectively $\rho_+$ defined in \eqref{rhoP_j4} and $\rho_-$ defined in \eqref{rhoM_j4}; the purple and the yellow line represent respectively $\rho_{\vartriangle}$ and $\rho_{\triangledown}$, defined in Section \ref{sec:probj4_e1}; the light-blue line and the pink line represent respectively $\rho_{\blacktriangle}$, defined in \eqref{rhoSaddleNodeP}, and $\rho_{\blacktriangledown}$, defined in \eqref{rhoSaddleNodeM}; finally the green line, the dark-red line, the light-green line and the red line represent respectively $\rho_{\diamond}$, $\rho_{\square}$, $\rho_{\diamond, \rm bis}$ and $\rho_{\square, \rm bis}$, defined in Section \ref{sec:probj4_E1bE2b}.}
	\label{fig: bifdiagramlambda0001}
\end{figure} 

\begin{figure}
	\centering
	\begin{subfigure}{0.325\textwidth}
		\includegraphics[width=\textwidth]{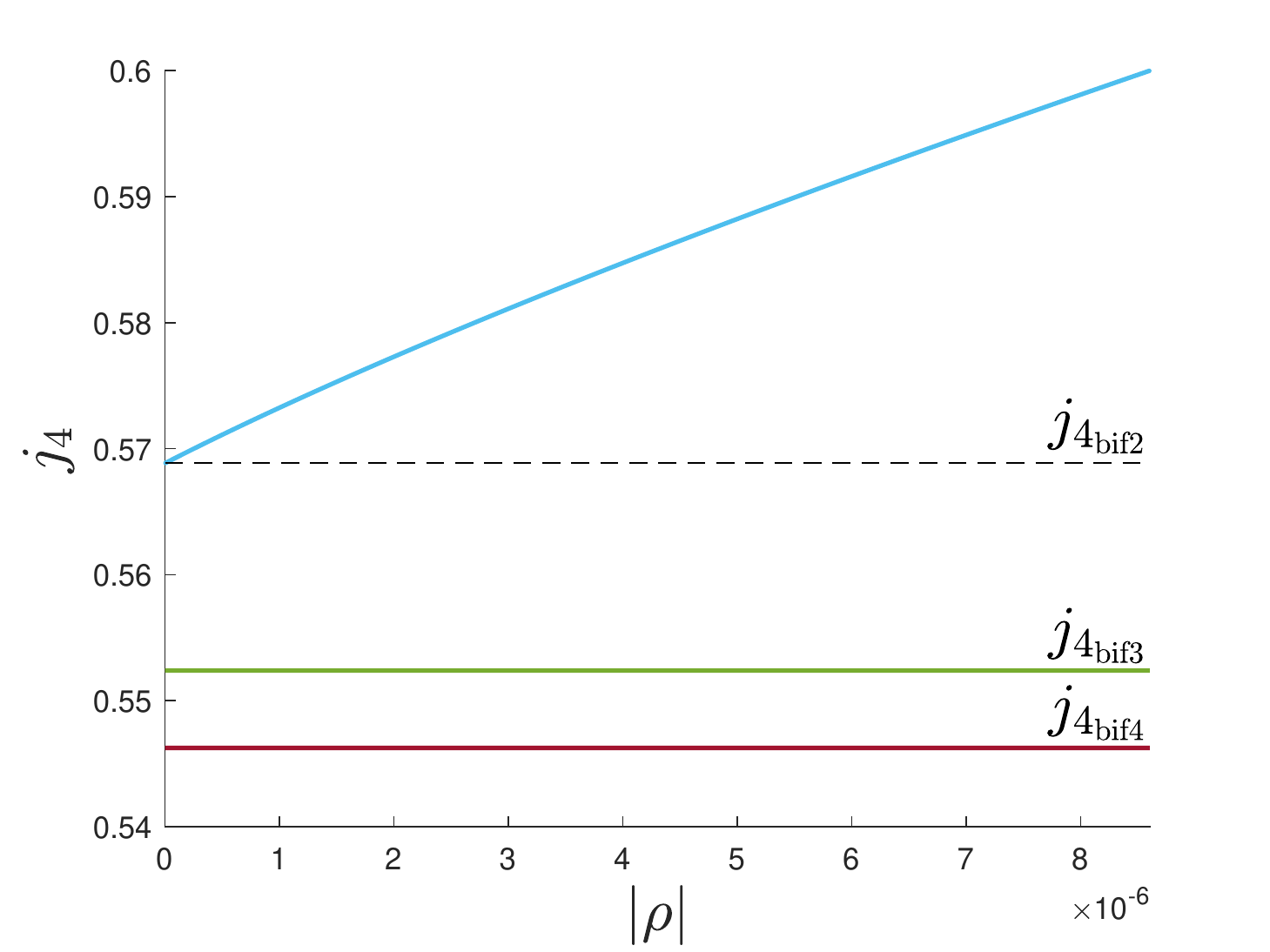}
	\end{subfigure}
	\begin{subfigure}{0.325\textwidth}
		\includegraphics[width=\textwidth]{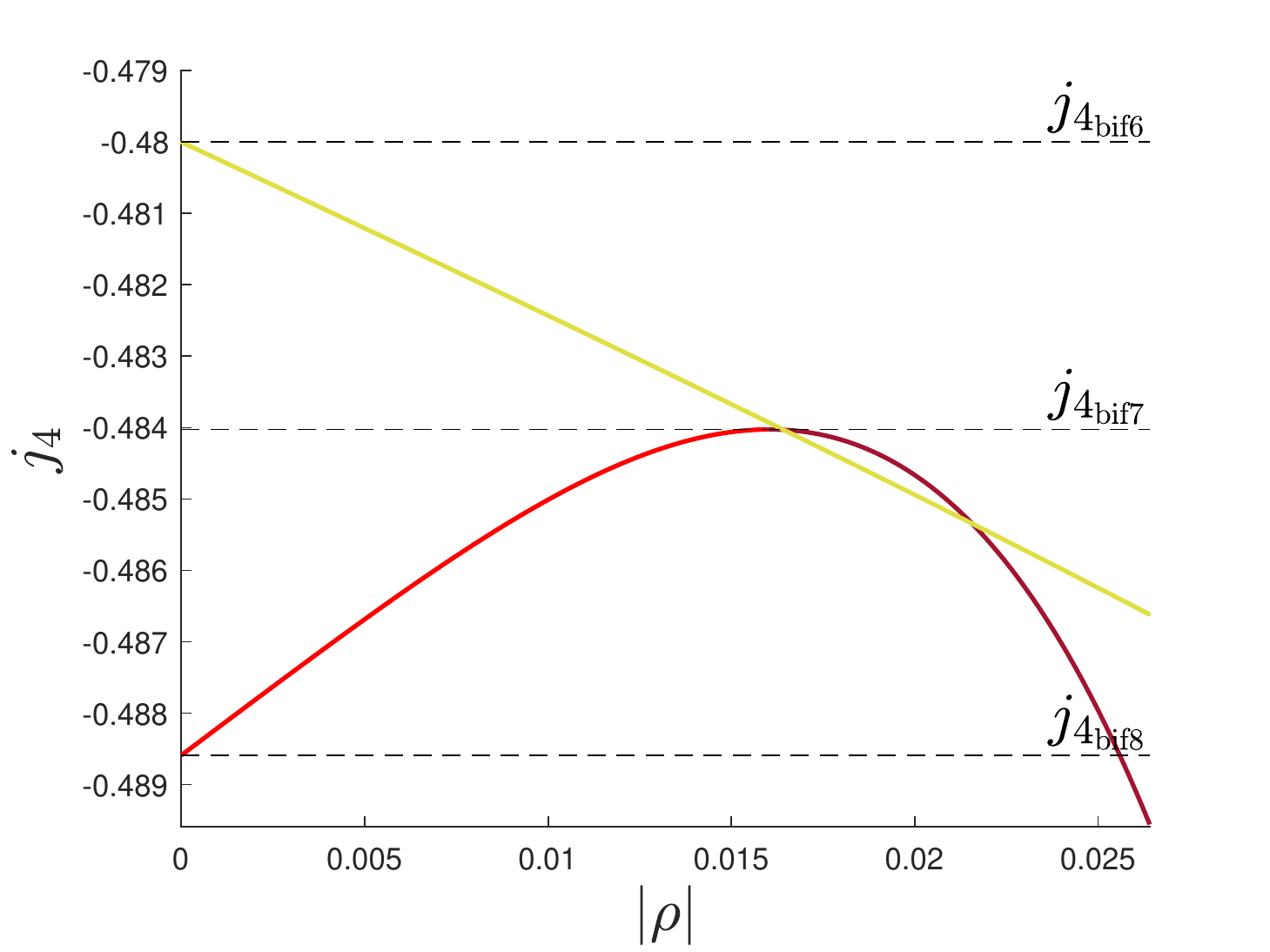}
	\end{subfigure}
	\begin{subfigure}{0.325\textwidth}
		\includegraphics[width=\textwidth]{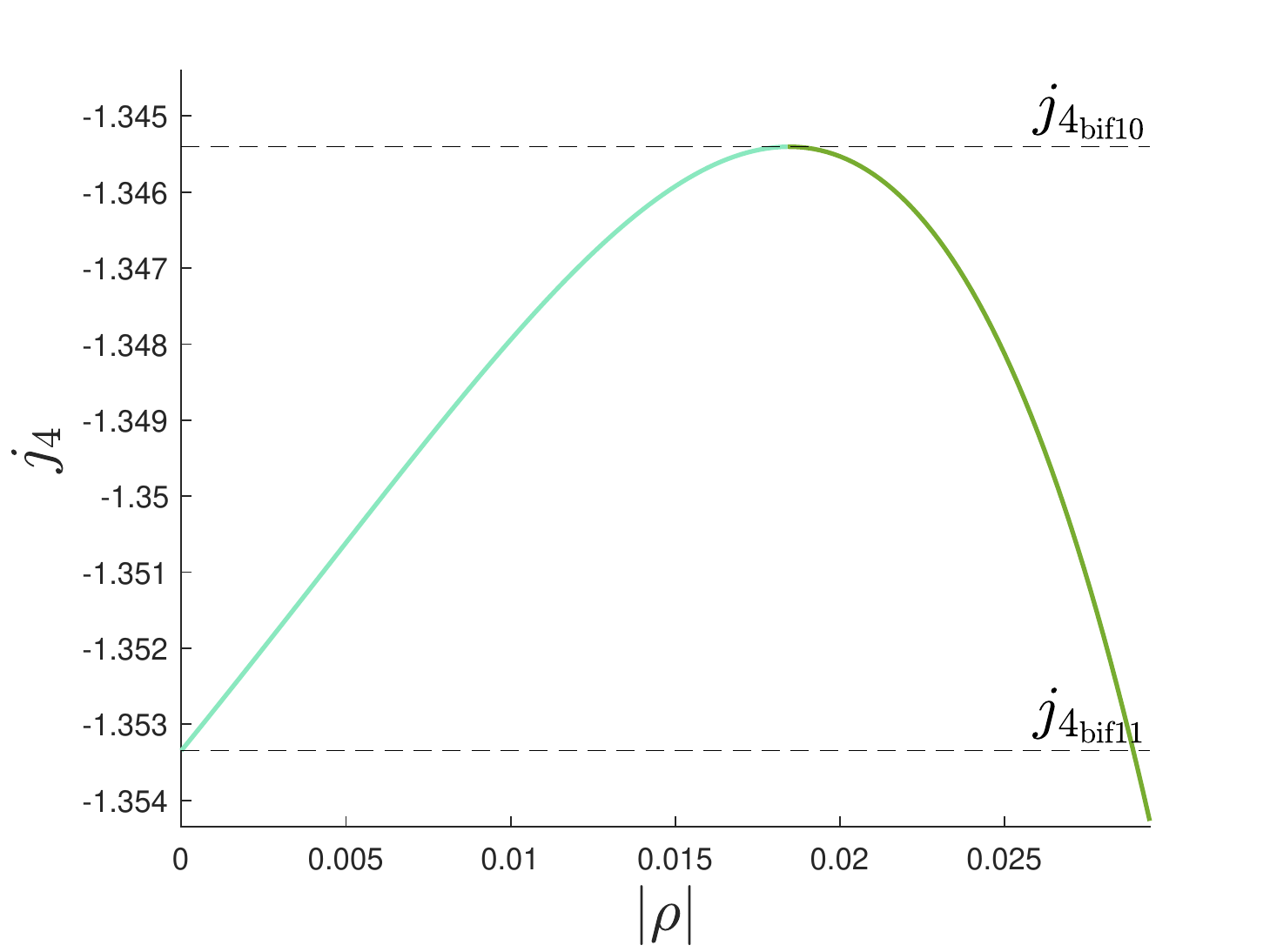}
	\end{subfigure}
	\caption{We show some enlargements of interesting regions of the bifurcation diagram in Fig.\ref{fig: bifdiagramlambda0001}.}
	\label{fig: bifdiagramlambda0001_en}
\end{figure} 

\begin{table}
	\centering
	\caption{Sequence of bifurcations and existing equilibrium points for different ranges of $\jj4\in[-6,6]$. Here, $\lambda=0.001$. Moreover, 
		$j_{4_{\rm bif 1}}\sim0.9972$, $j_{4_{\rm bif 2}}\sim 0.5695$, $j_{4_{\rm bif 3}}\sim 0.552$,  $j_{4_{\rm bif 4}}\sim 0.546$, $j_{4_{\rm bif 5}}\sim 0.2755$, $j_{4_{\rm bif 6}}=-12/25$, $j_{4_{\rm bif 7}}\sim-0.4840$, $j_{4_{\rm bif 8}}\sim -0.4886$, $j_{4_{\rm bif 9}}=-31/35$, $j_{4_{\rm bif 10}}\sim -1.3454$ and $j_{4_{\rm bif 11}}\sim -1.3533$. 
	}
	\label{tabbif}
	\begin{tabular}{|c|c|c|}
		\hline 
		$\jj4$ range	& bifurcations & existing equilibrium points\\
		\hline
		$j_{4_{\rm bif 1}}\le \jj4\le 6$ &  $\rho_-\ge\rho_+>\rho_{\blacktriangledown}>\rho_{	\blacktriangle}$ & $E_4$ for $\vert\rho\vert\le\rho_-$;  $E_3$ for $\vert\rho\vert\le\rho_+$; \\
		& &  $E_{6}$ for $\vert\rho\vert=\rho_{\blacktriangledown}$;  $E_{5}$for $\vert\rho\vert=\rho_{\blacktriangle}$;\\
		& &  $E_{8}$, $E_{10}$ for $\vert\rho\vert<\rho_{\blacktriangledown}$;\\
		& &  $E_{7}$, $E_{9}$ for $\vert\rho\vert<\rho_{\blacktriangle}$.\\
		\hline
		$j_{4_{\rm bif 2}}<\jj4<j_{4_{\rm bif 1}}$ & $\rho_+>\rho_->\rho_{\square}>\rho_{\diamond}>\rho_{\blacktriangledown}>\rho_{	\blacktriangle}$ & $E_3$ for $\vert\rho\vert\le\rho_+$;  $E_4$ for $\vert\rho\vert\le\rho_-$;\\
		& & $\bar{E}_{1,2}$ for $\rho_{\diamond}\le\vert\rho\vert\le \rho_{\square}$;\\
		& &  $E_{6}$ for $\vert\rho\vert=\rho_{\blacktriangledown}$;  $E_{5}$for $\vert\rho\vert=\rho_{\blacktriangle}$;\\
		& &  $E_{8}$, $E_{10}$ for $\vert\rho\vert<\rho_{\blacktriangledown}$;\\
		& &  $E_{7}$, $E_{9}$ for $\vert\rho\vert<\rho_{\blacktriangle}$.\\
		\hline	
		$j_{4_{\rm bif 3}}<\jj4\le j_{4_{\rm bif 2}}$ & $\jj4\gtrsim 0.553$: $\rho_+>\rho_->\rho_{\square}>\rho_{\diamond}\ge\rho_{\blacktriangledown}$ &   $E_3$ for $\vert\rho\vert\le\rho_+$;  $E_4$ for $\vert\rho\vert\le\rho_-$;\\
		& else:  $\rho_+>\rho_->\rho_{\square}>\rho_{\blacktriangledown}>\rho_{\diamond}$ & $\bar{E}_{1,2}$ for $\rho_{\diamond}\le\vert\rho\vert\le \rho_{\square}$; \\
		& & $E_{6}$ for $\vert\rho\vert=\rho_{\blacktriangledown}$;  \\
		& &  $E_{8}$, $E_{10}$ for $\vert\rho\vert<\rho_{\blacktriangledown}$.\\
		\hline
		$j_{4_{\rm bif 4}}<\jj4\le j_{4_{\rm bif 3}}$ &
		$\jj4\gtrsim 0.547$: $\rho_+>\rho_->\rho_{\square}\ge\rho_{\blacktriangledown}$, & $E_3$ for $\vert\rho\vert\le\rho_+$;  $E_4$ for $\vert\rho\vert\le\rho_-$; \\
		& else
		$\rho_+>\rho_->\rho_{\blacktriangledown}>\rho_{\square}$ &  $\bar{E}_{1,2}$ for $\vert\rho\vert\le \rho_{\square}$; $E_{6}$ for $\vert\rho\vert=\rho_{\blacktriangledown}$; \\
		& & $E_{8}$, $E_{10}$ for $\vert\rho\vert<\rho_{\blacktriangledown}$.\\
		\hline
		$j_{4_{\rm bif 5}}<\jj4\le j_{4_{\rm bif 4}}$ & $\rho_+>\rho_->\rho_{\blacktriangledown}$ & $E_3$ for $\vert\rho\vert\le\rho_+$;  $E_4$ for $\vert\rho\vert\le\rho_-$; \\
		& &  $E_{6}$ for $\vert\rho\vert=\rho_{\blacktriangledown}$;\\
		& & $E_{8}$, $E_{10}$ for $\vert\rho\vert<\rho_{\blacktriangledown}$.
		\\
		\hline 
		$j_{4_{\rm bif 6}}\le\jj4\le j_{4_{\rm bif 5}}$ & $\rho_+>\rho_-$ & $E_3$ for $\vert\rho\vert\le\rho_+$;  $E_4$ for $\vert\rho\vert\le\rho_-$. \\
		\hline
		$j_{4_{\rm bif 7}}<\jj4< j_{4_{\rm bif 6}}$ & $\rho_+>\rho_->\rho_{\triangledown}$ & $E_3$ for $\vert\rho\vert\le\rho_+$;  $E_4$ for $\vert\rho\vert\le\rho_-$; \\
		& & 	$E_{12}$ for $\vert\rho\vert\le\rho_{\triangledown}$.\\
		\hline
		$j_{4_{\rm bif 8}}<\jj4\le j_{4_{\rm bif 7}}$ &  $\jj4\gtrsim-0.4803$: $\rho_+>\rho_->\rho_{\triangledown}\ge\rho_{\square}\ge\rho_{\square,\rm bis}$ &  $E_3$ for $\vert\rho\vert\le\rho_+$;  $E_4$ for $\vert\rho\vert\le\rho_-$;  \\
		& else:
		$\rho_+>\rho_->\rho_{\square}>\rho_{\triangledown}>\rho_{\square,\rm bis}$ & $E_{12}$ for $\vert\rho\vert\le\rho_{\triangledown}$; \\
		& & ${\bar{E}_{1,2}}$ for $\rho_{\square,\rm bis}\le\vert\rho\vert\le\rho_{\square}$. \\
		\hline
		$j_{4_{\rm bif 9}}\le\jj4\le j_{4_{\rm bif 8}}$ & $\jj4\gtrsim-0.4853$: $\rho_+>\rho_->\rho_{\square}\ge\rho_{\triangledown}$ & $E_3$ for $\vert\rho\vert\le\rho_+$;  $E_4$ for $\vert\rho\vert\le\rho_-$; 
		\\
		& else: $\rho_+>\rho_->\rho_{\triangledown}>\rho_{\square}$ &  $E_{12}$ for $\vert\rho\vert\le\rho_{\triangledown}$;\\
		& & ${\bar{E}_{1,2}}$ for $\vert\rho\vert\le\rho_{\square}$. \\
		\hline
		$j_{4_{\rm bif 10}}<\jj4< j_{4_{\rm bif 9}}$ & $\jj4\gtrsim-0.919$: $\rho_+>\rho_->\rho_{\triangledown}>\rho_{\square}\ge \rho_{\triangle}$& $E_3$ for $\vert\rho\vert\le\rho_+$;  $E_4$ for $\vert\rho\vert\le\rho_-$;  \\
		& else: $\rho_+>\rho_->\rho_{\triangledown}>\rho_{\triangle}>\rho_{\square}$ & $E_{12}$ for $\vert\rho\vert\le\rho_{\triangledown}$; $E_{11}$ for $\vert\rho\vert\le\rho_{\triangle}$\\
		& & ${\bar{E}_{1,2}}$ for $\vert\rho\vert\le\rho_{\square}$. \\
		\hline
		$j_{4_{\rm bif 11}}<\jj4\le j_{4_{\rm bif 10}}$  & 
		$\rho_+>\rho_->\rho_{\triangledown}>\rho_{\triangle}>\rho_{\square}>\rho_{\diamond}\ge\rho_{\diamond, \rm bis}$ & $E_3$ for $\vert\rho\vert\le\rho_+$;  $E_4$ for $\vert\rho\vert\le\rho_-$;\\
		& & $E_{12}$ for $\vert\rho\vert\le\rho_{\triangledown}$; $E_{11}$ for $\vert\rho\vert\le\rho_{\triangle}$;\\
		& & ${\bar{E}_{1,2}}$ for $\rho_{\diamond}\le\vert\rho\vert\le\rho_{\square}$ \\
		& &
		and for $\vert\rho\vert\le\rho_{\diamond,\rm bis}$\\
		\hline
		$-6\le\jj4\le j_{4_{\rm bif 11}}$ & $\rho_+>\rho_->\rho_{\triangledown}>\rho_{\triangle}>\rho_{\square}>\rho_{\diamond}$ & $E_3$ for $\vert\rho\vert\le\rho_+$;  $E_4$ for $\vert\rho\vert\le\rho_-$;\\
		& & $E_{12}$ for $\vert\rho\vert\le\rho_{\triangledown}$; $E_{11}$ for $\vert\rho\vert\le\rho_{\triangle}$;\\
		& & ${\bar{E}_{1,2}}$ for $\rho_{\diamond}\le\vert\rho\vert\le\rho_{\square}$.\\
		\hline			
	\end{tabular}	
\end{table}
We show in Fig.\ref{fig: bifdiagramlambda0001} the bifurcation diagram, where the colour lines represent the values of $\vert\rho\vert$ for which a bifurcation occurs. Some enlargements of interesting regions of the diagram are given in Fig.\ref{fig: bifdiagramlambda0001_en}. In Table \ref{tabbif} we summarise the bifurcations sequence and list the existing points for different ranges of $\jj4\in[-6,6]$. Through a stability analysis based on the Poincaré-Hopf theorem, we obtain that
\begin{itemize}
	\item $\vert\rho\vert=\rho_-$ and $\vert\rho\vert=\rho_+$ are pitchfork bifurcations which cause a variation of stability of $E_2$ and affect the existence and the stability of the equilibrium points $E_3$ and $E_4$;
	\item $\vert\rho\vert=\rho_{
		\diamond}$, $\vert\rho\vert=\rho_{
		\diamond, \rm bis}$,$\vert\rho\vert=\rho_{
		\square}$ and $\vert\rho\vert=\rho_{
		\square, \rm bis}$ are pitchfork bifurcations, influencing the stability of $E_3$ and $E_4$ and the existence of $\bar{E}_1$ and $\bar{E}_2$: when $\bar{E}_1$ and $\bar{E}_2$ exist, they are unstable, while $E_3$ and $E_4$ are stable;
	\item $\vert\rho\vert=\rho_{\vartriangle}$ and $\vert\rho\vert=\rho_{\triangledown}$ are pitchfork bifurcations affecting the stability of $E_1$ and the existence and stability of $E_{11}$ and $E_{12}$; 
	\item $\vert\rho\vert=\rho_{\blacktriangle}$ and $\vert\rho\vert=\rho_{\blacktriangledown}$ are saddle-node bifurcations; they have no consequence on the stability of existing equilibrium solutions, but give rise to an even number of equilibrium points of type $E_+$ and $E_-$, half of which are stable, while the other half is unstable. 
\end{itemize}

\begin{figure}
	\centering
	\begin{subfigure}{1\textwidth}
		\centering
		\includegraphics[width=0.4\textwidth]{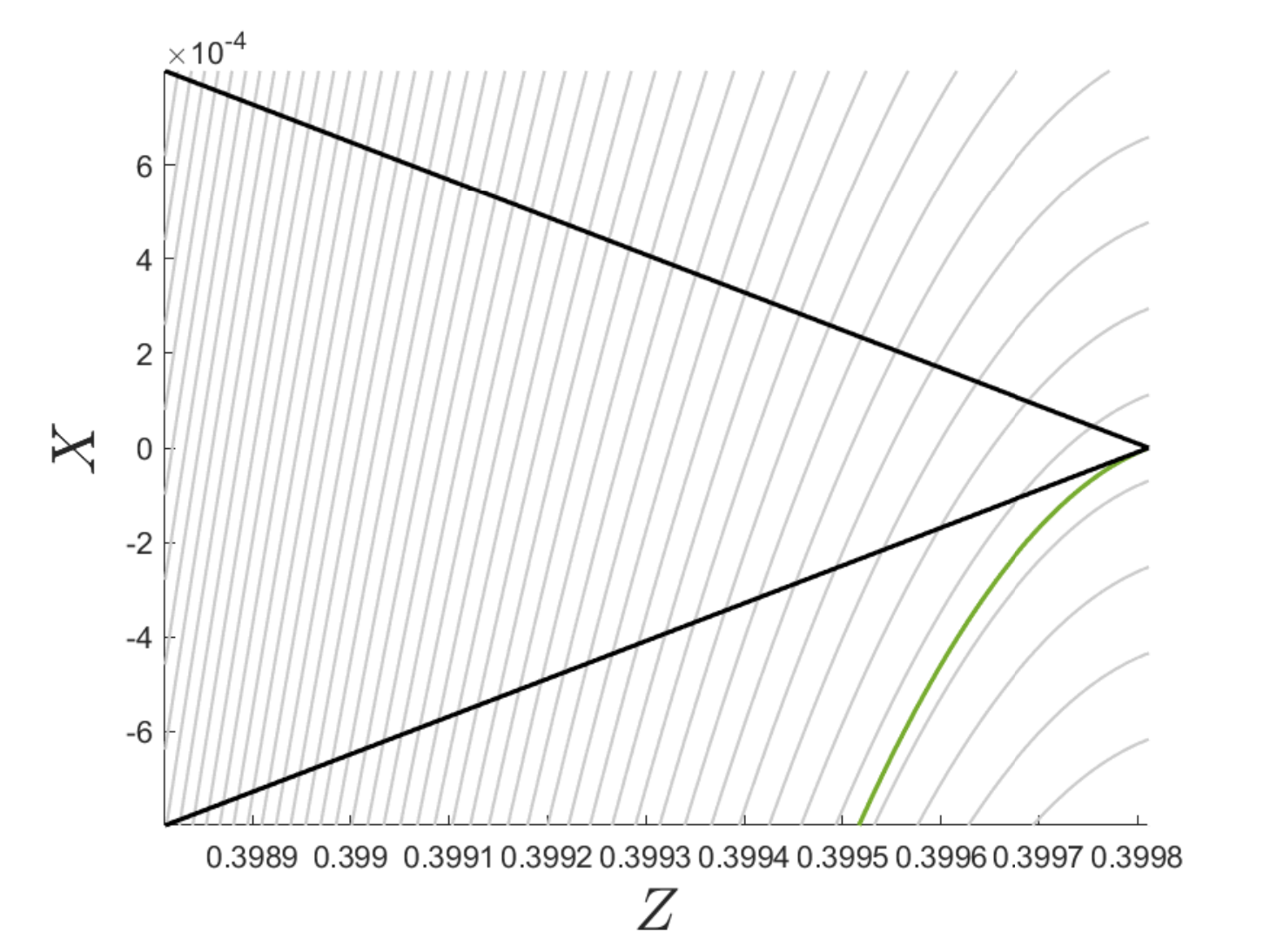}
		\includegraphics[width=0.4\textwidth]{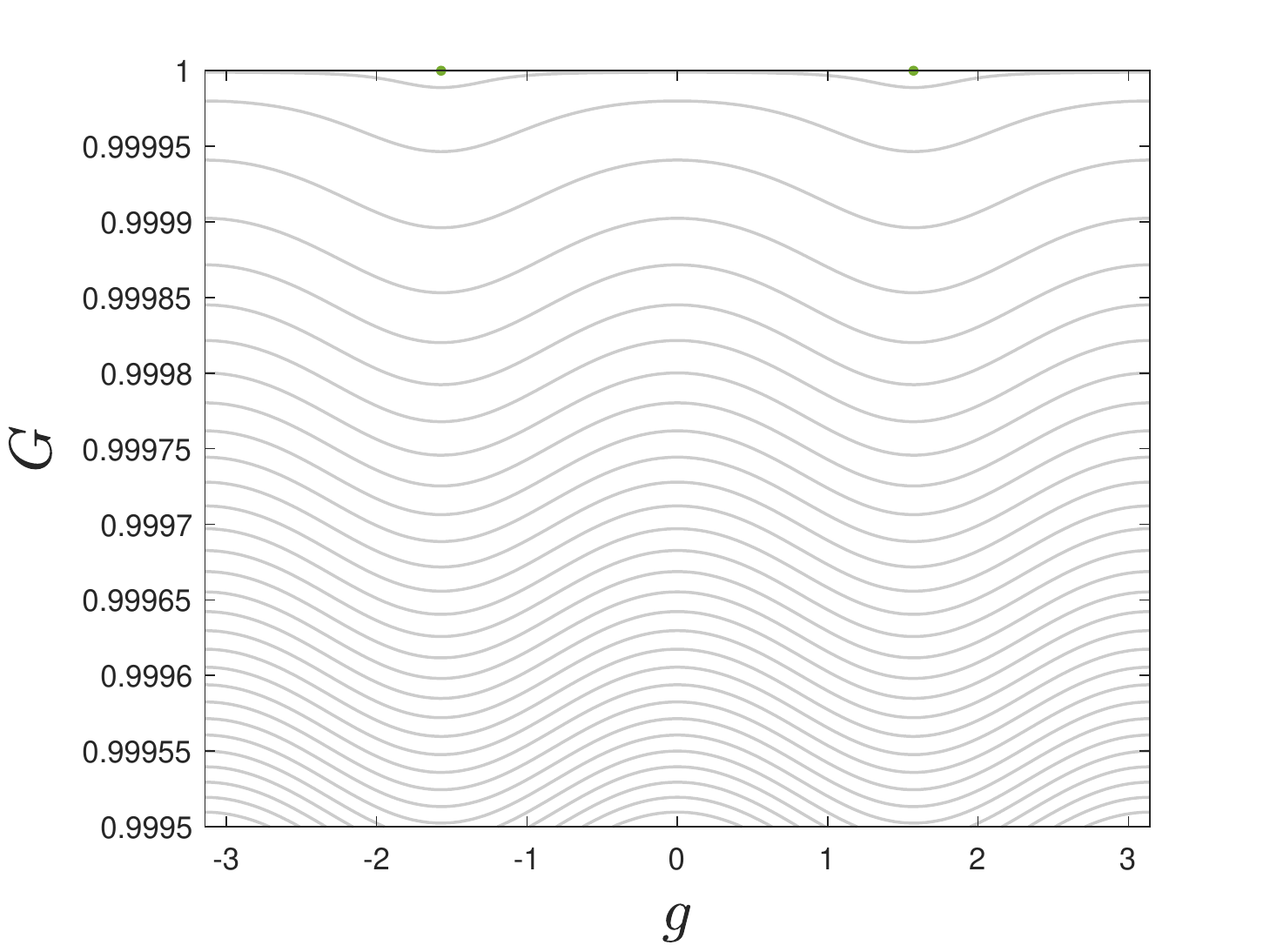}
		\subcaption{$\vert\rho\vert=\rho_-$, $\rho_-\sim0.44763$}
	\end{subfigure}
	\begin{subfigure}{1\textwidth}
		\centering
		\includegraphics[width=0.4\textwidth]{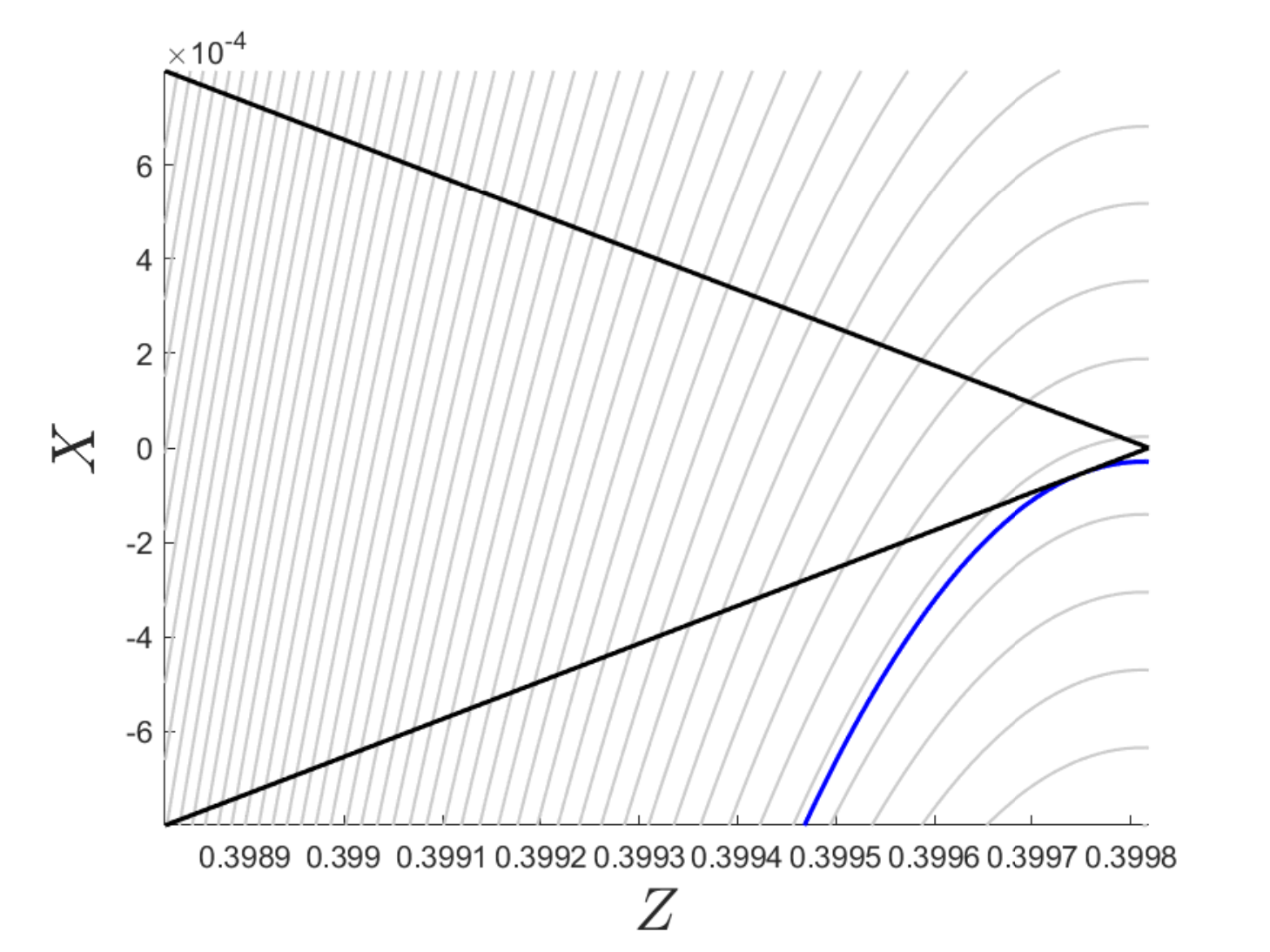}
		\includegraphics[width=0.4\textwidth]{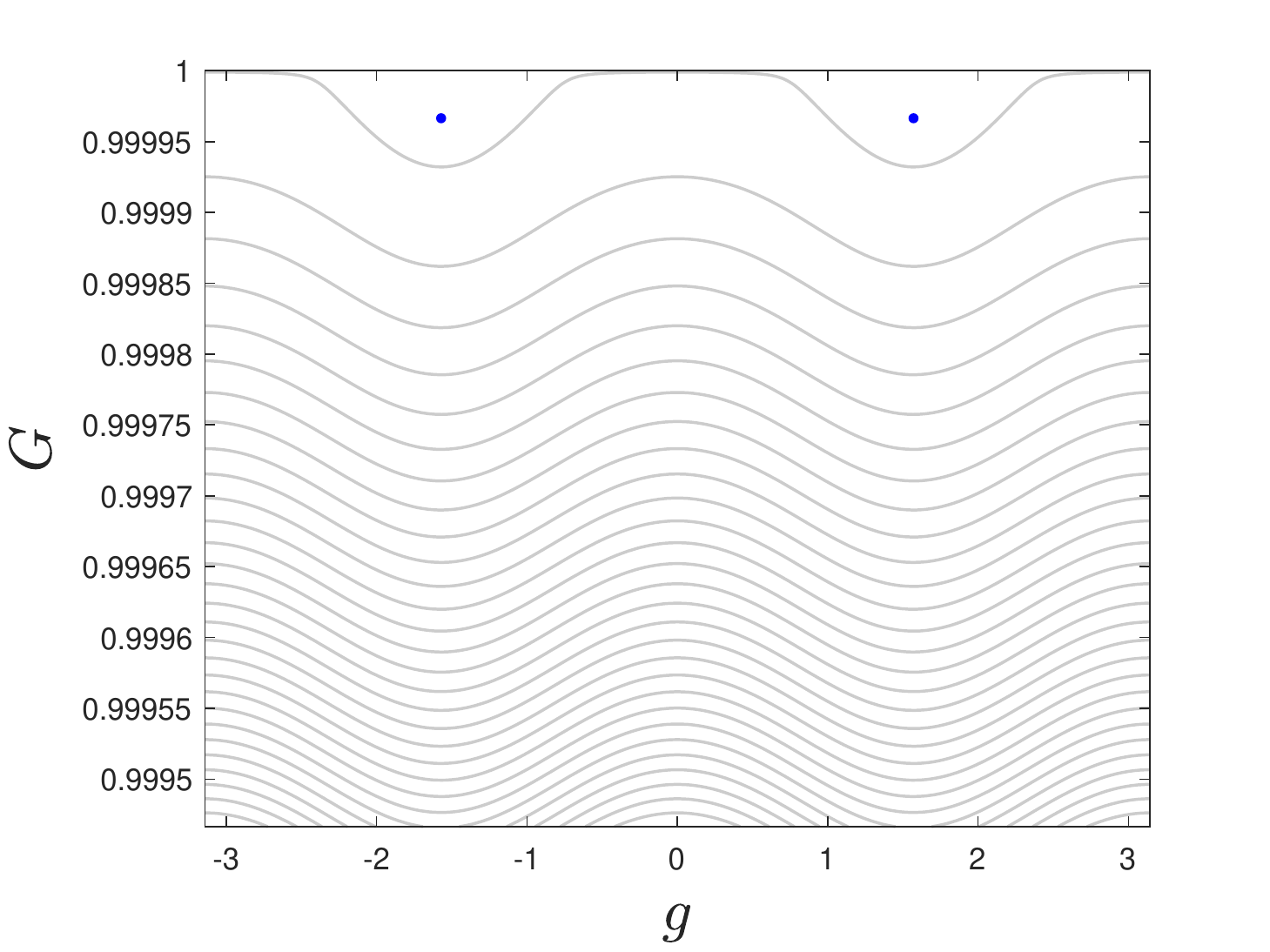}
		\subcaption{$\vert\rho\vert=0.4472$}
	\end{subfigure}
	\begin{subfigure}{1\textwidth}
		\centering
		\includegraphics[width=0.4\textwidth]{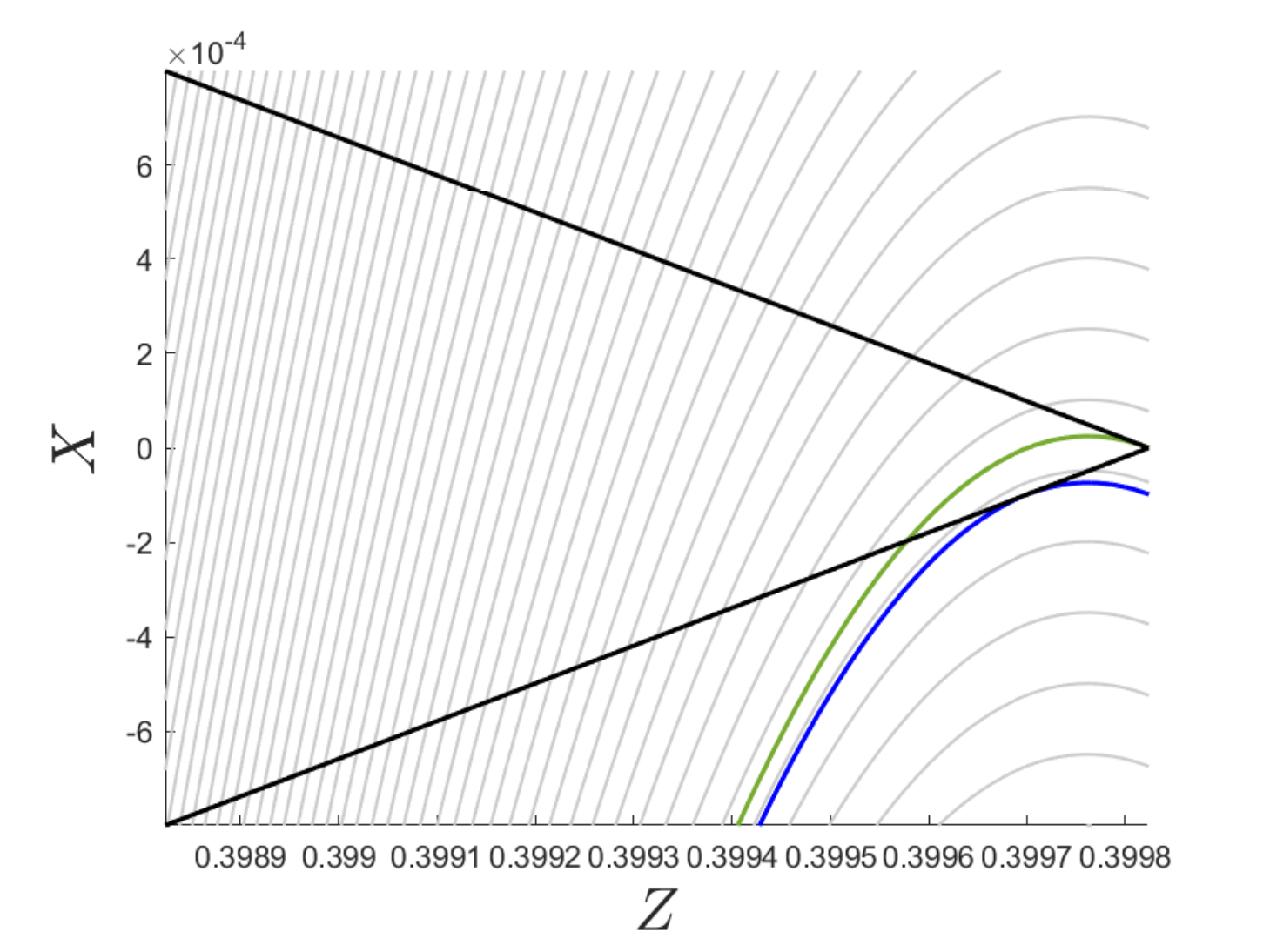}
		\includegraphics[width=0.4\textwidth]{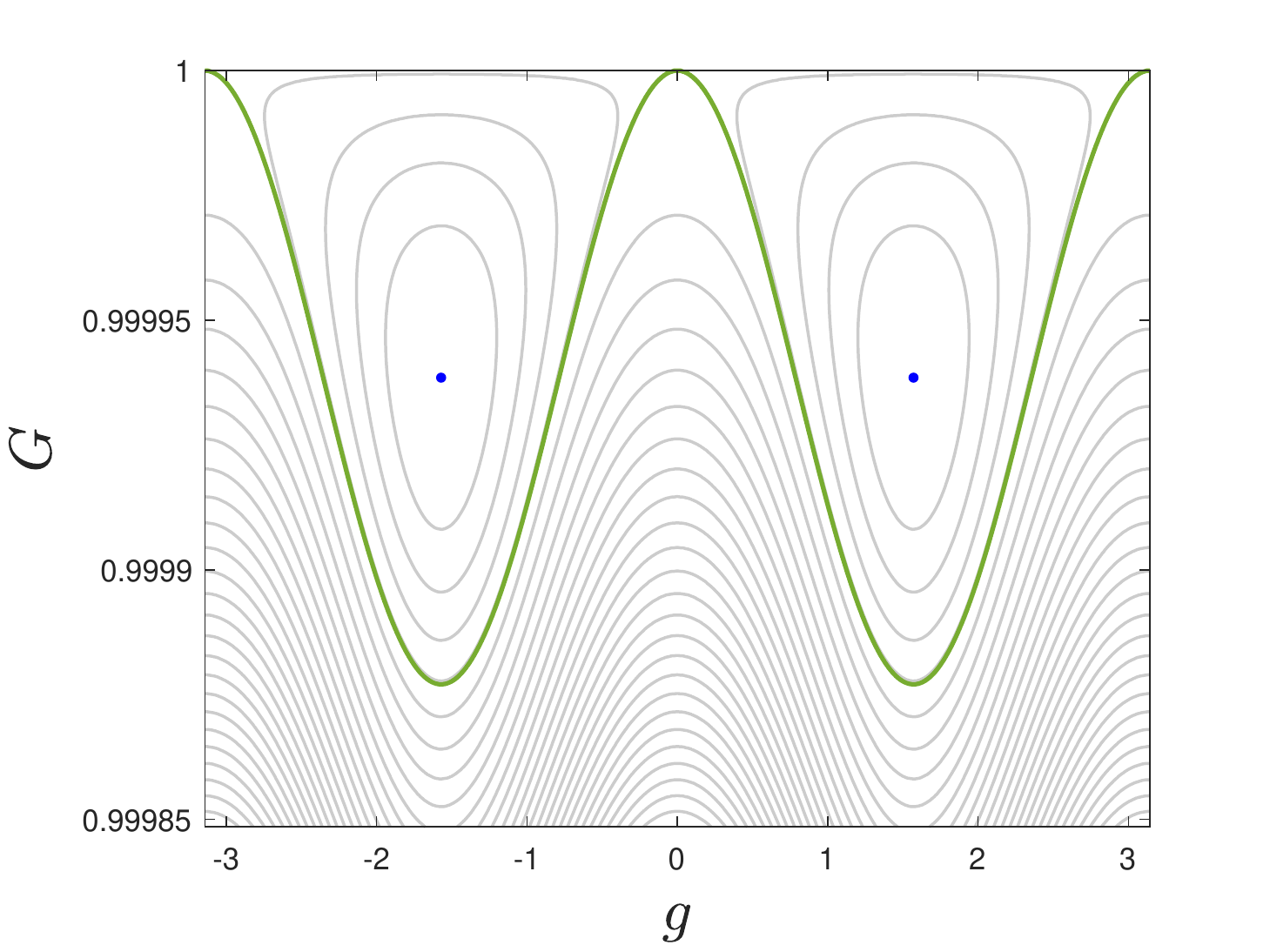}
		\subcaption{$\vert\rho\vert=\rho_+$, $\rho_+\sim0.44761$}
	\end{subfigure}
	\begin{subfigure}{1\textwidth}
		\centering
		\includegraphics[width=0.4\textwidth]{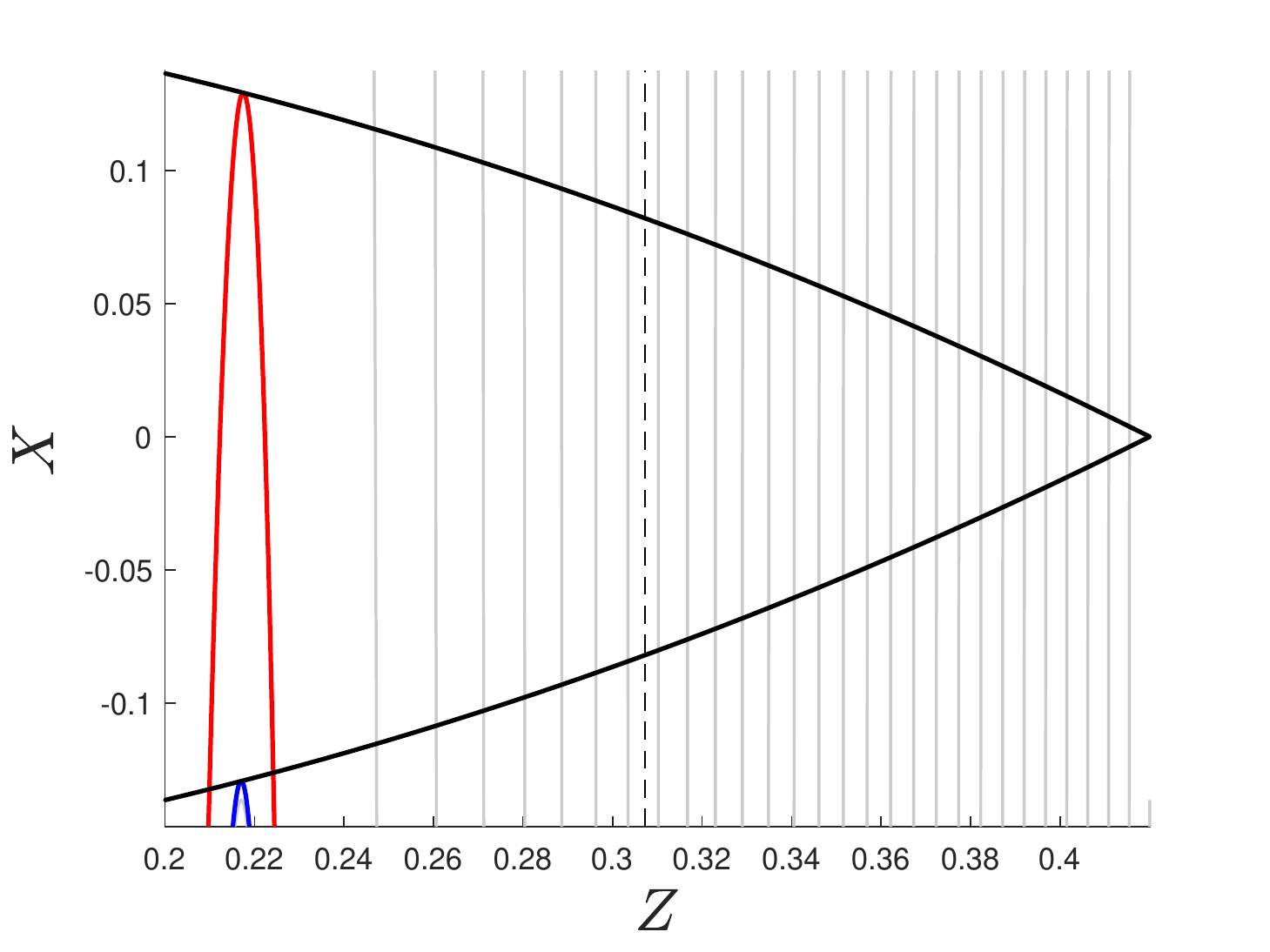}
		\includegraphics[width=0.4\textwidth]{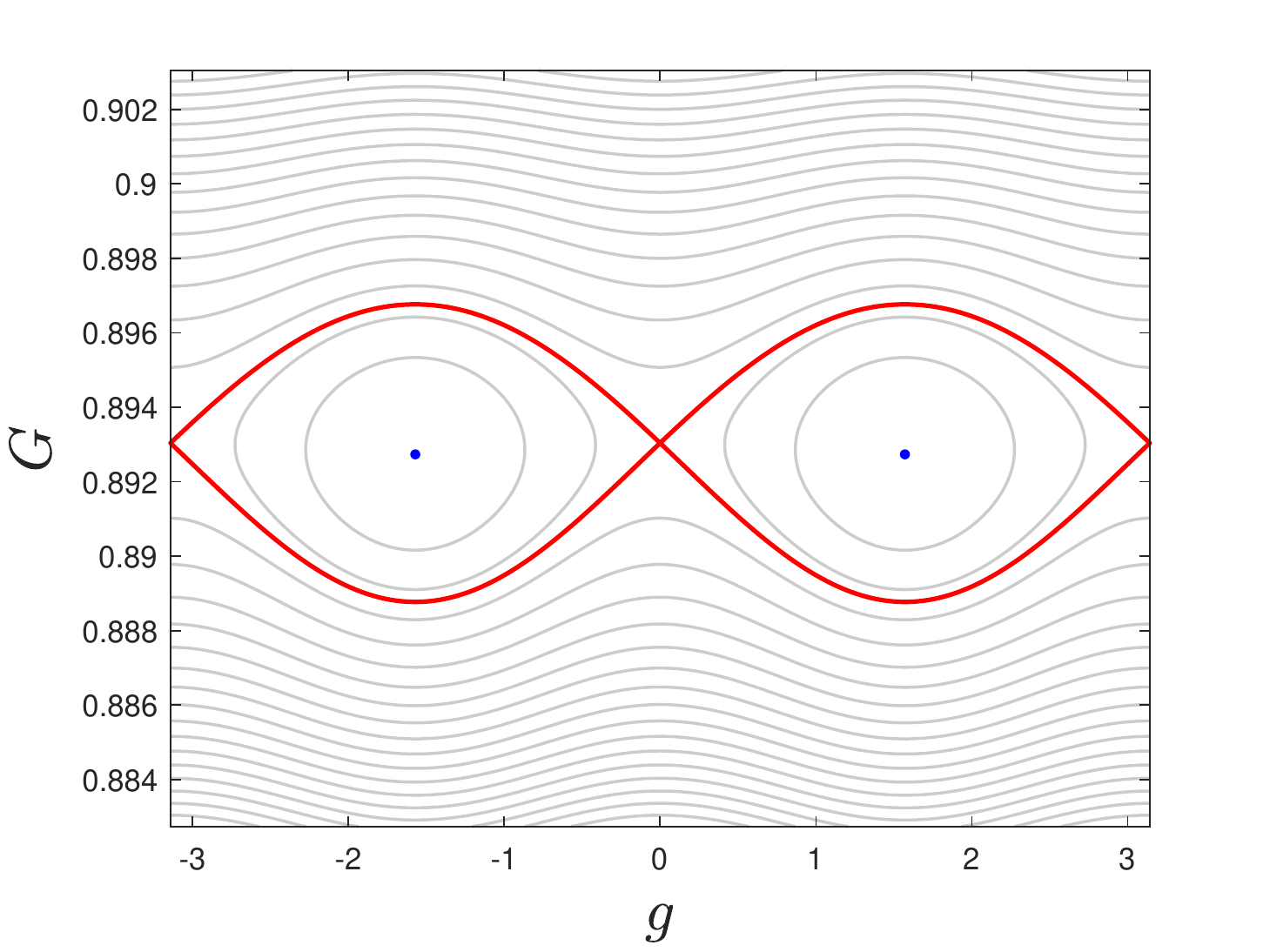}
		\subcaption{$\vert\rho\vert=0.4$}
		\label{fig:j413_bif1_casod}
	\end{subfigure}
	\caption{Level curves for the $J_4$-problem with $\jj4=1.3$ for four different values of $\vert\rho\vert$. On the left, the level curves are represented on the $(Z,X)$ plane: enlargements of the regions with the equilibrium points are performed. The black line represents the contour $\mathcal{C}$ of the \textit{lemon} space. The dashed black line corresponds to $Z=\bar{Z}$, at which the level curves have a singularity. The coloured line are the level curves tangent to $\mathcal{C}$ at the equilibrium points: they are green if the equilibrium point is degenerate, red if it is unstable and blue if it is stable. On the right, the level curves are shown on corresponding enlargements in the $(g,G)$ plane.}
	\label{fig:j413_bif1}
\end{figure}

\begin{figure}
	\centering
	\begin{subfigure}{1\textwidth}
		\centering
		\includegraphics[width=0.4\textwidth]{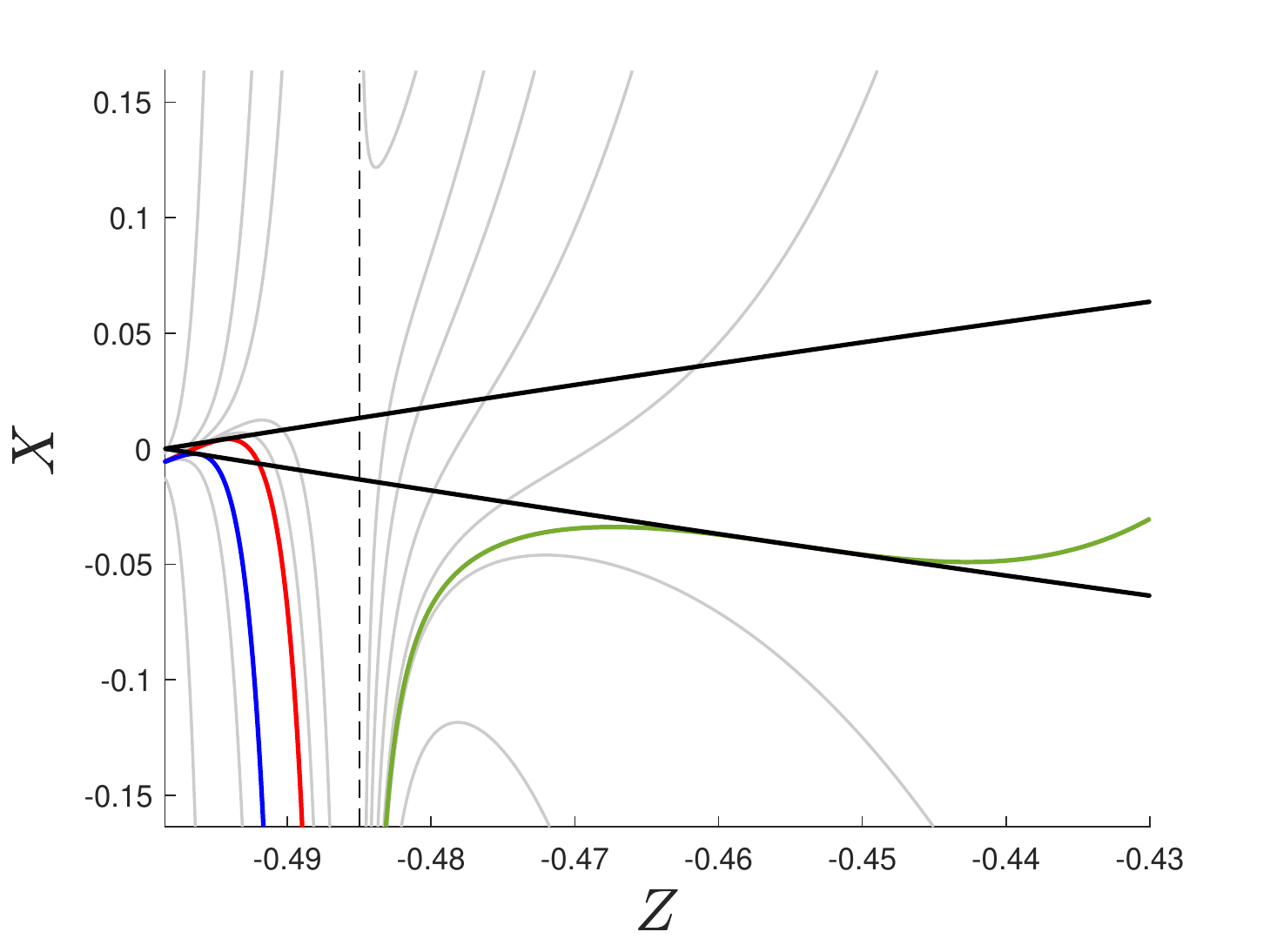}
		\includegraphics[width=0.4\textwidth]{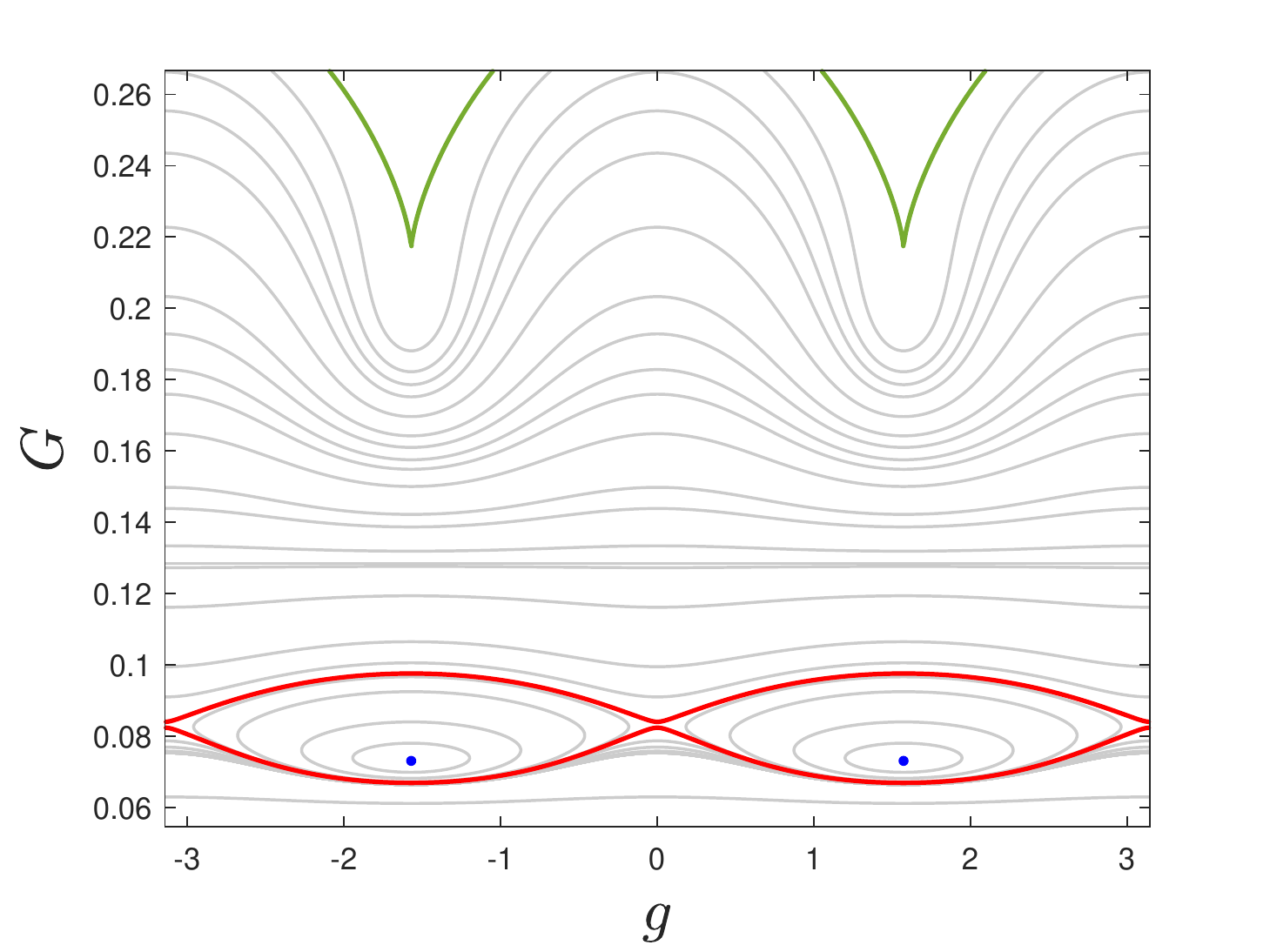}
		\subcaption{$\vert\rho\vert=\rho_{\blacktriangledown}$, $\rho_{\blacktriangledown}\sim 0.054542$}
	\end{subfigure}
	\begin{subfigure}{1\textwidth}
		\centering
		\includegraphics[width=0.4\textwidth]{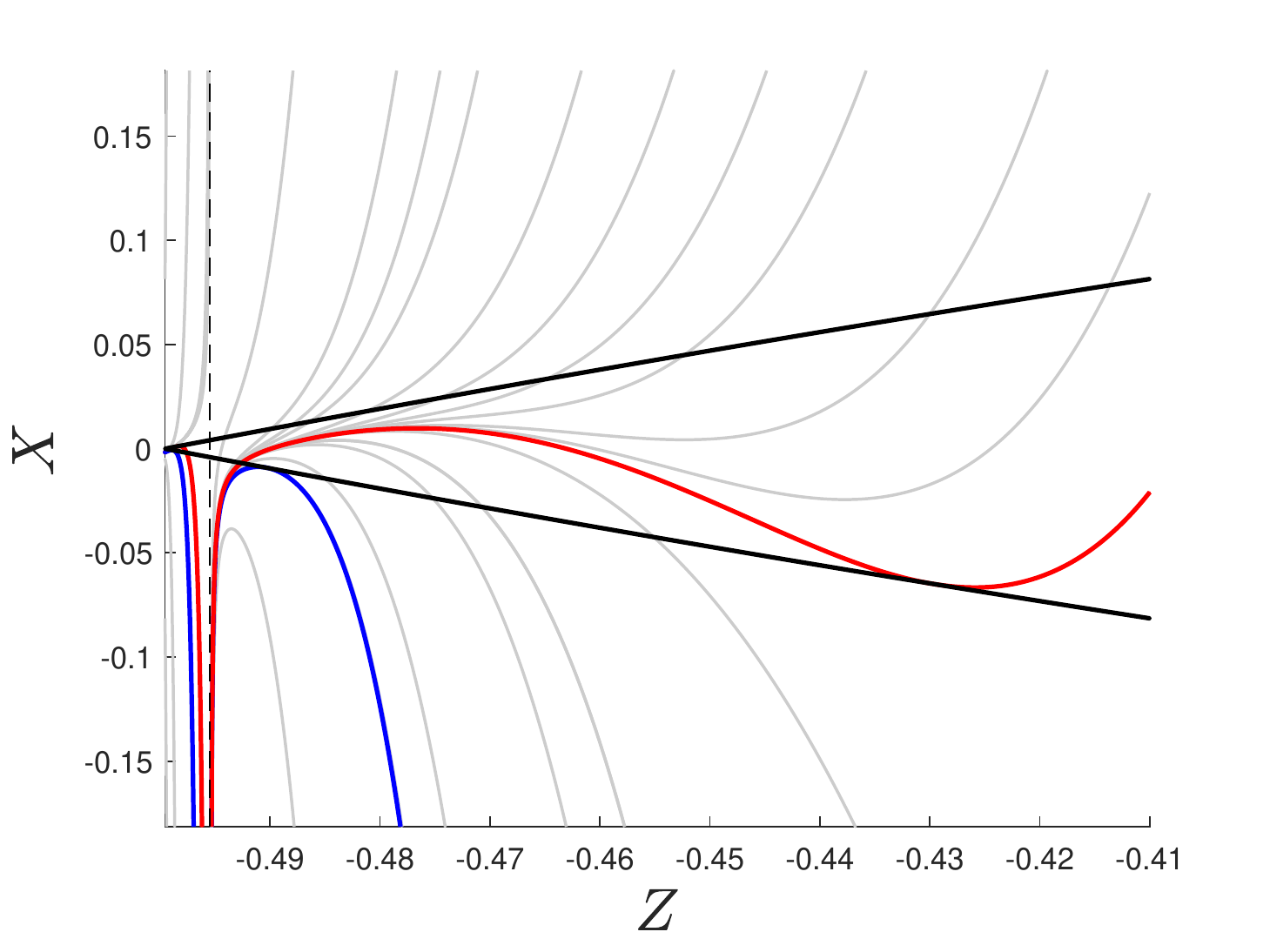}
		\includegraphics[width=0.4\textwidth]{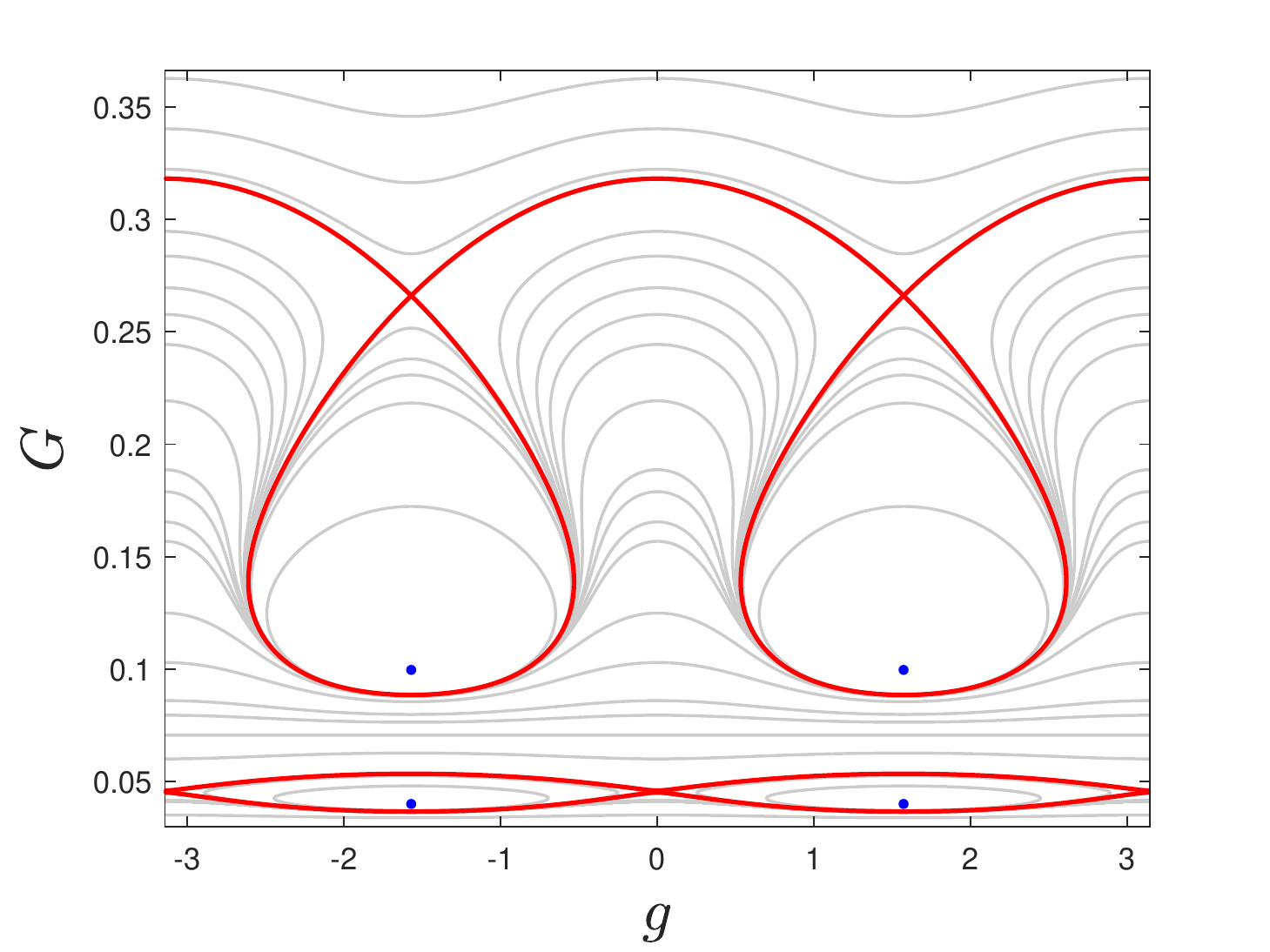}
		\subcaption{$\vert\rho\vert=0.03$}
	\end{subfigure}
	\begin{subfigure}{1\textwidth}
		\centering
		\includegraphics[width=0.4\textwidth]{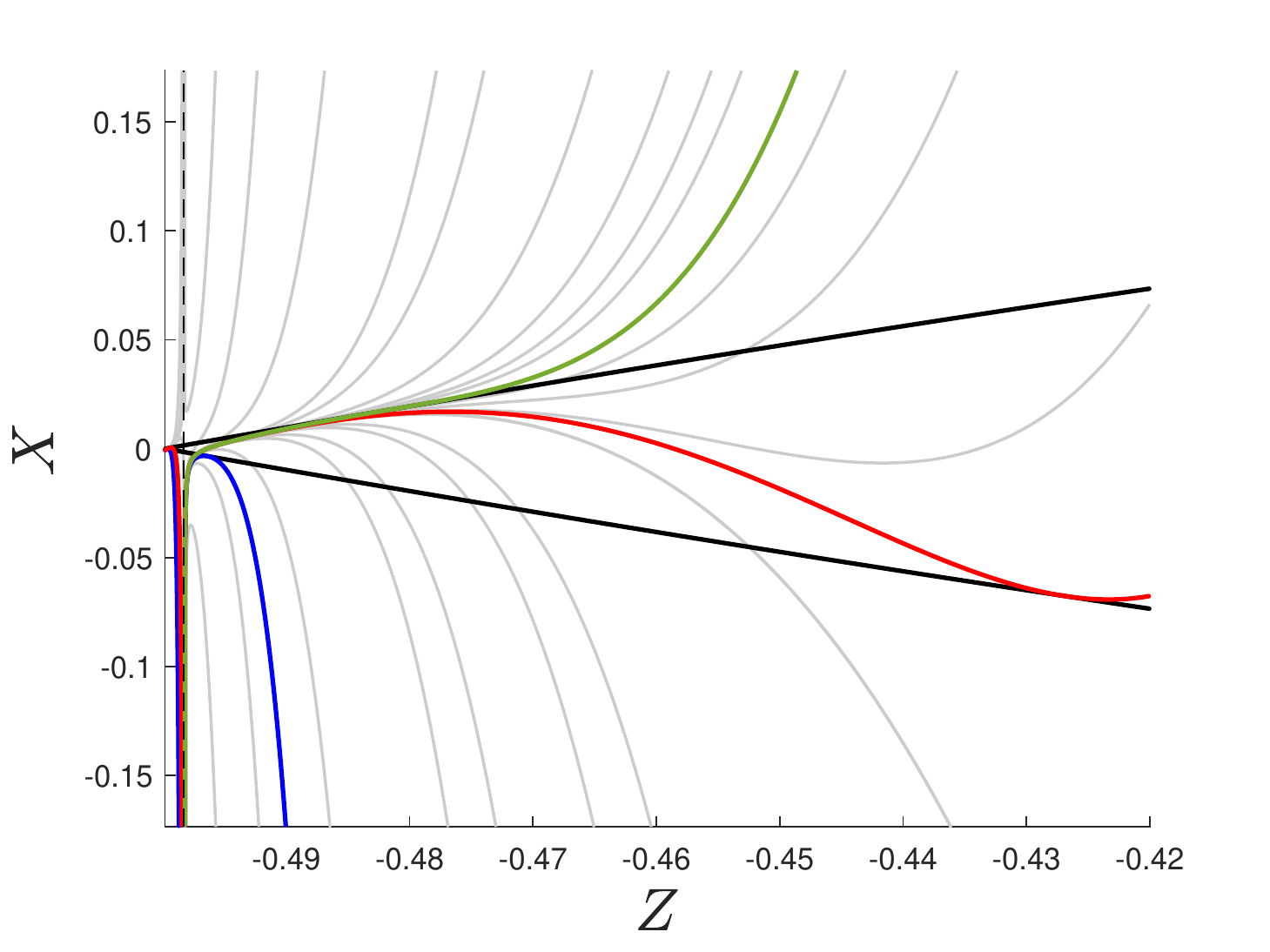}
		\includegraphics[width=0.4\textwidth]{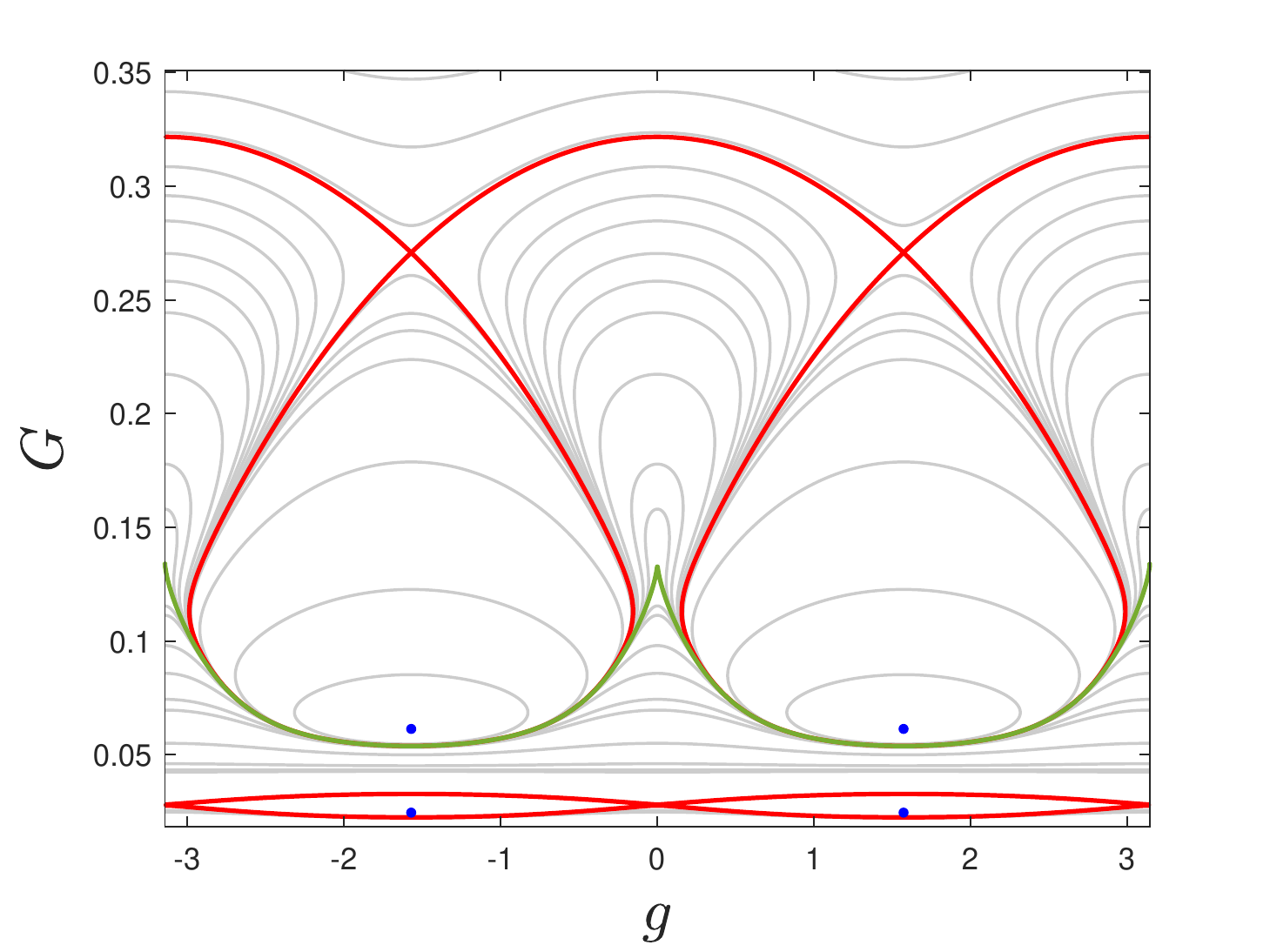}
		\subcaption{$\vert\rho\vert=\rho_{\blacktriangle}$, $\rho_{\blacktriangle}\sim0.018379$}
	\end{subfigure}
	\begin{subfigure}{1\textwidth}
		\centering
		\includegraphics[width=0.4\textwidth]{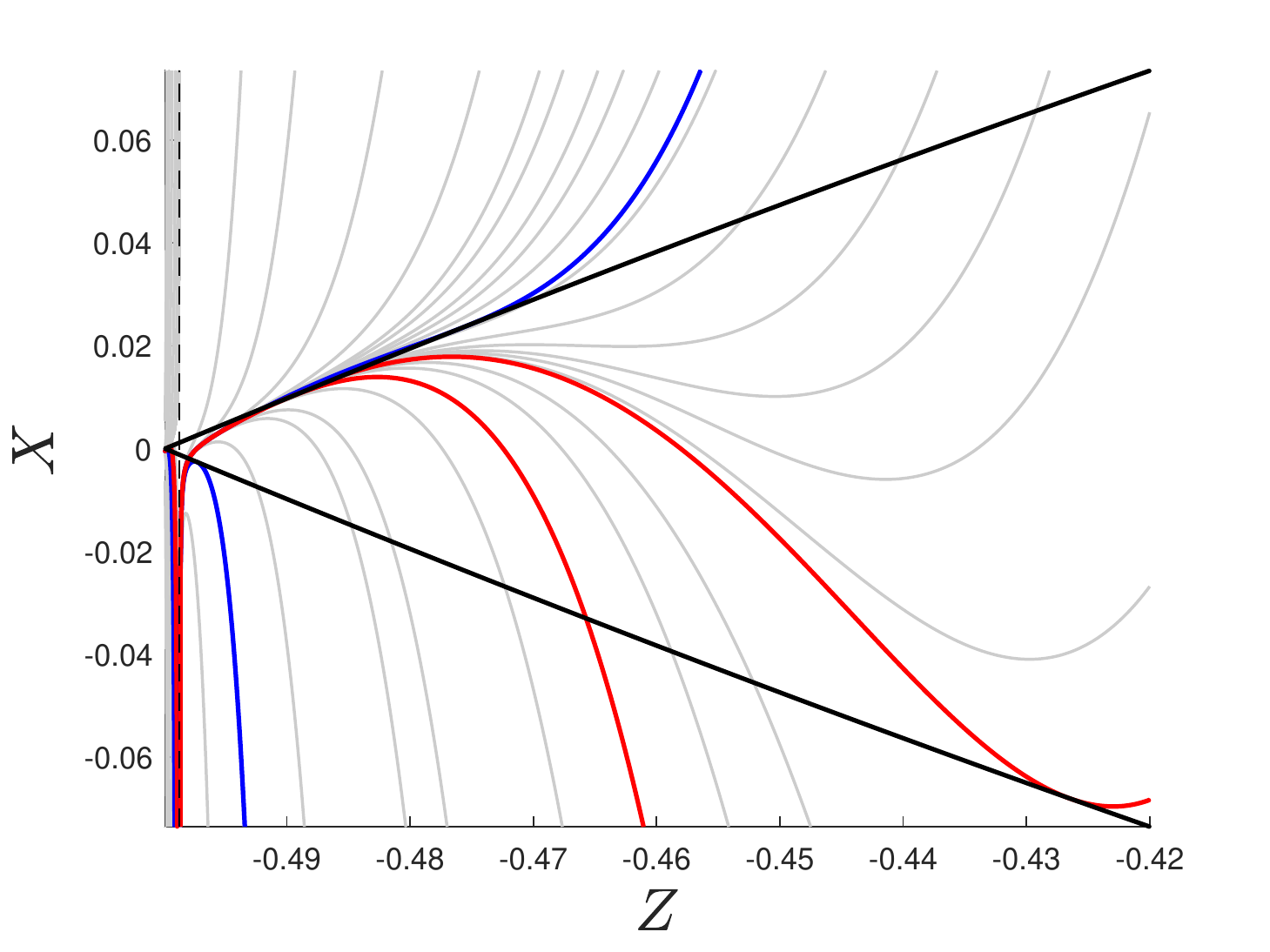}
		\includegraphics[width=0.4\textwidth]{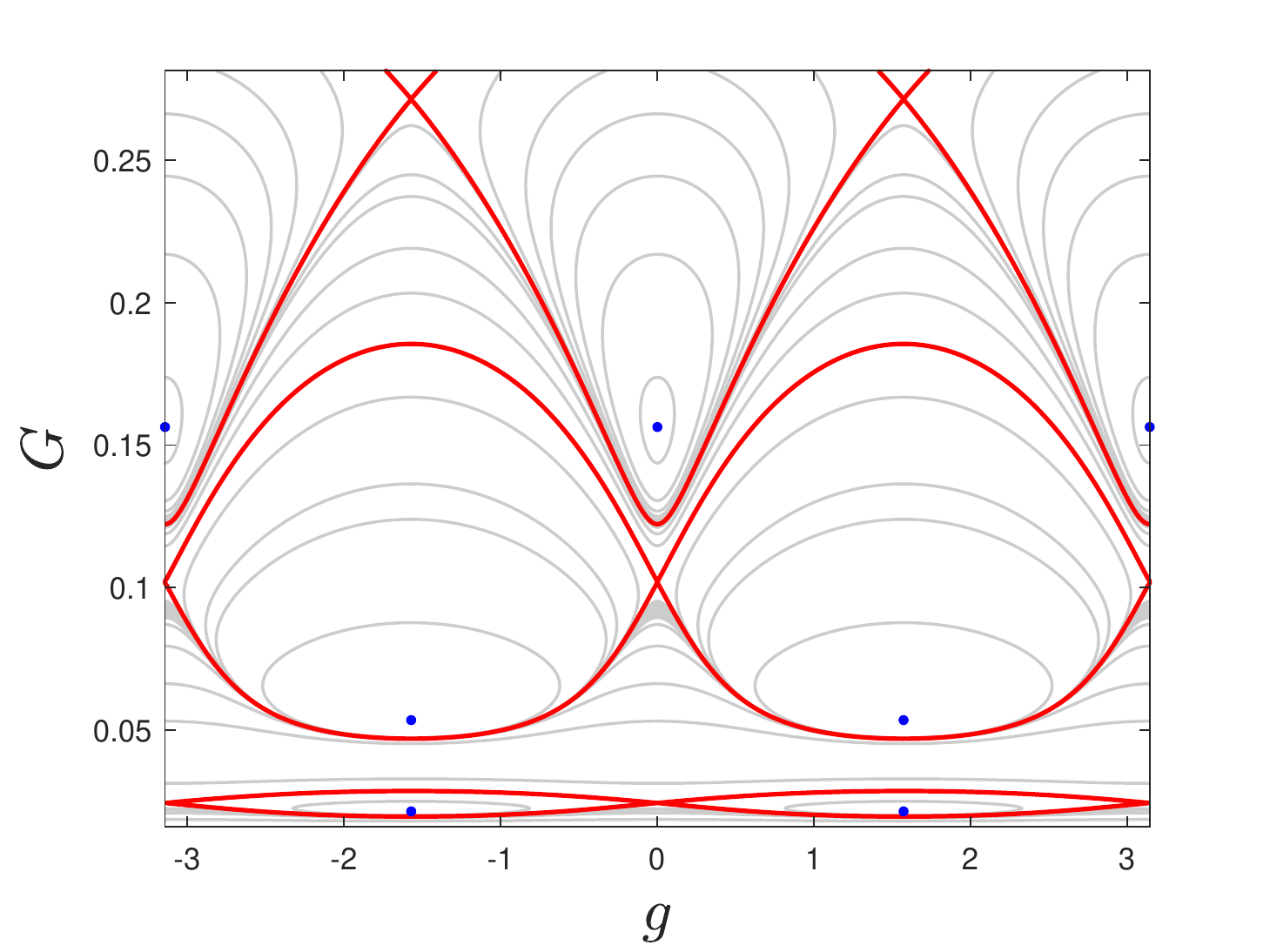}
		\subcaption{$\vert\rho\vert=0.016$}
	\end{subfigure}
	\caption{Level curves for the $J_4$-problem with $\jj4=1.3$ for four different values of $\vert\rho\vert$ in the $(Z,X)$ and the $(g,G)$ planes. Enlargements of the regions containing the equilibrium points are performed. The same colour code employed in Fig.\ref{fig:j413_bif1} is used.}
	\label{fig:j413_bif2}
\end{figure}

To explain how to read the bifurcation diagram, let us fix a value of $\jj4$ in the range $(j_{4_{\rm bif 1}},6)$, which is of interest for the Earth ($\jj4\sim1.3$) and Mars ($\jj4\sim 4$). It holds $\rho_{-}>\rho_{+}>\rho_{\blacktriangledown}>\rho_{\blacktriangle}$. For each $\vert\rho\vert$ $E_1$ is stable. Moreover,
\begin{itemize}
	\item for $\vert\rho\vert>\rho_{-}$, $E_2$ is stable;
	\item at $\vert\rho\vert=\rho_{-}$, there is a bifurcation: $E_2$ is degenerate and $E_4$ coincides with $E_2$;
	\item for $\rho_{+}<\vert\rho\vert<\rho_{-}$, $E_2$ is unstable and $E_4$ is stable;
	\item at $\vert\rho\vert=\rho_{+}$, there is a bifurcation: $E_2$ is degenerate and coincides with $E_3$; $E_4$ is stable;
	\item for $\rho_{\blacktriangledown}<\vert\rho\vert<\rho_{+}$, $E_2$ and $E_4$ are stable, while $E_3$ is unstable;
	\item for $\vert\rho\vert=\rho_{\blacktriangledown}$, $E_2$ and $E_4$ are stable and $E_3$ is unstable; there also exists the equilibrium point $E_6$ which is degenerate; 
	\item for $\rho_{\blacktriangle}<\vert\rho\vert<\rho_{\blacktriangledown}$, $E_2$ and $E_4$ are stable and $E_3$ is unstable; there exist the equilibrium points $E_{8}$ and $E_{10}$: one of them is stable, the other is unstable; 
	\item for $\vert\rho\vert=\rho_{\blacktriangle}$, $E_2$ and $E_4$ are stable; $E_3$ is stable; one between $E_{8}$ and $E_{10}$ is stable, while the other is unstable;  there also exists the equilibrium point $E_5$ which is degenerate;
	\item  $\vert\rho\vert<\rho_{\blacktriangle}$, $E_2$ and $E_4$ are stable; $E_3$ is stable; one between $E_{8}$ and $E_{10}$ is stable, while the other is unstable; there exist the equilibrium points $E_7$ and $E_9$: one of them is stable, the other is unstable.
\end{itemize}
In Fig.\ref{fig:j413_bif1}, we show the level curves in a neighbourhood of the bifurcations $\vert\rho\vert=\rho_-$ and $\vert\rho\vert=\rho_+$. It is interesting to compare the phase portrait in Fig. \ref{fig:j413_bif1_casod} 
with the one shown in Fig.\ref{fig:j2ex} for the $\J2$-problem: the concavities of the colour curves tangent to the contour of the lemon space are opposite. Indeed, in this case, $E_3$ is unstable and $E_4$ is stable. In Fig.\ref{fig:j413_bif2}, we show the level curves in a neighbourhood of the two bifurcations $\vert\rho\vert=\rho_{\blacktriangledown}$ and  $\vert\rho\vert=\rho_{\blacktriangle}$.

\begin{figure}
	\centering
	\begin{subfigure}{1\textwidth}
		\centering
		\includegraphics[width=0.4\textwidth]{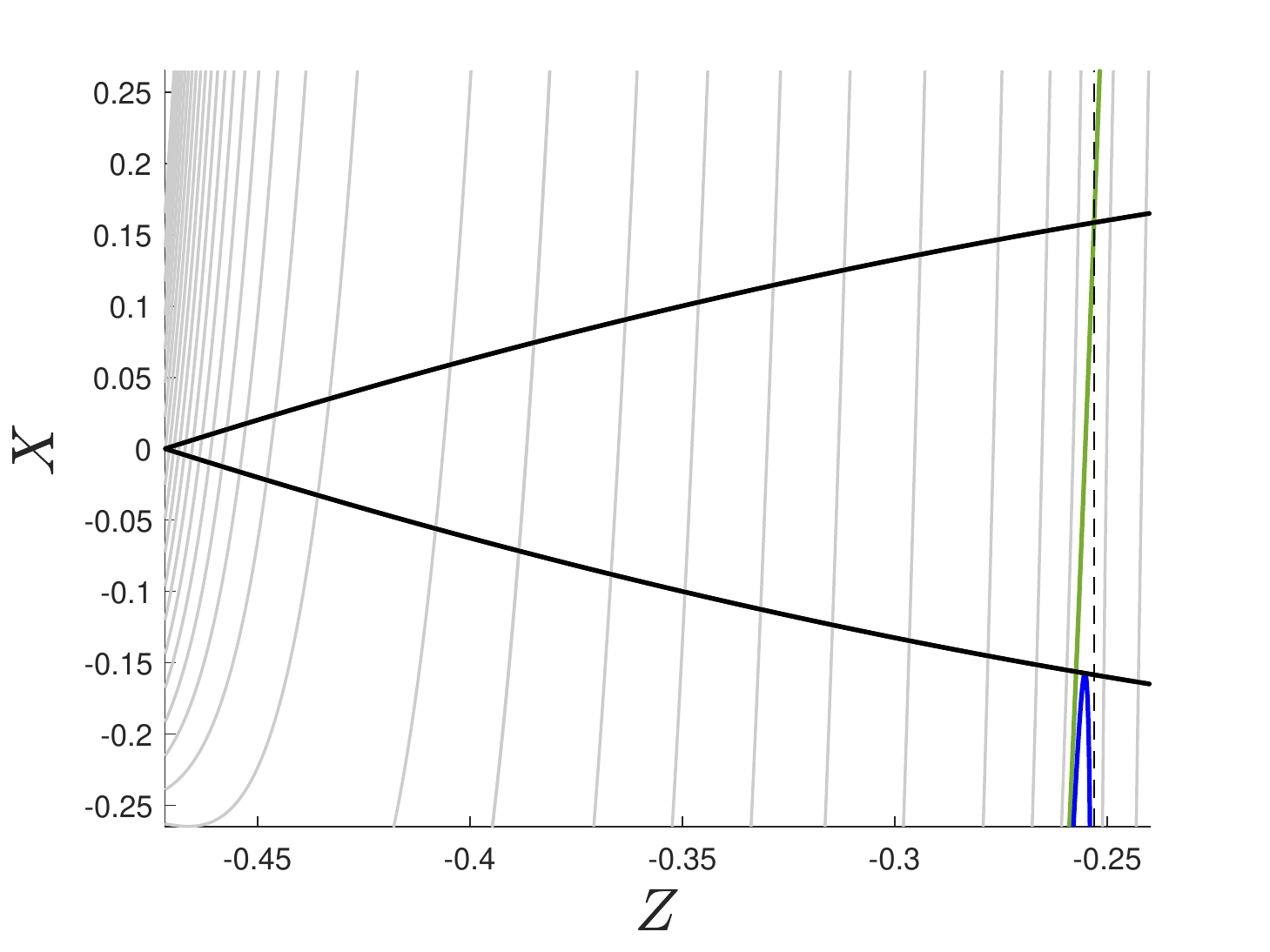}
		\includegraphics[width=0.4\textwidth]{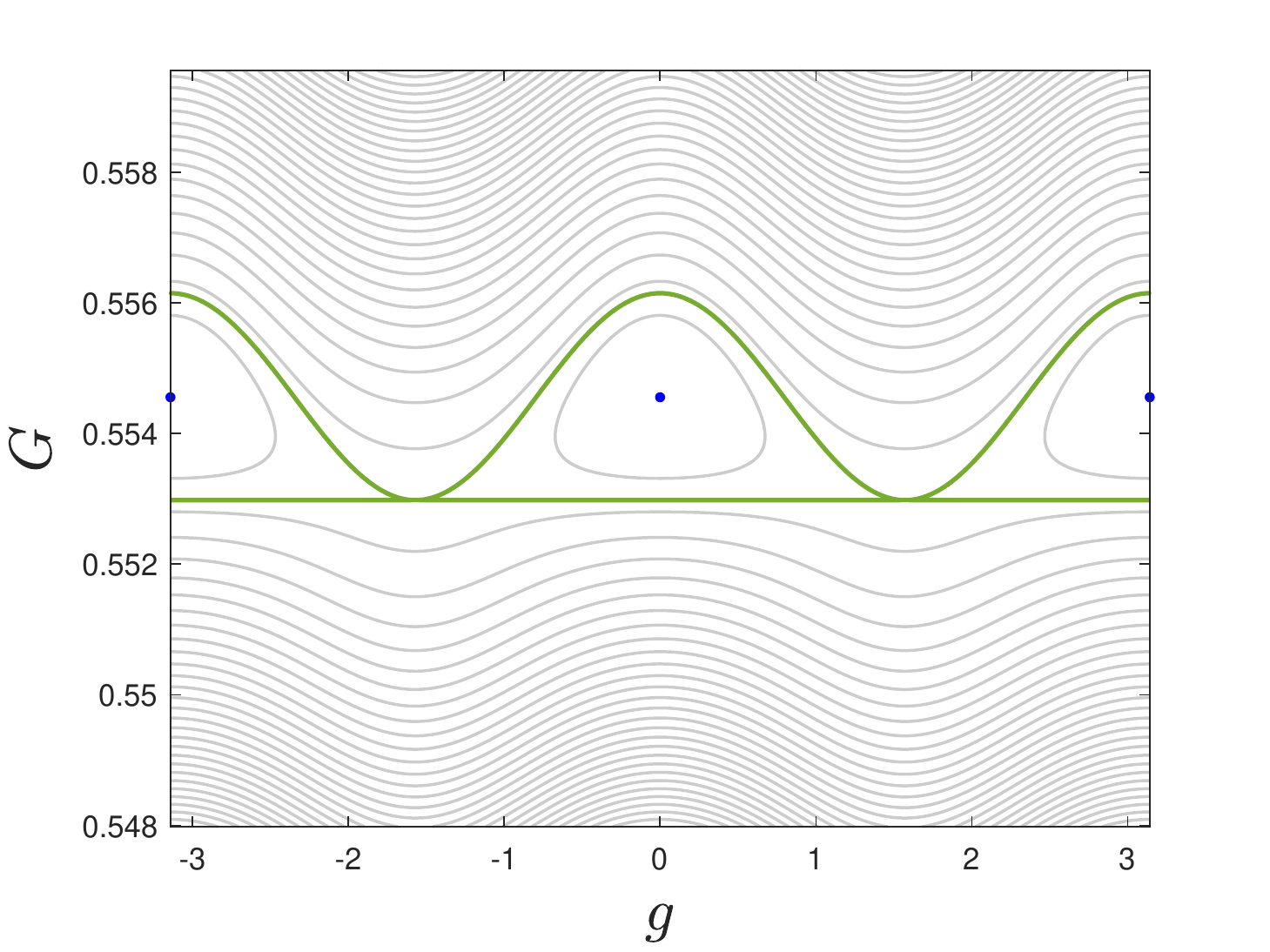}
		\subcaption{$\vert\rho\vert=\rho_{\square}$, $\rho_{\square}\sim 0.25067$}
	\end{subfigure}
	\begin{subfigure}{1\textwidth}
		\centering
		\includegraphics[width=0.4\textwidth]{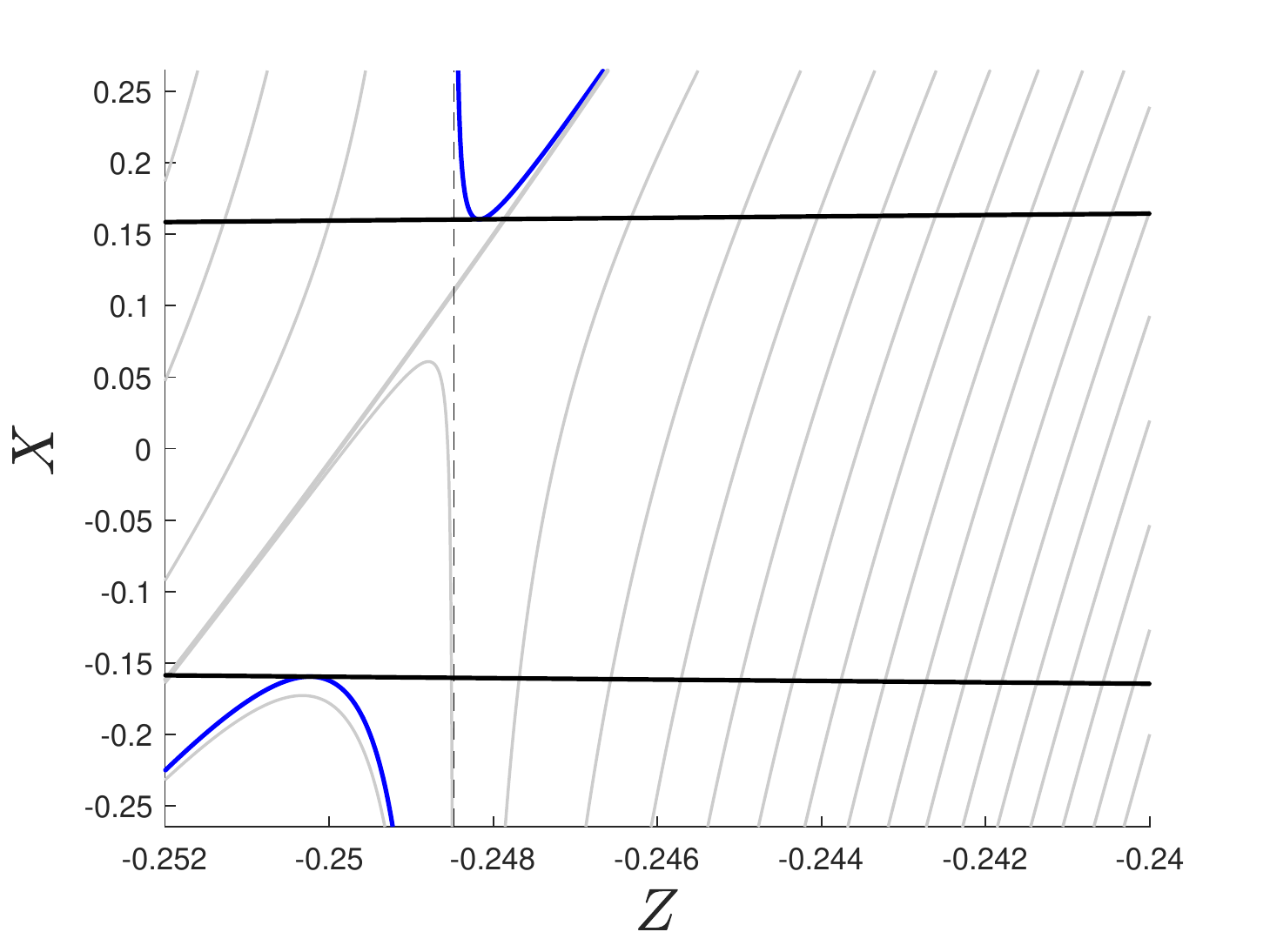}
		\includegraphics[width=0.4\textwidth]{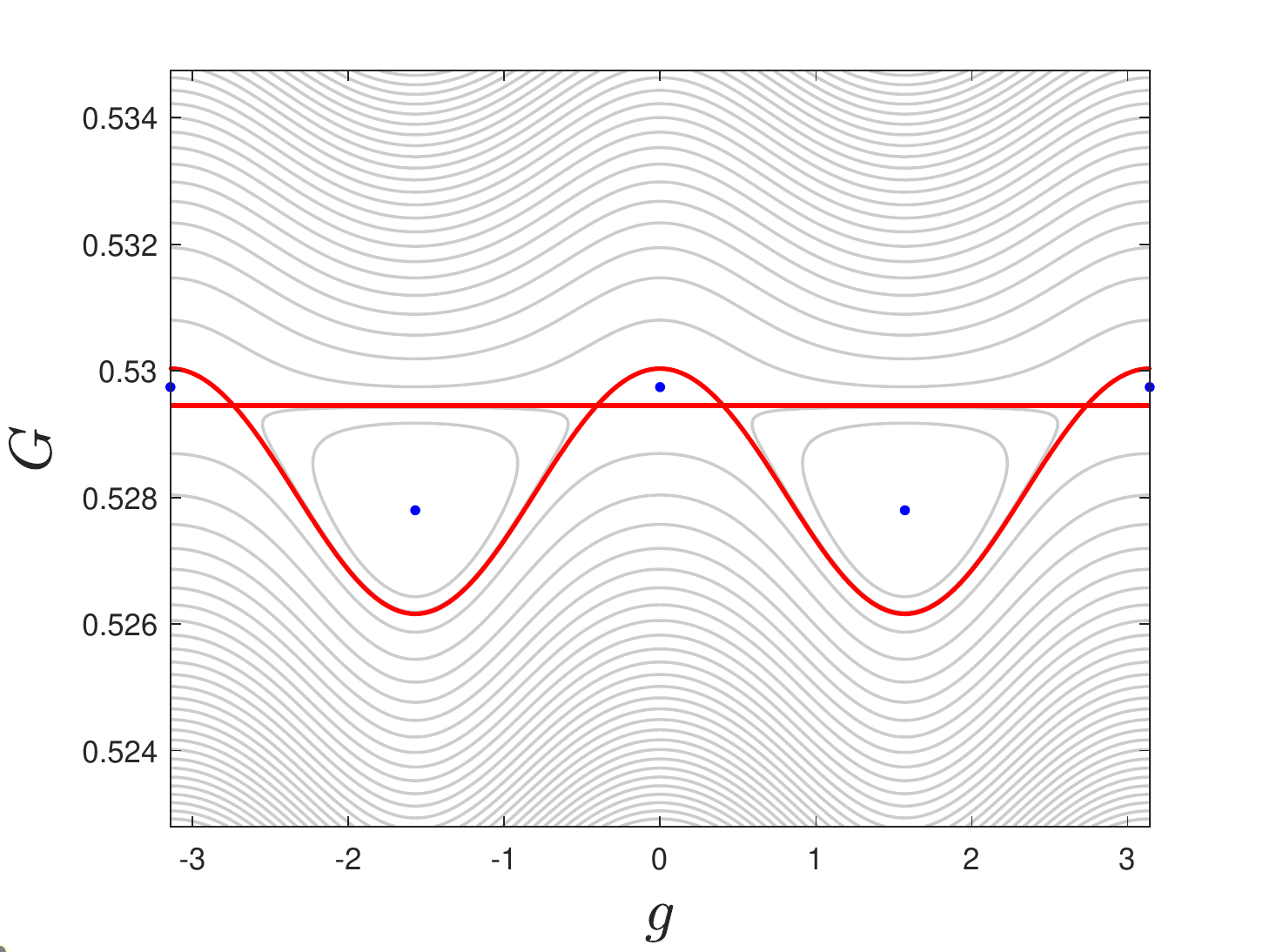}
		\subcaption{$\vert\rho\vert=0.24$}
	\end{subfigure}
	\begin{subfigure}{1\textwidth}
		\centering
		\includegraphics[width=0.4\textwidth]{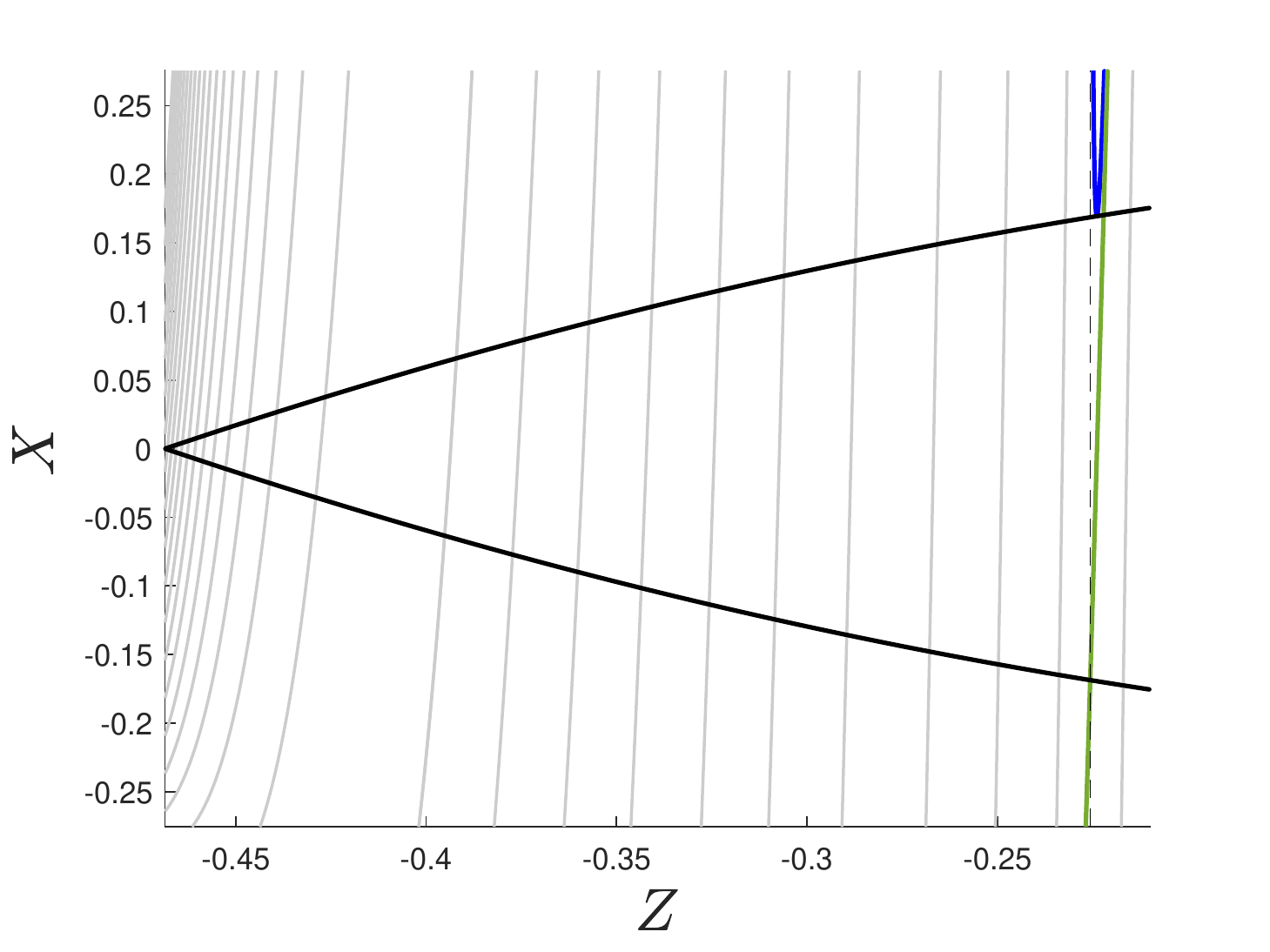}
		\includegraphics[width=0.4\textwidth]{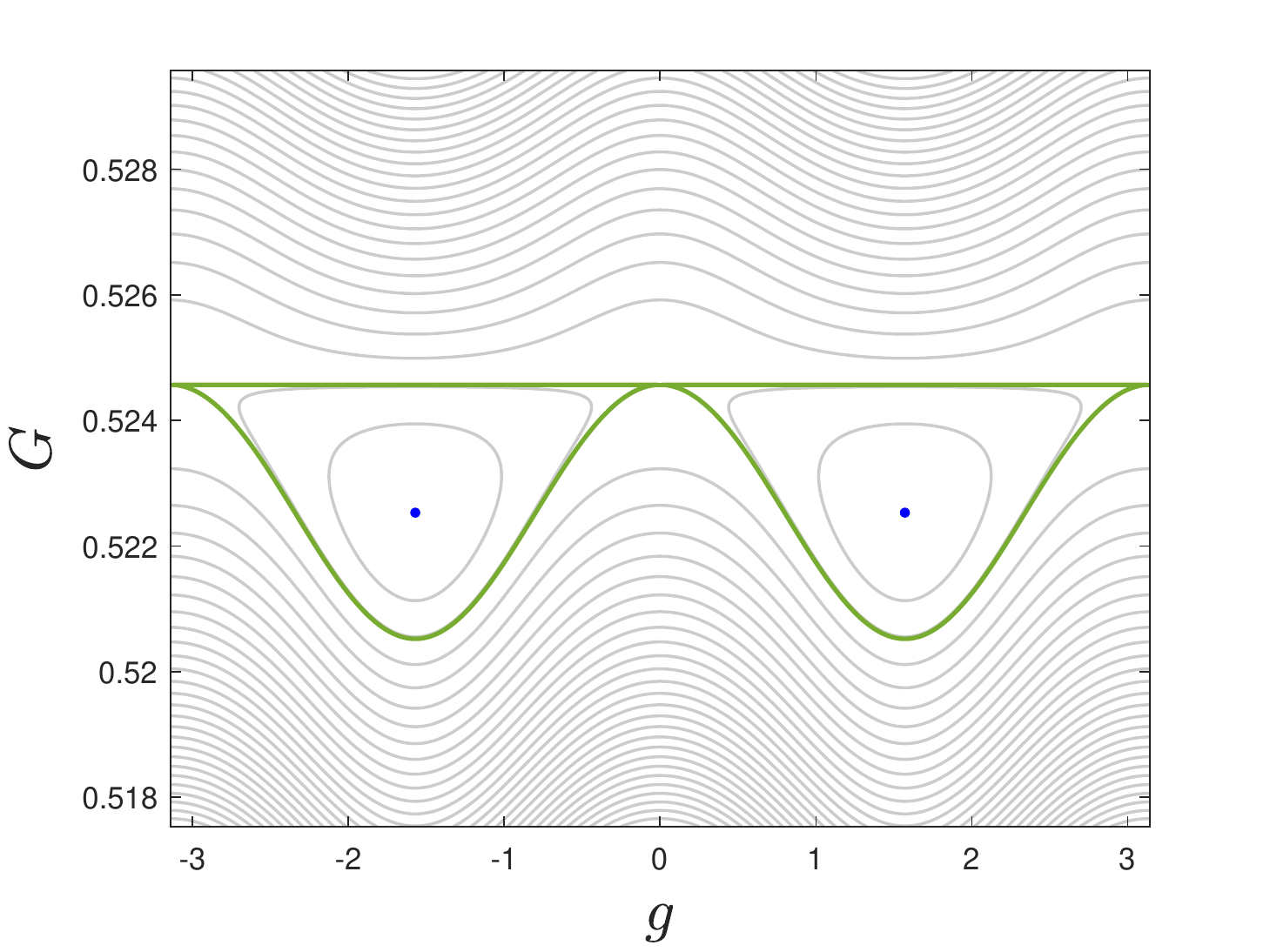}
		\subcaption{$\vert\rho\vert=\rho_{\diamond}$, $\rho_{\diamond}\sim 0.23779$}
	\end{subfigure}
	\begin{subfigure}{1\textwidth}
		\centering
		\includegraphics[width=0.4\textwidth]{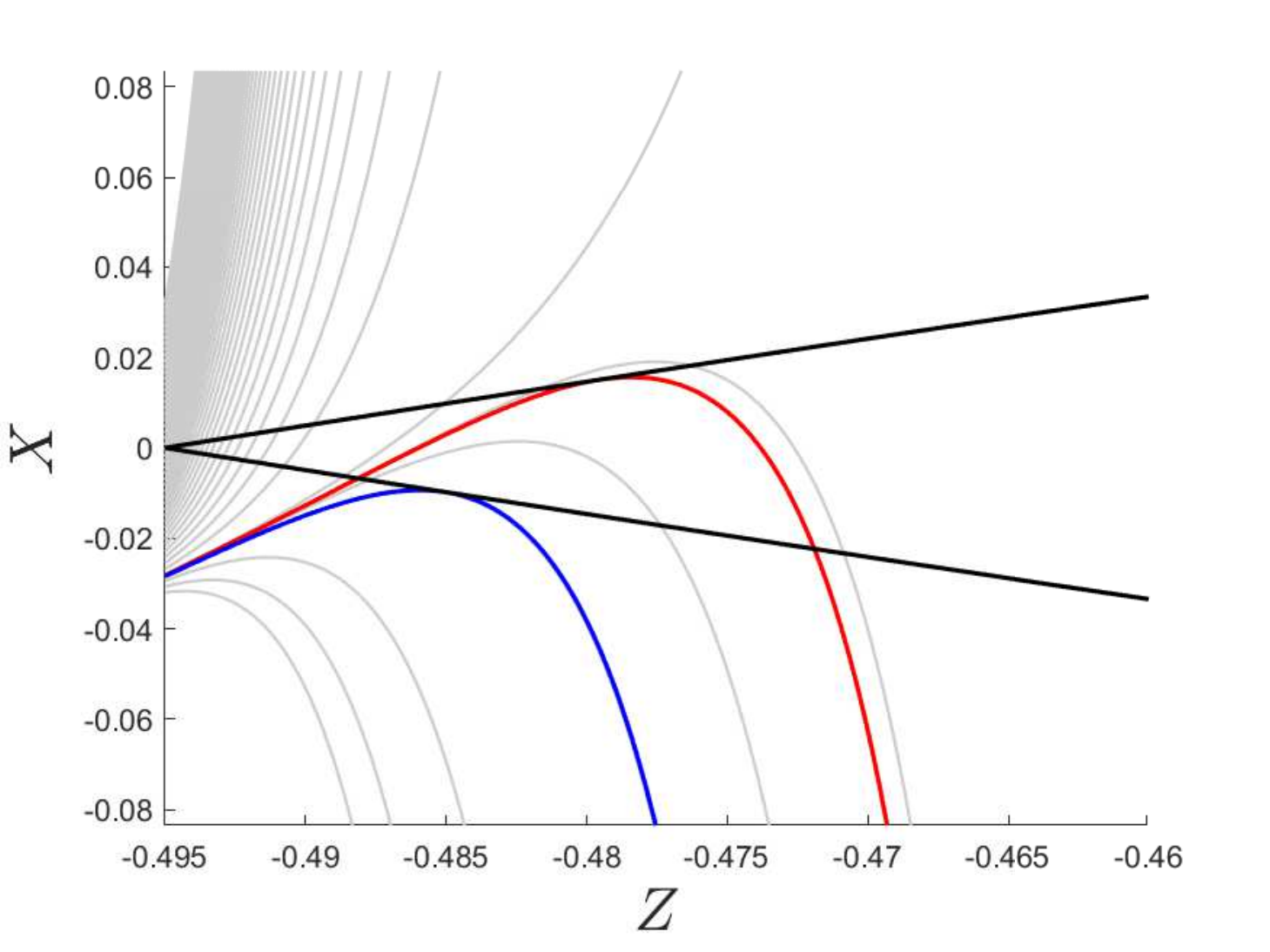}
		\includegraphics[width=0.4\textwidth]{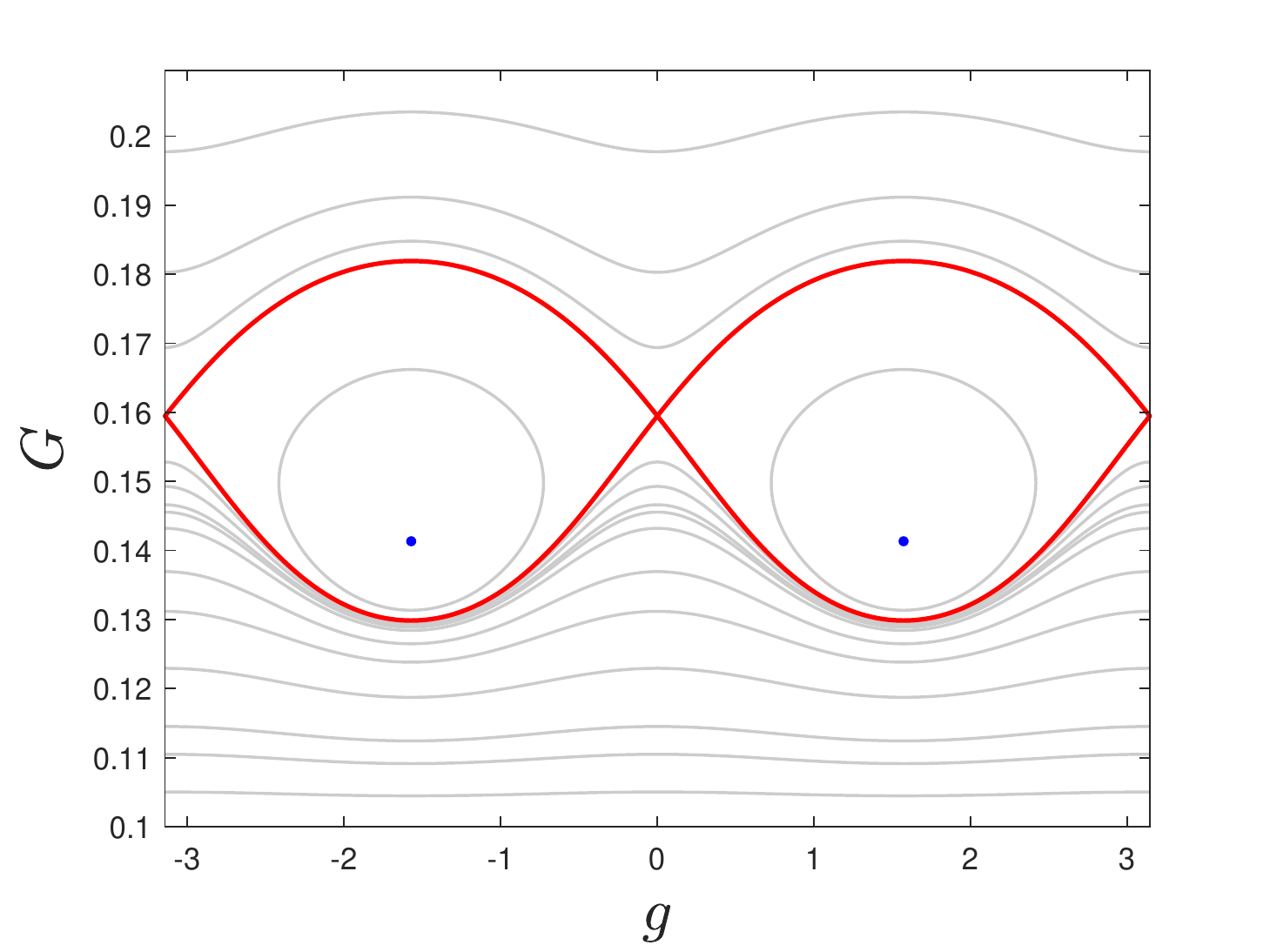}
		\subcaption{$\vert\rho\vert=0.1$}
	\end{subfigure}
	\caption{Level curves for the $J_4$-problem with $\jj4=0.95$ for four different values of $\vert\rho\vert$ in the neighbourhood of the bifurcations $\vert\rho\vert=\rho_{\square}$ and $\vert\rho\vert=\rho_{\diamond}$ on the $(Z,X)$ and the $(g,G)$ planes. Enlargements of the regions containing the equilibrium points are performed. The same colour code employed in Fig.\ref{fig:j413_bif1} is used.}
	\label{fig:j4095_bifYnot0}
\end{figure}

Another significant range of values of $\jj4$ is $(j_{4_{\rm bif 2}},j_{4_{\rm bif 1}})$. Here, it holds $\rho_{+}>\rho_{-}>\rho_{\square}>\rho_{\diamond}>\rho_{\blacktriangledown}>\rho_{\blacktriangle}$. When $\vert\rho\vert>\rho_{\square}$ the dynamical evolution is similar to that occurring in the $\J2$-problem. Instead, for $\vert\rho\vert<\rho_{\diamond}$, it has the same features of the one obtained for $\jj4>{\jj4}_{\rm bif 1}$ when $\vert\rho\vert<\rho_{+}$. The link between these two situations is {established} by the bifurcations $\vert\rho\vert=\rho_{\square}$ and 
$\vert\rho\vert=\rho_{\diamond}$, which cause a variation in the stability of the equilibrium points $E_3$ and $E_4$: 
\begin{itemize}
	\item for $\rho_{\square}<\vert\rho\vert<\rho_{-}$, $E_2$ and $E_3$ are stable, while $E_4$ is unstable; 
	\item for $\vert\rho\vert=\rho_{\square}$, $E_2$ and $E_3$ are stable; 
	the equilibrium points $\bar{E}_1$ and $\bar{E}_2$ coincide with $E_4$ and are degenerate; 
	\item for $\rho_{\diamond}<\vert\rho\vert<\rho_{\square}$ $E_2$, $E_3$ are stable; $\bar{E}_1$ and $\bar{E}_2$ are unstable, while $E_4$ is stable;  
	\item for $\vert\rho\vert=\rho_{\diamond}$  $E_2$ and $E_4$ are stable;  $\bar{E}_1$ and $\bar{E}_2$ coincide with $E_3$ and are degenerate;  
	\item for $\rho_{\blacktriangledown}<\vert\rho\vert<\rho_{\diamond}$, $E_2$ and $E_4$ are stable and $E_3$ is unstable.
\end{itemize}
In Fig.\ref{fig:j4095_bifYnot0}, we show the levels curves in a neighbourhood of the bifurcations $\vert\rho\vert=\rho_{\square}$, and  $\vert\rho\vert=\rho_{\diamond}$. At the bifurcation $\vert\rho\vert=\rho_{\square}$, in the $(Z,X)$ plane there is a level curve intersecting the contour of the \textit{lemon} space at $Z=\bar{Z}$: the intersection point is $E_4$, coinciding with  $\bar{E}_1$ and $\bar{E}_2$. In all the range of values of $\vert\rho\vert$ such that $\bar{E}_1$ and $\bar{E}_2$ exist, there is a level curve for which $Z=\bar{Z}$ is not a singularity. When $\vert\rho\vert=\rho_{\diamond}$ the intersection point is $E_3$. 

Note that for values of $\jj4$ lower and higher than $j_{4_{\rm bif 1}}$, the stability of $E_3$ and $E_4$ is different when they appear after the occurrence of the bifurcations $\vert\rho\vert=\rho_+$ and  $\vert\rho\vert=\rho_-$. A similar result was also found by \citet{Coffey1994}. Here, the authors argued that this change of stability occurs at $\jj4=1$, i.e. when we deal with the so called \textit{Vinti} problem. Instead, we observe that the variation of the stability occurs at $j_{4_{\rm bif 1}}$ given by \eqref{j4V}, which depends on $\lambda$.


To conclude, let us remark once again that the above analysis is general and it does not care about particular physical limitations. For example, one can notice that for low $\vert\rho\vert$, the value of $G$ characterising the equilibrium points is typically small. This implies a large eccentricity. There is then the risk that the resulting distance of the pericentre is smaller than the central body's radius. In such a case, the resulting equilibrium cannot physically exist. For example, if we consider the case of Mars, the equilibrium points resulting from the bifurcations $\vert\rho\vert=\rho_{\blacktriangledown}$ and $\vert\rho\vert=\rho_{\blacktriangle}$ do not exist for $\lambda=0.001$.

\subsection{The $\J2$-problem with relativistic terms}
We study now the zonal problem containing both the $J_2$ and the relativistic terms. From \eqref{KP}, \eqref{NFZ1}, \eqref{HZ21} and \eqref{NFZ2}, the closed form is
\[ \begin{split} \mathcal{K}_{c}& =-\frac{\mu^2}{2 L^2}+\frac{\mu^4  J_2 R_{P}^2 (G^2 - 3H^2)}{4 G^5 L^3 }+ \frac{\mu^4}{c^2 L^4 G }\left(5G-8L\right)+\frac{3 \mu^6 J_2^2 R_{P}^4}{128 L^5 G^{11} }  \big[ -5 G^6 - 4 G^5 L \\ & \hspace{5mm}+ 24 G^3 H^2 L - 36 G H^4 L - 35 H^4 L^2 +  G^4 (18 H^2 + 5 L^2) \\ & \hspace{5mm}- 5 G^2 (H^4 + 2 H^2 L^2) +  2 (G^2 - 15 H^2) (G^2 - L^2) (G^2 - H^2) \cos 2 g \big] \\ & \hspace{5mm}  +\frac{\mu^6 J_2 R_{P}^2}{8 c^2 L^5 G^7}\big[(G^2 - 3 H^2) (6 L^2 - 5 G^2)-6(G^2 - 3 H^2) (4 G^2 - 3 GL - 5 L^2) \\ & \hspace{5mm}  - 9  (L^2 - G^2) (G^2 - H^2) \cos 2 g\big]. \end{split} \]
We neglect here the $J_4$ terms to make evident the effects of the relativistic contribution.

We adopt the same non-dimensional system described in Section \ref{sec:J2}. Let us set 
\[
j_C = \frac{1}{\lambda c^2},
\]
with $\lambda$ defined in \eqref{lambdadef}. We recall that $\lambda$ was considered of the same order as the book-keeping parameter $\epsilon$. Since the normalisation of the initial Hamiltonian was performed by assuming $c^{-2}$ of order $\epsilon$ as well (see Section \ref{sec:NF}), $\jjC$ should have a value in the neighbourhood of $1$ or lower. If this was not the case, the book-keeping scheme used to compute the closed form would not be suitable anymore. Let us also remark that in the adimensional system the value of $c$ and, thus, that of $\jjC$ depend on the units of length and time, i.e. on the semi-major axis of the orbit of interest. 
	
	\noindent We introduce $\lambda$ and $\jjC$ in the Hamiltonian. Then, we neglect the constant terms and we perform the time transformation \eqref{timetrans}. Also for this problem the resulting normalised Hamiltonian has the same structure of \eqref{normalformstructure}, with
	\[
	\begin{split}
	g(Z,\bm{a}) &=  \frac {-5{{\rho}}^{2}+2Z+1}{\sqrt{2} \left( {{\rho}}^{2}+2Z+1
		\right) ^{\frac{5}{2}}}+\jjC\frac{3}{8}\,\frac{5\sqrt{2\rho^2+4Z+2}-16}{\sqrt{2\rho^2+4Z+2}} -\frac{3\lambda}{16\sqrt{2}\left(\rho^{2}+2Z+1\right)^{\frac{11}{2}}}\Big( 40{Z}^{3} \\ & \quad + \left( -84{{\rho}}^{2}+20 \right) {Z}^{2}+ \left( -
	74{{\rho}}^{4}-44{{\rho}}^{2}-10 \right) Z-11{{\rho}}^{
		6}+273{{\rho}}^{4}-{{\rho}}^{2}-5 \\
	& \quad  +4\sqrt {2\,{{\rho}}^{2}+4Z+2} \left(-5{{\rho}}^{2}+2Z+1\right) ^{2}\Big) +  \frac{\jjC\lambda(-5\rho^2+2Z+1)}{2\sqrt{2}\left(\rho^{2}+2Z+1\right)^{\frac{7}{2}}}(-29\rho^2\\& \quad +18\sqrt{2\rho^2+4Z+2}-58Z+43),\\
	f(Z,\bm{a}) & =  -\frac{3}{2}\,\lambda {\frac { \left( -29\,{{\rho}}^{2}+2\,Z+1 \right) }{ \sqrt{2}\left( {
				{\rho}}^{2}+2\,Z+1 \right) ^{\frac{11}{2}}}}-\frac{18\lambda\jjC }{ \sqrt{2}\left( {
			{\rho}}^{2}+2Z+1 \right) ^{\frac{7}{2}}},
			\end{split}
	\]
	and $\bm{a}=(\rho; \lambda,j_C)$. 
	If we neglect the terms of first order in $\lambda$, we find two potential equilibrium solutions at
	\begin{equation}
	Z = \frac{1}{8}\,\frac{-4j_C\rho^2-4\jjC+1\pm 	\sqrt{-80\jjC\rho^2+1}}{\jjC}, \qquad \forall X,Y.
	\label{rel_order0}
	\end{equation}
	In the following, we make some considerations about the problem considering $\lambda\in(0,1)$. For this problem, we perform a qualitative analysis. We find out that for $\jjC\ll 1$, the sequence of bifurcations is the same as in the $J_2$ problem. On the contrary for higher values of $\jjC$, the dynamical evolution is more complex and depends on the values of $\lambda$ and $\jjC$. The existence of a pair of equilibrium points of type $E_+$, one stable and the other unstable, is triggered by a saddle-node bifurcation. The unstable point can become stable following a pitchfork bifurcation, which affects the existence of the equilibrium points $\bar{E}_1$ and $\bar{E}_2$. The stable one can disappear following a pitchfork bifurcation, which changes the stability of the equilibrium point $E_2$. A similar sequence of bifurcations occurs also concerning the equilibrium points of type $E_-$. If none of the bifurcations affecting the stability of $E_2$ occur, this point is always stable. The equilibrium point $E_1$ is always stable.

	\subsubsection{About the stability of $E_1$}
	We have
	\begin{equation}
	s_+(-\ZMAX;\bm{a}) = \frac {(8\rho^6-110\lambda\rho^4+84\vert \rho\vert^3\lambda+186\lambda\rho^2)\jjC+8\rho^4-7\lambda\rho^2+12\lambda\vert\rho\vert+31\lambda}{\lambda(12\jjC\rho^2-7)},
	\end{equation}
	and
	\begin{equation}
	s_-(-\ZMAX;\bm{a}) = \frac {(4\rho^6-61\lambda\rho^4+42\vert\rho\vert^3\lambda+99\lambda\rho^2)\jjC+4\rho^4+6\lambda\vert\rho\vert+12\lambda}{\lambda(12\jjC\rho^2-7)}.
	\end{equation}
	It holds
	\[
	\rho^2(-110\rho^2+84\vert \rho\vert+186)>0, \quad \forall \rho \in(0,1)
	\] 
	and 
	\[
	\rho^2(-61\rho^2+42\vert\rho\vert+99)>0, \quad \forall \rho \in(0,1).
	\]
	Thus, if $\rho^2>7/12\jjC$, both $s_+(-\ZMAX;\bm{a})$ and $s_-(-\ZMAX;\bm{a})$ are positive; instead, if $\rho^2<7/12\jjC$ they are both negative. From \eqref{careq_E1}, we can conclude that $E_1$ is always stable. Moreover, $E_1$ never coincides with an equilibrium point of either type $E_+$ or $E_-$.  
	
	\subsubsection{About the stability of $E_2$}
	We have
	\begin{equation}
	s_+(\ZMAX;\bm{a}) = \frac {425\lambda\rho^4+(1672\jjC\lambda-146\lambda+80)\rho^2-392\jjC\lambda+64\jjC+9\lambda-16}{2(-15\rho^2+24\jjC+1)\lambda}.
	\label{sPE3_c}
	\end{equation}
	It holds $s_+(\ZMAX;\bm{p})=0$ for $\vert\rho\vert=\tilde{\rho}_{+}$, with
	\begin{equation}
	\tilde{\rho}^2_{+} = \frac{-836\jjC\lambda+73 \lambda-40+4\sqrt{43681\jjC^2\lambda^2+2784\jjC\lambda^2+2480\jjC\lambda+94\lambda^2+60\lambda+100}}{425\lambda},
	\label{rhoP_c}
	\end{equation}
	which is positive, thus admissible, if either $\lambda\ge\frac{8}{49}$ or  $\lambda<\frac{8}{49}$ and $\jjC <\frac{16-9\lambda}{64-392\lambda}$. 
	We also have
	\begin{equation}
	s_-(\ZMAX;\bm{a}) = \frac {365\lambda\rho^4+(1768\jjC\lambda-82\lambda+80)\rho^2-488\jjC\lambda+64\jjC+5\lambda-16}{2(-15\rho^2+24\jjC+1)\lambda},
	\label{sPE4_c}
	\end{equation}
	and $s_-(\ZMAX;\bm{a})=0$ for $\vert\rho\vert=\tilde{\rho}_{-}$, with
	\begin{equation}
	\tilde{\rho}^2_{-} = \frac{-884\jjC\lambda+41 \lambda-40+4\sqrt{48841\jjC^2\lambda^2+6602\jjC\lambda^2+2960\jjC\lambda-9\lambda^2+160\lambda+100}}{365\lambda};
	\label{rhoM_c}
	\end{equation}
	$\tilde{\rho}^2_{-}>0$ if either $\lambda\ge\frac{8}{61}$ or  $\lambda<\frac{8}{61}$ and $\jjC <\frac{16-5\lambda}{64-488\lambda}$. 
	Let us remark that for $\lambda<\frac{8}{61}$ it holds $\frac{16-5\lambda}{64-488\lambda}>\frac{16-9\lambda}{64-392\lambda}$. Thus, if $\tilde{\rho}_+>0$ is an admissible solutions, also $\tilde{\rho}_-$ is admissible. 
	
	For each $\lambda$ and $\jjC$ such that both $\tilde{\rho}_-$ and $\tilde{\rho}_+$ are admissible zeros of $s_+(\ZMAX,\bm{a})$ and $s_-(\ZMAX,\bm{a})$, it holds $\tilde{\rho}_->\tilde{\rho}_+$ if $	\jjC>\jCL$, with
	\[
	\jCL = \frac{397\lambda-180+\sqrt{142321\lambda^2-1800\lambda+32400}}{\lambda}.
	\]
	Let us now consider equation \eqref{careq_E2}. If $\lambda$ and $\jjC$ are such that neither  $\tilde{\rho}_-$ and $\tilde{\rho}_+$ are admissible zeros, then $E_2$ is always stable. Also for $\jjC=\jCL$, $E_2$ is always stable, except when $\vert\rho\vert=\tilde{\rho}_+=\tilde{\rho}_-$: in this case, it is degenerate. If $\lambda$ and $\jjC$ are such that $\tilde{\rho}_-$ is an admissible solution, while $\tilde{\rho}_+^2\le0$, $E_2$ is stable for $\vert\rho\vert>\tilde{\rho}_-$, it is degenerate at $\vert\rho\vert=\tilde{\rho}_-$ and is unstable for $\vert\rho\vert<\tilde{\rho}_-$. Finally, if both $\tilde{\rho}_-$ and $\tilde{\rho}_+$ are admissible solutions, $E_2$ is unstable when the value of $\vert\rho\vert$ lies between $\tilde{\rho}_-$ and $\tilde{\rho}_+$, it is degenerate if either $\vert\rho\vert=\tilde{\rho}_-$ or $\vert\rho\vert=\tilde{\rho}_+$ and it is stable for all the other values of $\vert\rho\vert$. When 
	$\vert\rho\vert=\tilde{\rho}_+$, $E_2$ coincides with an equilibrium point of type $E_+$. When $\vert\rho\vert=\tilde{\rho}_-$ it coincides with an equilibrium point of type $E_-$. 
	
	\subsubsection{About the existence of the equilibrium points of type $E_+$ and $E_-$}
	To discuss the existence of equilibrium points of type $E_+$ and $E_-$, we use here the same strategy adopted for the problems previously analysed. 
	
	We have
	\[
	s_+\left(G^2-\frac{1+\rho^2}{2};\bm{a}\right) = \frac{\tilde{S}_{+}(G;\bm{a})}{4G^2\lambda(24G^4\jjC+G^2-15\rho^2)},
	\]
	with
	\begin{equation*}
	\begin{split}
	\tilde{S}_+(G;\bm{a}) = & (-225 G^2\lambda+360 G\lambda+715\lambda)\rho^4+(-2080G^6\jjC\lambda+1728 G^5\jjC\lambda\\
	&+160 G^6+3696G^4\jjC\lambda+98 G^4\lambda-192G^3\lambda-198G^2\lambda)\rho^2\\
	&+128G^{10}\jjC
	+320G^8\jjC\lambda-384G^7\jjC\lambda-32G^8-720G^6\jjC\lambda\\
	&+15G^6\lambda+24G^5\lambda-21G^4\lambda.
	\end{split}
	\label{S_cP}
	\end{equation*}
	We obtain $\tilde{S}_+(G;\bm{a})=0$ for  $\rho^2=\tilde{\rho}^2_{E_{+_{{1,2}}}}$, with
	\begin{equation}
	\tilde{\rho}^2_{E_{+_{{1,2}}}} = \frac{G^2}{5\lambda}\frac{\tilde{A}_+\pm4\sqrt{\tilde{B}_+}}{\tilde{C}_+},
	\label{rho2sol_cP}
	\end{equation}
	\begin{align*}
	\tilde{A}_+ = & -(-1040G^4\jjC+864G^3\jjC+1848G^2\jjC+49G^2-96G-99)\lambda-80G^4,
	\\
	\tilde{B}_+ = &  \frac{\tilde{A}_+^2 +5\lambda \tilde{C}_+\tilde{D}_+}{16}, \qquad 
	\tilde{C}_+ = -45G^2+72G+143,\\
	\tilde{D}_+ = &    (-320G^4\jjC+384G^3\jjC+720G^2\jjC-15G^2-24G+21)\lambda-128G^6\jjC+32G^4        .
	\end{align*}
	Note that for $G^2<1/4\jjC$,  $\tilde{D}_+>0$; instead, for $G^2>1/4\jjC$, $\tilde{A}_+<0$. Thus, $\forall \lambda$, $\forall \jjC$, $\forall G$, $\tilde{\rho}^2_{E_{+_{{2}}}}<0$ and it is not admissible as solution. While $\tilde{\rho}^2_{E_{+_{{1}}}}>0$ if $G$, $\lambda$ and $\jjC$ are such that $\tilde{D}_+>0$. Since $5\lambda\tilde{C}_+-\tilde{A}_+>0$ and $16\tilde{B}_+-(5\lambda\tilde{C}_+-\tilde{A}_+)^2<0$, 
	it holds $\tilde{\rho}^2_{E_{+_{{1}}}}<G^2$. 
	When admissible,  $\tilde{\rho}^2_{E_{+_{{1}}}}$ is generally not monotone with respect to $G$. However, for $\jjC=0$ it is equal to the same solution found for the $\J2$-problem, i.e. $\tilde{\rho}^2_{E_{+_{{1}}}}={\rho}^2_{E_{+_{{2}}}}$ (see Section \ref{sec:j2prob_EpEm}). As a consequence, we expect that for sufficiently small values of $\jjC$,  $\tilde{\rho}^2_{E_{+_{{1}}}}$ is an increasing function of $G$ in the range of interest, i.e. $G\in(0,1]$. In this case, for $\vert\rho\vert<\tilde{\rho}_+$, there exists only one equilibrium point of type $E_+$. Instead, for higher values of $\jjC$, such that $\tilde{\rho}^2_{E_{+_{{1}}}}$ is not monotone, the outcome is different. Let us call ${\rho}_{\blacklozenge}$ the value of $\vert\rho\vert$ such that
	\[
	{\rho}_{\blacklozenge} = \sqrt{\max_{G} \tilde{\rho}^2_{E_{+_{{1}}}}}.
	\]
	We have that
	\begin{itemize}
		\item for $\vert\rho\vert>{\rho}_{\blacklozenge}$, there is no equilibrium point of type $E_+$;
		\item for $\vert\rho\vert={\rho}_{\blacklozenge}$, we have one equilibrium solution, which we call $E_{13}$;
		\item for $\vert\rho\vert<{\rho}_{\blacklozenge}$ there exist multiple equilibrium solutions, typically two which we call $E_{15}$ and $E_{17}$.
	\end{itemize}
	Let us suppose that the $Z$ coordinate of $E_{17}$ is larger than that of $E_{15}$. When $\lambda\ge\frac{8}{49}$ or when $\lambda<\frac{8}{49}$ and $\jjC <\frac{16-9\lambda}{64-392\lambda}$, $E_2$ coincides with $E_{17}$ for $\vert\rho\vert=\tilde{\rho}_+$. Thus, for $\vert\rho\vert<\tilde{\rho}_+$, the number of equilibrium solutions reduces to one: there will exist only $E_{15}$.
	
	In conclusion, we can infer that {reducing the value of} $\jjC$, the value $G=G_{\blacklozenge}$, corresponding to the maximum point of $\vert\tilde{\rho}^2_{E_{+_{{1}}}}\vert$, increases. For a fixed $\lambda$, it exists a value of $\jjC$ such that $G_{\blacklozenge}=1$, i.e. for which ${\rho}_{\blacklozenge}=\tilde{\rho}_+$. Thus, for lower values of $\jjC$, the bifurcation $\vert\rho\vert={\rho}_{\blacklozenge}$ disappears and the only existing equilibrium point of type $E_+$ is $E_{15}$ for $\vert\rho\vert<\tilde{\rho}_+$. 
	
	As far as the equilibrium points of type $E_-$, we have
	\[
	s_-\left(G^2-\frac{1+\rho^2}{2};\bm{a}\right) = \frac{\tilde{S}_{-}(G;\bm{a})}{4G^2\lambda(24G^4\jjC+G^2-15\rho^2)},
	\]
	with
	\begin{equation*}
	\begin{split}
	\tilde{S}_-(G;\bm{a}) = & (315G^2\lambda+360G\lambda+55\lambda)\rho^4+(-2560G^6\jjC\lambda+1728G^5\jjC\lambda+160G^6\\
	&+4368G^4\jjC\lambda-350G^4\lambda-192G^3\lambda+378G^2\lambda)\rho^2+128G^{10}\jjC\\
	&+608G^8\jjC\lambda-384G^7\jjC\lambda-32G^8-1200G^6\jjC\lambda+35G^6\lambda\\
	&+24G^5\lambda-49G^4\lambda.
	\end{split}
	\label{S_cP}
	\end{equation*}
	It holds $\tilde{S}_-(G;\bm{p})=0$ if $\rho^2=\tilde{\rho}^2_{E_{-_{1,2}}}$, with
	\begin{equation}
	\rho^2=\tilde{\rho}^2_{E_{-_{1,2}}} = \frac{G^2}{5\lambda}\frac{\tilde{A}_-\pm4\sqrt{\tilde{B}_-}}{\tilde{C}_-},
	\label{rho2sol_cM}
	\end{equation}
	\begin{align*}
	\tilde{A}_- = & (1280G^4\jjC-864G^3\jjC-2184G^2\jjC+175G^2+96G-189)\lambda-80G^4,
	\\
	\tilde{B}_+ = &  \frac{\tilde{A}_-^2 +5\lambda \tilde{C}_-\tilde{D}_-}{16},
	\qquad 
	\tilde{C}_- =  63G^2+72G+11,\\
	\tilde{D}_- = &    (-608G^4\jjC+384G^3\jjC+1200G^2\jjC-35G^2-24G+49)\lambda-128G^6\jjC+32G^4.
	\end{align*}
	One can observe that for $G^2<1/4\jjC$,  $\tilde{D}_->0$ and that for $G^2\ge1/4\jjC$,  $\tilde{A}_-<0$. Thus, $\forall \lambda$, $\forall \jjC$ and $\forall G$, $\tilde{\rho}^2_{E_{-_{2}}}<0$. Instead, for $G$, $\jjC$ and $\lambda$ such that $\tilde{D}_->0$, $\tilde{\rho}^2_{E_{-_{1}}}>0$. Since $5\lambda \tilde{C}_+-\tilde{A}_+>0$ and $16\tilde{B}_+-(5\lambda \tilde{C}_+-\tilde{A}_+)^2<0$, it also holds
	$\tilde{\rho}^2_{E_{-_{1}}}<G^2$. Thus, there exist values of $G$, $\jjC$ and $\lambda$ such that $\tilde{\rho}^2_{E_{-_{1}}}$ is an admissible solution. As $\tilde{\rho}^2_{E_{+_{1}}}$, in general the function  $\tilde{\rho}^2_{E_{-_{1}}}$ is not monotone with respect to  $G$. We find an outcome similar to  the one obtained for the equilibrium points of type $E_+$. Let us consider sufficiently high values of $\jjC$ such that  $\tilde{\rho}^2_{E_{-_{1}}}$ is not monotone and let us set
	\[
	\rho_{\blacksquare} =  \sqrt{\max_{G} \tilde{\rho}^2_{E_{-_{1}}}}.
	\]
	We have that
	\begin{itemize}
		\item for $\vert\rho\vert>\rho_{\blacksquare}$, there is no equilibrium point of the type of $E_-$;
		\item for $\vert\rho\vert=\rho_{\blacksquare}$, we have one equilibrium solution, which we call $E_{14}$;
		\item for $\vert\rho\vert<\rho_{\blacksquare}$ there exist multiple equilibrium solutions, typically two which we call $E_{16}$ and $E_{18}$.
	\end{itemize}
	Suppose that $E_{18}$ has a larger $Z$ coordinate than $E_{16}$. When $\lambda\ge\frac{8}{61}$ or when $\lambda<\frac{8}{61}$ and $\jjC <\frac{16-5\lambda}{64-488\lambda}$, at $\vert\rho\vert=\tilde{\rho}_-$ $E_{18}$ coincides with $E_{2}$ and for $\vert\rho\vert<\tilde{\rho}_-$ it disappears. For a fixed $\lambda$, by considering decreasing values of $\jjC$ the value of $G$, $G=G_{\blacksquare}$, corresponding to the maximum point of $\tilde{\rho}^2_{E_{-_{1}}}$, increases. Below the value of $\jjC$ for which $\rho_{\blacksquare}=\tilde{\rho}_-$, $G_{\blacksquare}$ does not belong to the admissible range of values for $G$. In these cases, there only exists the equilibrium point $E_{16}$ for $\vert\rho\vert<\tilde{\rho}_-$.  
	
	\subsubsection{About the existence of $\bar{E}_1$ and $\bar{E}_2$}
	The coordinates $\bar{X}$ and $\bar{Z}$ of the two equilibrium points of type $E_+$ are
	\[
	\bar{Z} = \frac{1}{48}\,\frac{-24\jjC\rho^2+\sqrt{1440\jjC\rho^2+1}-24\jjC-1}{\jjC},
	\]
	and
	\[
	\begin{split}
	\bar{X} =& \frac{1}{20736}\frac{1}{\jjC^3\lambda\left((720\jjC\rho^2+1)\sqrt{1440\jjC\rho^2+1}-1440\jjC\rho^2-1\right)}\Bigg(+144\sqrt{3}\jjC^2\lambda\big((103680\jjC^2\rho^4\\
	&+3312\jjC\rho^2+5)\sqrt{1440\jjC\rho^2+1}-1192320\jjC^2\rho^4-6912\jjC\rho^2-5\big)\sqrt{\frac{\sqrt{1440\jjC\rho^2+1}-1}{\jjC}}\\
	&+(361428480\jjC^4\lambda\rho^4+27552960\jjC^3\lambda\rho^4+3903552\jjC^3\lambda\rho^2-3369600\jjC^2\rho^4+19584\jjC^2\lambda\rho^2\\
	&+1080\jjC^2\lambda-17280\jjC\rho^2-51\jjC\lambda-14)\sqrt{1440\jjC\rho^2+1}-5244134400\jjC^4\lambda\rho^6\\
	&-2892049920\jjC^4\lambda\rho^4+559872000\jjC^3\rho^6-54872640\jjC^3\lambda\rho^4-4681152\jjC^3\lambda\rho^2+12182400\jjC^2\rho^4\\
	& +17136\jjC^2\lambda\rho^2-1080\jjC^2\lambda+27360\jjC\rho^2+51\jjC\lambda+14 \Bigg).
	\end{split}
	\]
	To have $\bar{Z}\in[\ZMAX,\ZMAX]$, $\rho^2<\min\left(\frac{24\jjC+1}{15},\frac{7}{12\jjC}\right)$.
	Let us set $\mathcal{Y}=\bar{Y}_{1,2}^2$. In general, for given $\jjC$ and $\lambda$, it can exists a subset of values of $\rho$ such that $\mathcal{Y}>0$, i.e. such that $\bar{E}_1$ and $\bar{E}_2$ exist. The endpoints of this range are values of $\rho$ for which $\bar{E}_1$ and $\bar{E}_2$ coincide with either an equilibrium point of type $E_+$ or $E_-$. Let us call $\rho_{\diamond}$ the value of $\vert\rho\vert$ such that $\bar{E}_1$ and $\bar{E}_2$ coincide with an equilibrium point of type $E_+$ and $\rho_{\square}$ the the value of $\vert\rho\vert$ such that they coincide with an equilibrium point of type $E_-$. We can conclude that necessarily  $\rho_{\diamond}<\rho_{\blacklozenge}$ and $\rho_{\square}<\rho_{\blacksquare}$. For $\jjC\rightarrow 0$ we obtain instead the same outcome found for the $\J2$-problem: for sufficiently small values of $\jjC$, there does not exist any value of $\rho$ for which $\bar{E}_1$ and $\bar{E}_2$ exist. 
	
	\subsubsection{About the stability of the equilibrium points of type $E_+$ and $E_-$ and of $\bar{E}_1$ and $\bar{E}_2$}
	
	Let us consider value of $\jjC$ sufficiently high, such that $E_{15}$, $E_{16}$, $E_{17}$ and $E_{18}$ exist. We can assume that these equilibrium points are close to the equilibrium solutions \eqref{rel_order0} of the problem at order zero in $\lambda$. With this hypothesis, we can estimate their stability. To this aim, we need to assume $\jjC\rho^2<1/80$. At order zero in $\lambda$ we obtain the same equations for the equilibrium points $E_{15}$ and $E_{16}$, i.e.
	\[ \begin{split} \frac{d^2\tilde{X}}{dZ^2}\pm\frac{d^2\hat{X}}{dZ^2}\sim & 16\sqrt{-80 j_C\rho^2+1}\Big(-144000 j_C^3\rho^6+28400\jjC^2\rho^4-880 j_C\rho^2+7+\\
	&\sqrt{-80\jjC\rho^2+1}(10000 j_C^2\rho^4-600 j_C\rho^2+7)\Big).
	\end{split} \] 
	
	The same holds for $E_{17}$ and $E_{18}$:
	\[ \begin{split} \frac{d^2\tilde{X}}{dZ^2}\pm\frac{d^2\hat{X}}{dZ^2}\sim & 16\sqrt{-80\jjC\rho^2+1}\Big(144000\jjC^3\rho^6-28400\jjC^2\rho^4+880\jjC\rho^2-7+\\
	&\sqrt{-80\jjC\rho^2+1}(10000\jjC^2\rho^4-600\jjC\rho^2+7)\Big); \end{split} \]
	
	From the equations we obtain that for $7/810<\jjC\rho^2\le 1/80$, $E_{15}$ and $E_{18}$ are unstable, while $E_{16}$ and $E_{17}$ are stable; instead for $\jjC\rho^2<7/810$, $E_{15}$ and $E_{17}$ are both stable, while $E_{16}$ and $E_{18}$ are both unstable. From this zero-order analysis and by applying the Poincaré-Hopf theorem we can infer the actual dynamical evolution: 
	\begin{itemize}
		\item for $\vert\rho\vert={\rho}_{\blacklozenge}$, there exists one equilibrium solution $E_{13}$ which is degenerate; 
		
		\item for $\rho_{\diamond}<\vert\rho\vert<{\rho}_{\blacklozenge}$, there exist $E_{15}$, which is unstable and $E_{17}$, which is stable; 
		
		\item for $\vert\rho\vert=\rho_{\diamond}$, $E_{15}$ coincides with $\bar{E}_1$ and $\bar{E}_2$ and it is degenerate; $E_{17}$ is stable;
		
		\item for $\vert\rho\vert<\rho_{\diamond}$, both $E_{15}$ and $E_{17}$ are stable. 
	\end{itemize}
	Something similar occurs concerning the equilibrium points $E_{16}$ and $E_{18}$:
	\begin{itemize}
		\item for $\vert\rho\vert=\rho_{\blacksquare}$, there is one equilibrium solution $E_{14}$ which is degenerate; 
		
		\item for $\rho_{\square}<\vert\rho\vert<\rho_{\blacksquare}$,  there exist the two equilibrium solutions $E_{16}$ which is stable and $E_{18}$ which is unstable; 
		
		\item for $\vert\rho\vert=\rho_{\square}$, $E_{16}$ coincides with $\bar{E}_1$ and $\bar{E}_2$ and it is degenerate; $E_{18}$ is unstable;
		
		\item for $\vert\rho\vert<\rho_{\square}$,  both $E_{17}$ and $E_{18}$ are unstable.
	\end{itemize}
	If $\rho_{\square}<\rho_{\diamond}$, $\bar{E}_1$ and $\bar{E}_2$  are unstable. On the contrary if $\rho_{\square}>\rho_{\diamond}$ $\bar{E}_1$ and $\bar{E}_2$  are stable. 
	Finally, if $\lambda\ge\frac{8}{49}$ or if  $\lambda<\frac{8}{49}$ and $\jjC <\frac{16-9\lambda}{64-392\lambda}$, for $\vert\rho\vert<\tilde{\rho}_+$ $E_{17}$ disappears while the stability of $E_{15}$ remains unaltered. Similarly if $\lambda\ge\frac{8}{61}$ or if $\lambda<\frac{8}{61}$ and $\jjC <\frac{16-5\lambda}{64-488\lambda}$, for $\vert\rho\vert<\tilde{\rho}_-$ $E_{18}$ disappears, while the stability of $E_{16}$ does not change. 
	\begin{figure}
		\centering
		\begin{subfigure}{1\textwidth}
			\centering
			\includegraphics[width=0.4\textwidth]{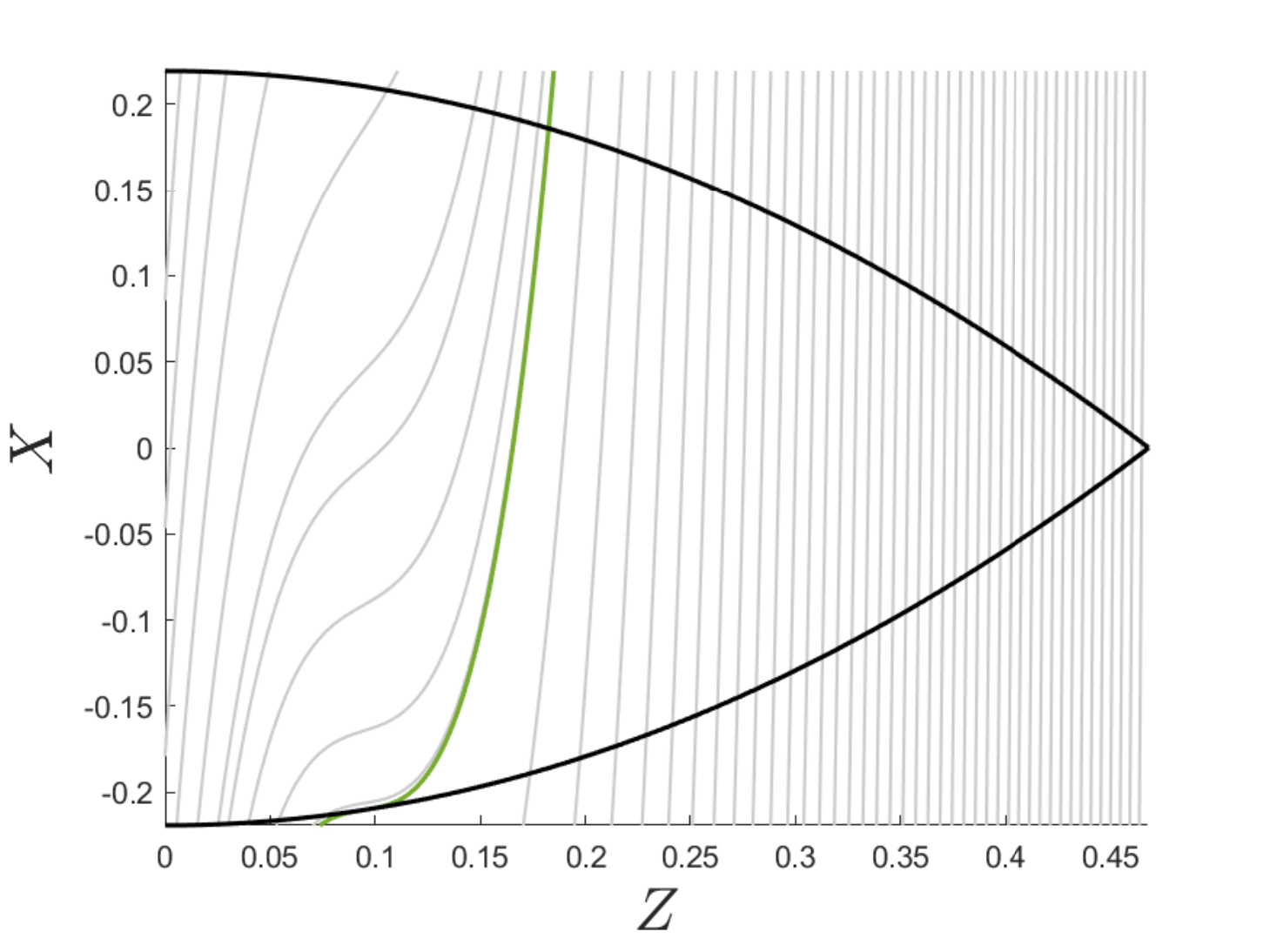}
			\includegraphics[width=0.4\textwidth]{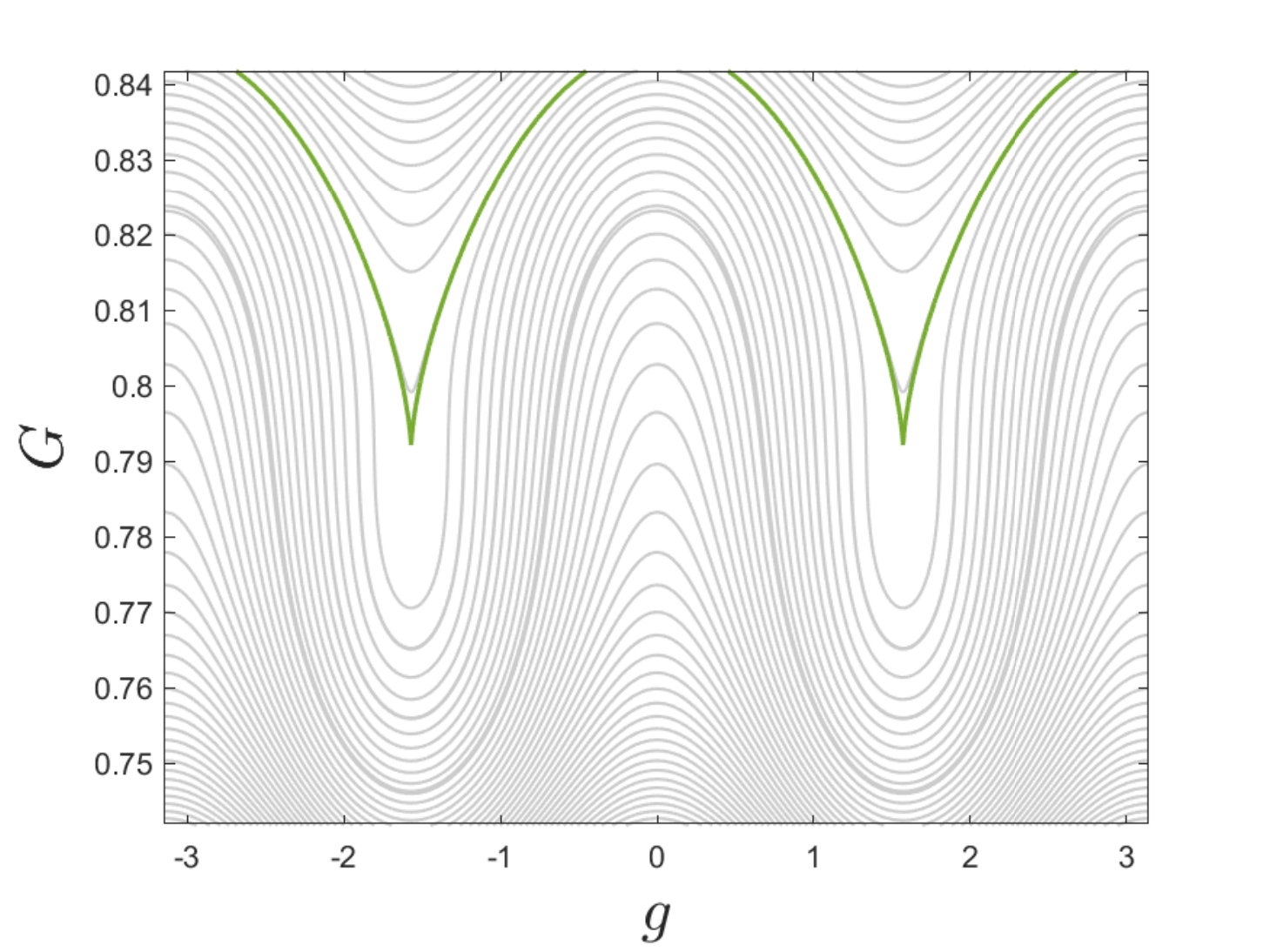}
			\caption{$\vert\rho\vert=\rho_{\blacksquare}$, $\rho_{\blacksquare}\sim 0.2518$}
			\label{fig:jcEx_1}
		\end{subfigure}
		\begin{subfigure}{1\textwidth}
			\centering
			\includegraphics[width=0.4\textwidth]{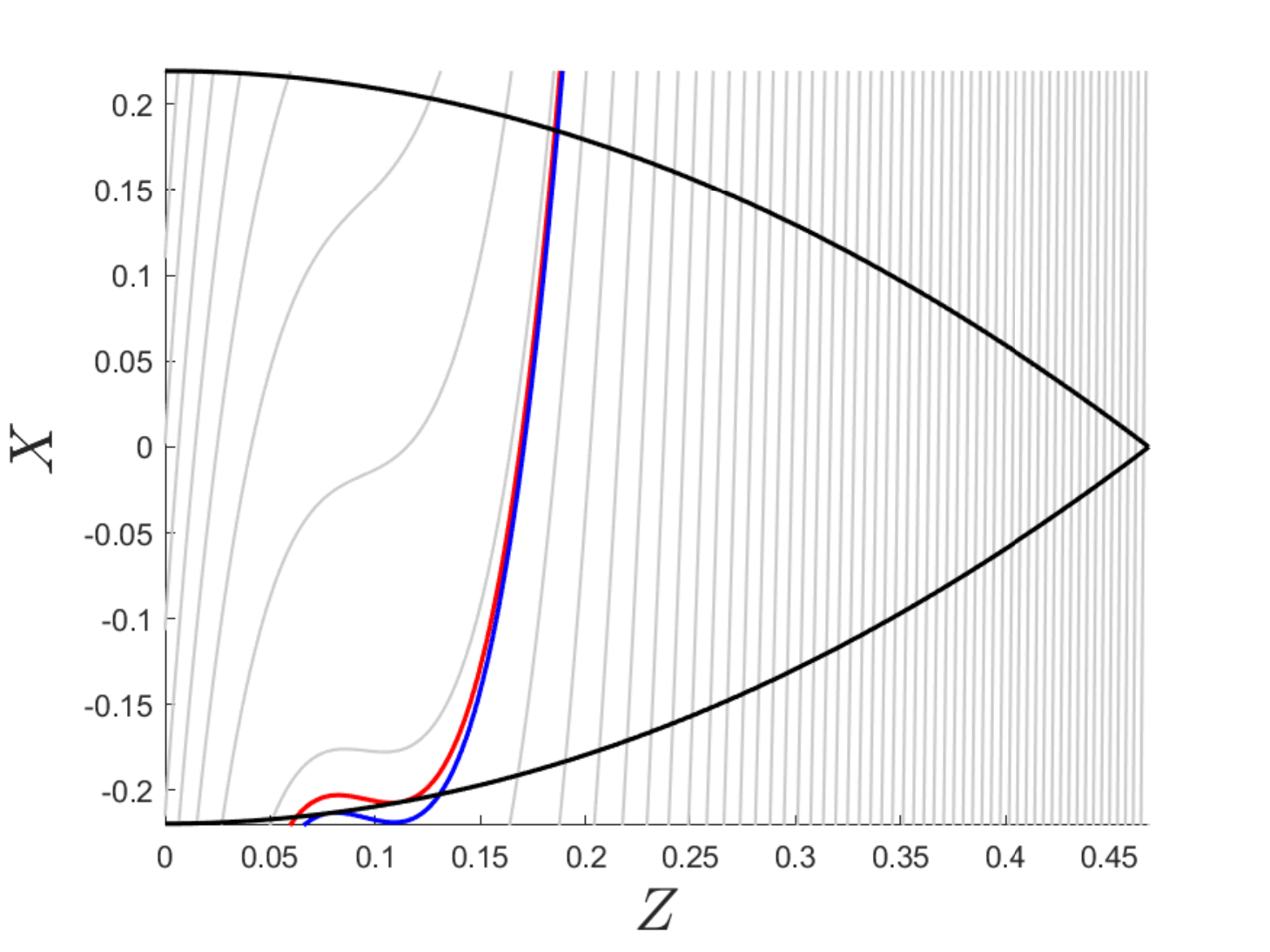}
			\includegraphics[width=0.4\textwidth]{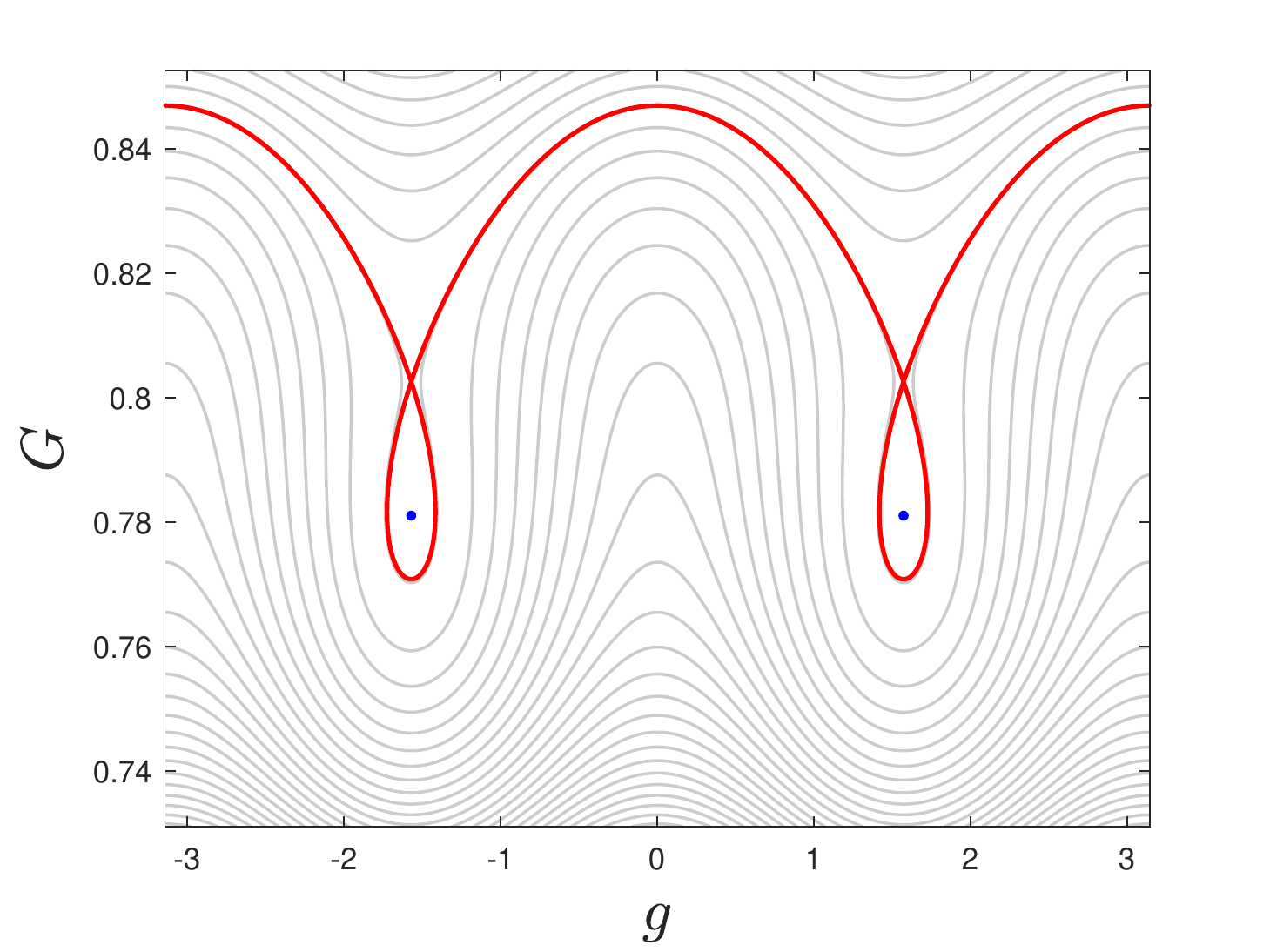}
			\caption{$\vert\rho\vert=0.2517$}
			\label{fig:jcEx_12}
		\end{subfigure}
		\begin{subfigure}{1\textwidth}
			\centering
			\includegraphics[width=0.4\textwidth]{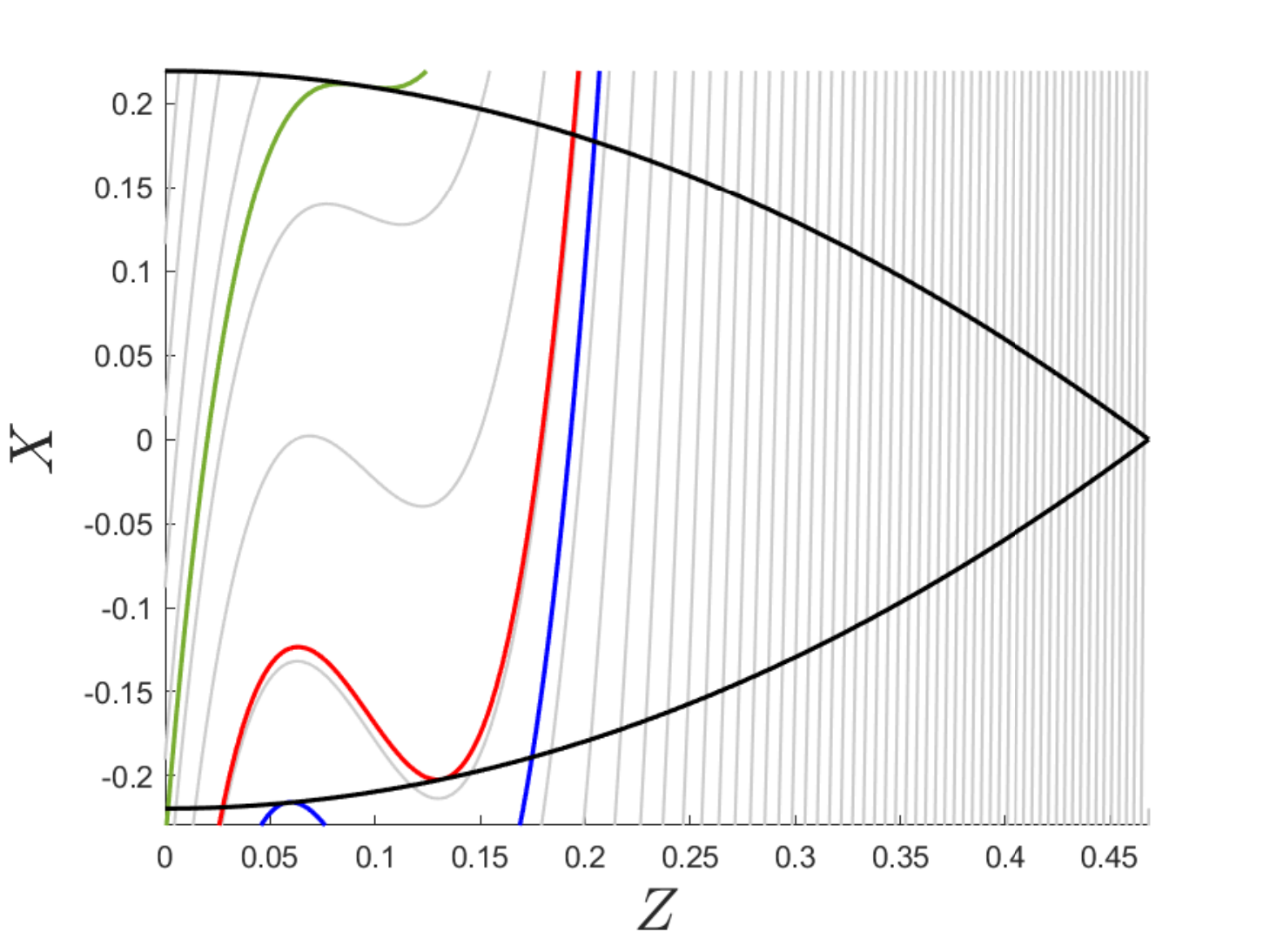}
			\includegraphics[width=0.4\textwidth]{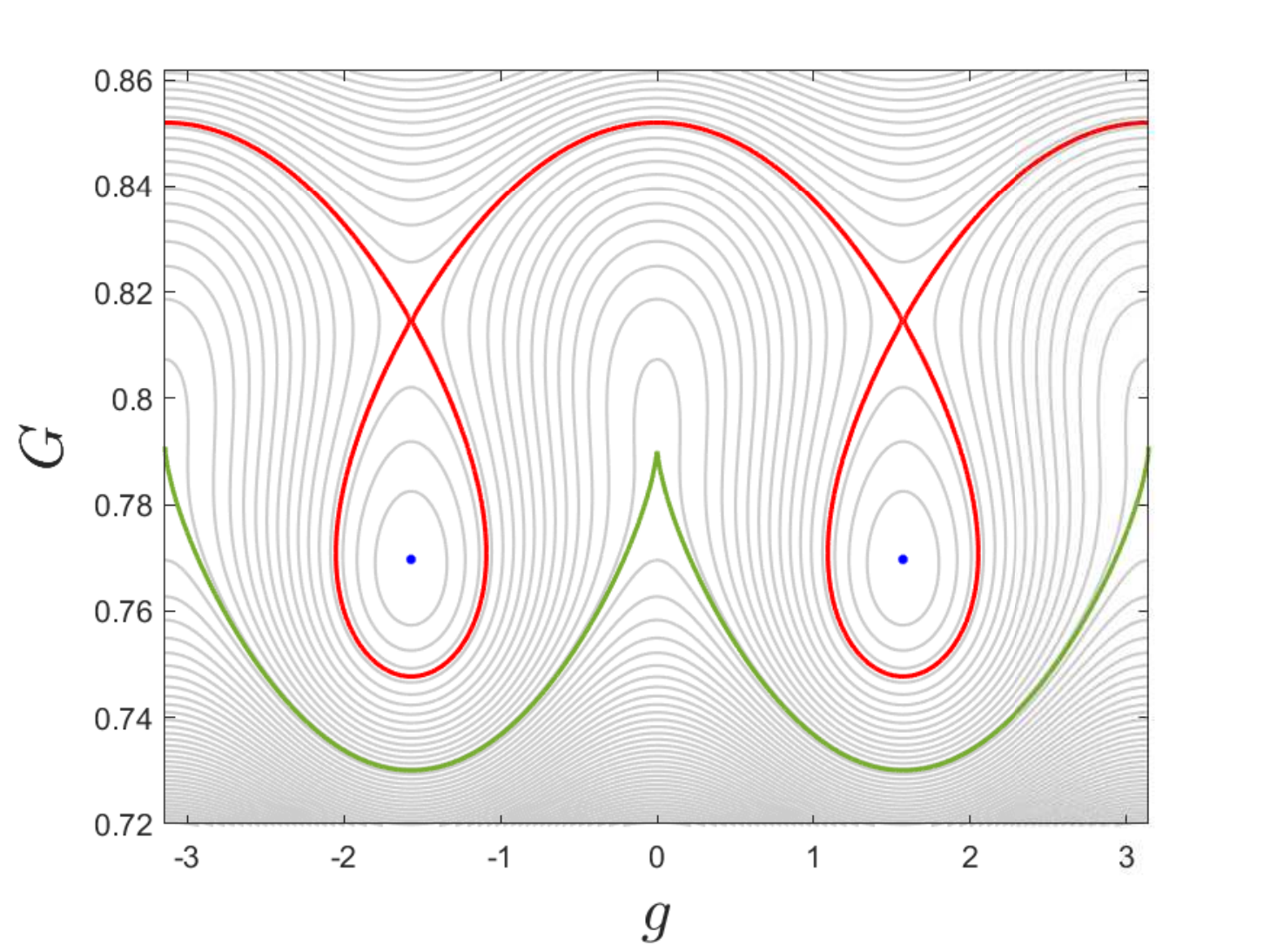}
			\caption{$\vert\rho\vert=\rho_{\blacklozenge}$, $\rho_{\blacklozenge}\sim 0.2514$}
			\label{fig:jcEx_2}
		\end{subfigure}
		\begin{subfigure}{1\textwidth}
			\centering
			\includegraphics[width=0.4\textwidth]{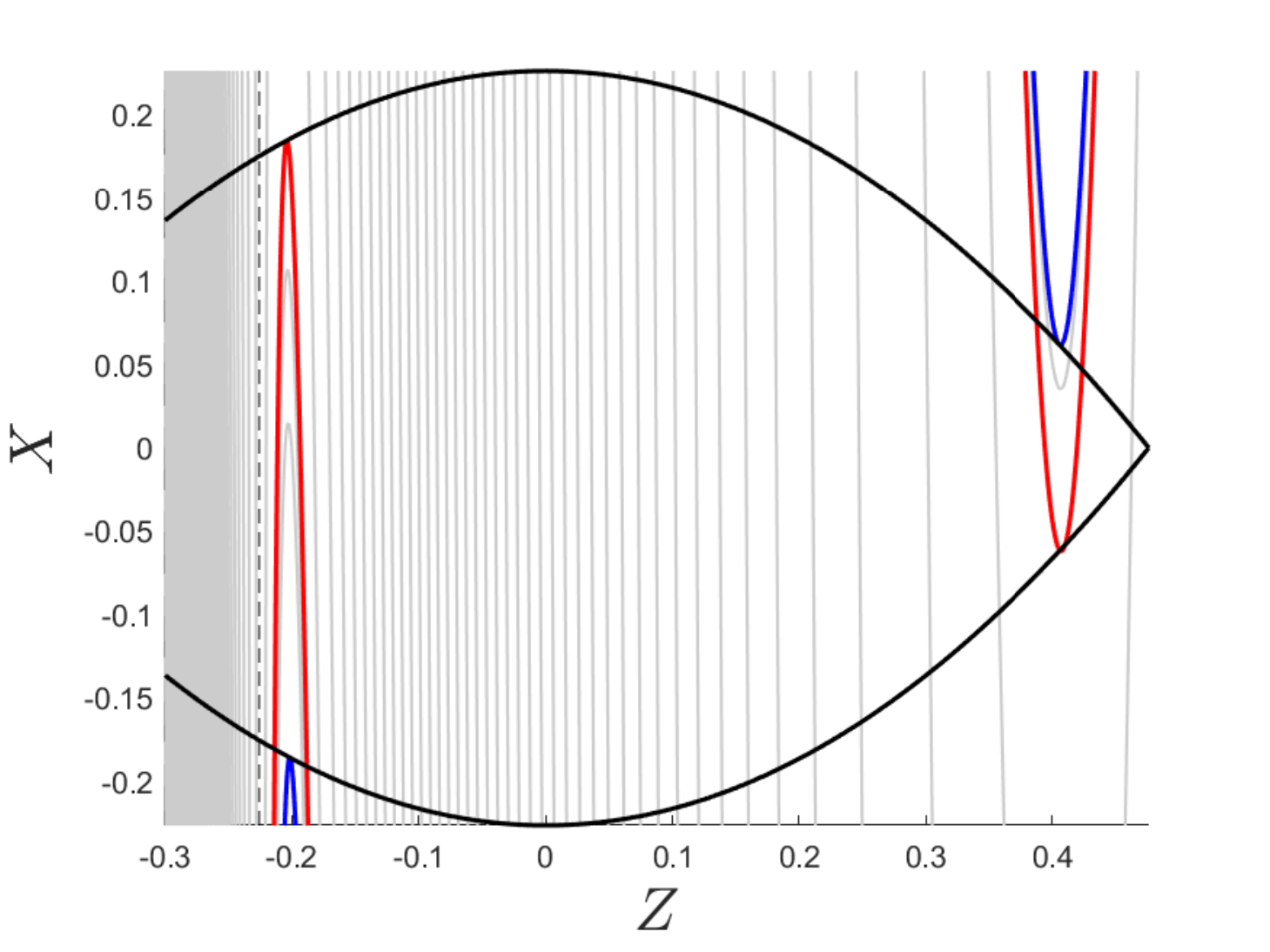}
			\includegraphics[width=0.4\textwidth]{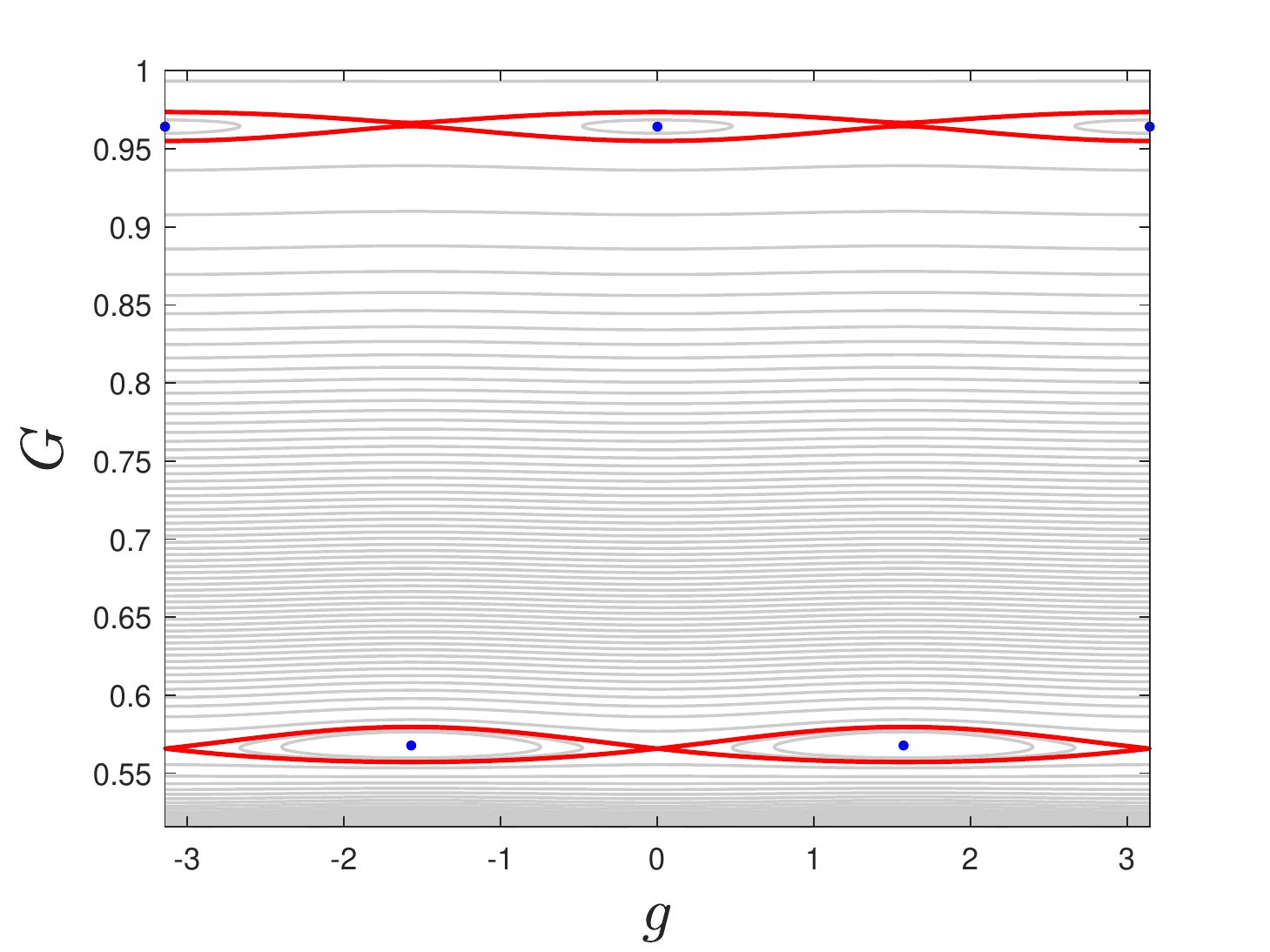}
			\caption{$\vert\rho\vert=0.22$}
			\label{fig:jcEx_23}
		\end{subfigure}
		\caption{Level curves for the $\J2$-problem with relativistic term with $\jjC=0.2$ and $\lambda=0.001$, for four different values of $\vert\rho\vert\in(\rho_{\diamond},\rho_{\blacksquare}]$. On the left, the levels curve are represented on the $(Z,X)$ plane. Enlargements of the  regions containing the equilibrium points are performed. On the right, the level curves are shown on corresponding enlargements in the $(g,G)$ plane. The same colour code used in Fig.\ref{fig:j413_bif1} is employed.}
	\end{figure}
	\begin{figure}
		\centering
		\begin{subfigure}{1\textwidth}
			\centering
			\includegraphics[width=0.4\textwidth]{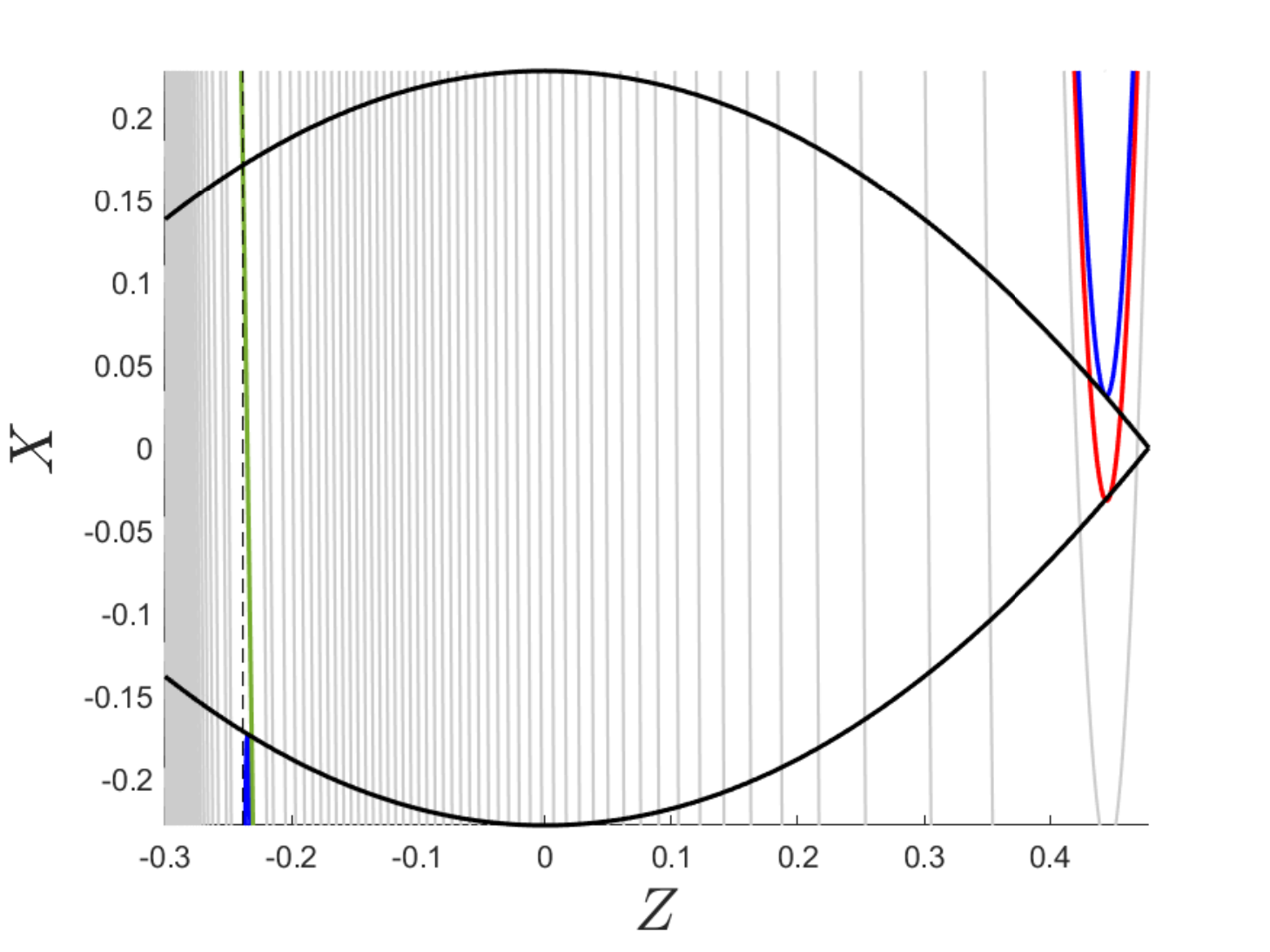}
			\includegraphics[width=0.4\textwidth]{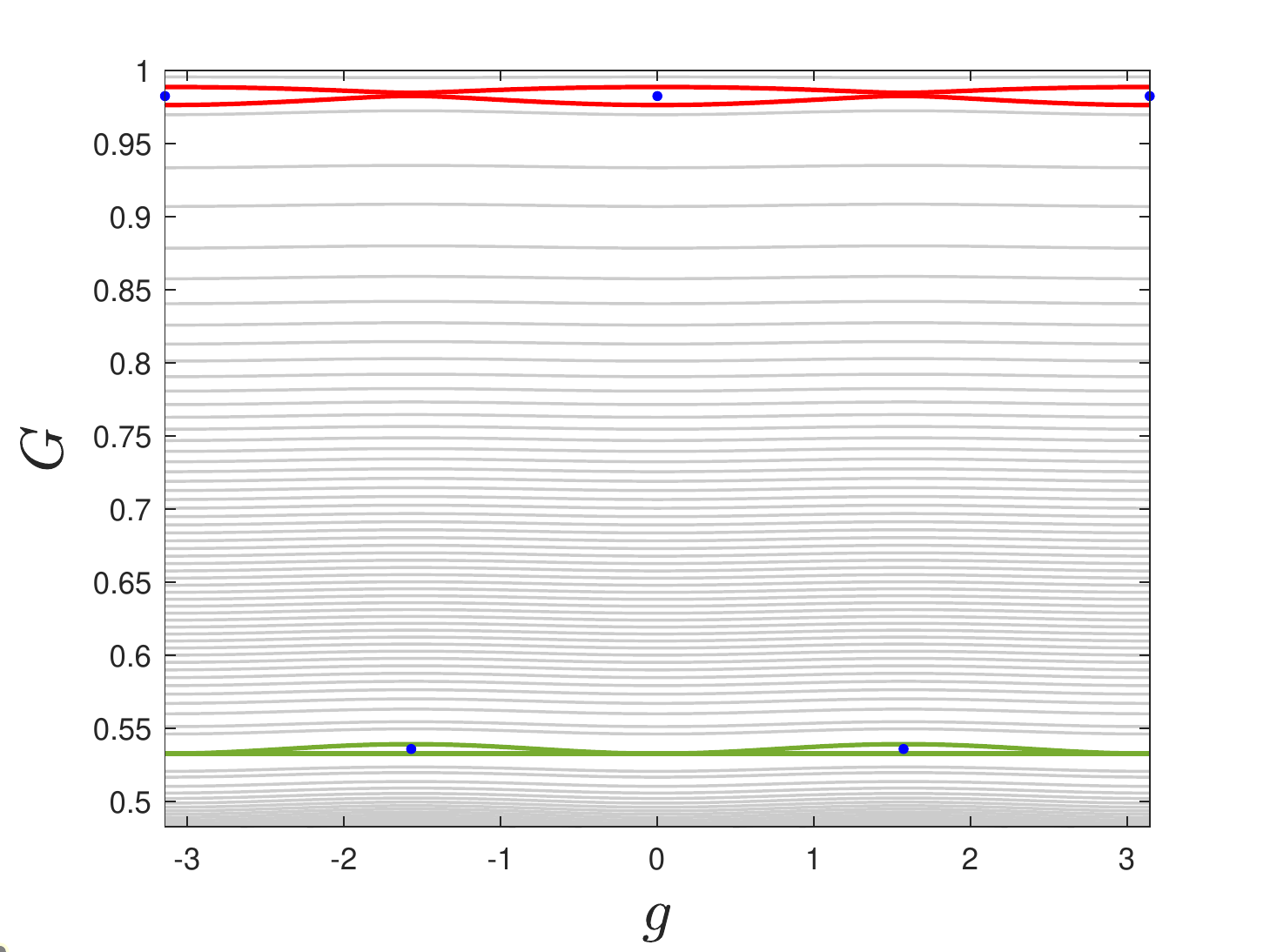}
			\caption{$\vert\rho\vert=\rho_{\diamond}$, $\rho_{\diamond}\sim 0.2114$}
			\label{fig:jcEx_3}
		\end{subfigure}
		\begin{subfigure}{1\textwidth}
			\centering
			\includegraphics[width=0.4\textwidth]{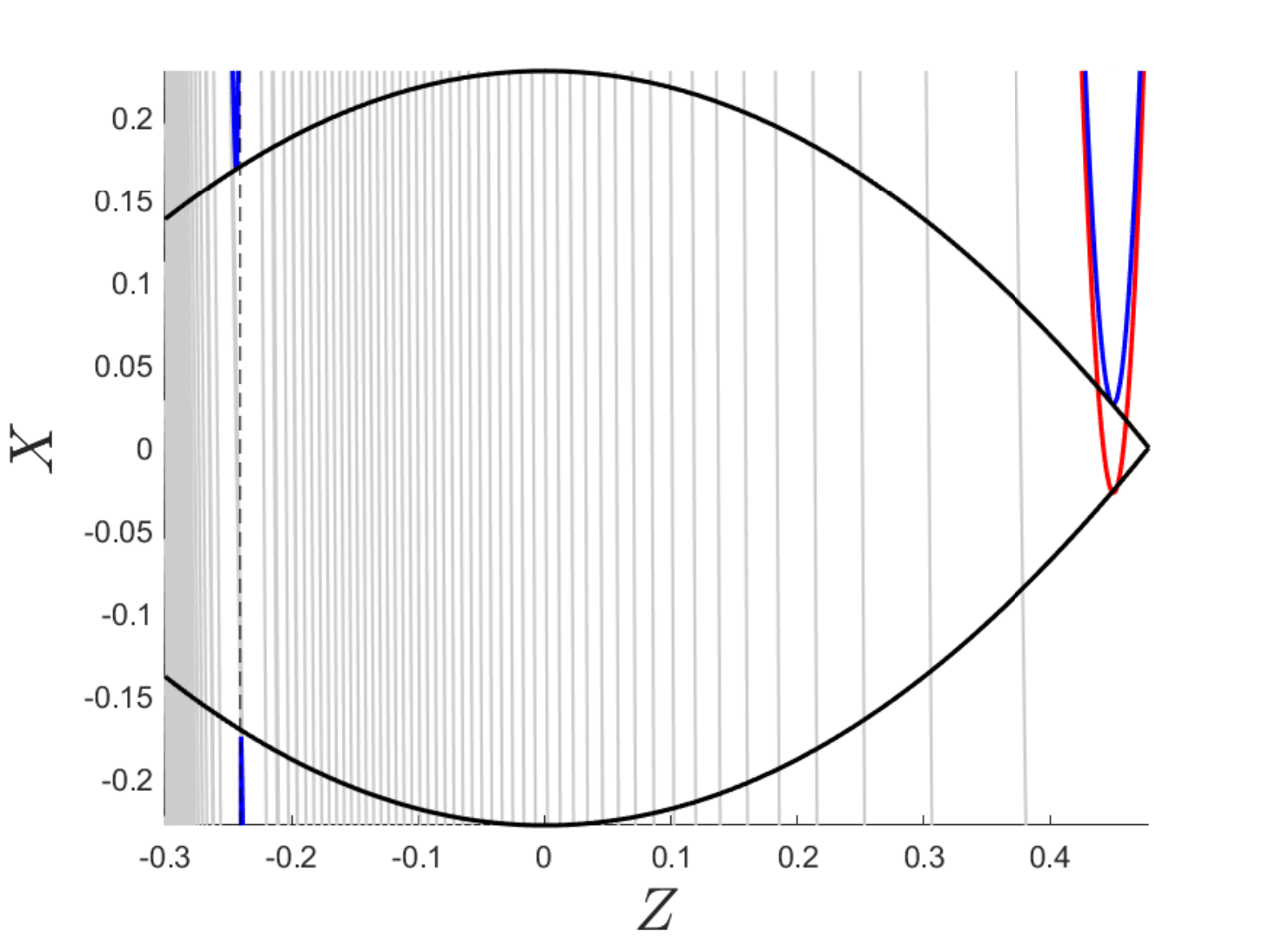}
			\includegraphics[width=0.4\textwidth]{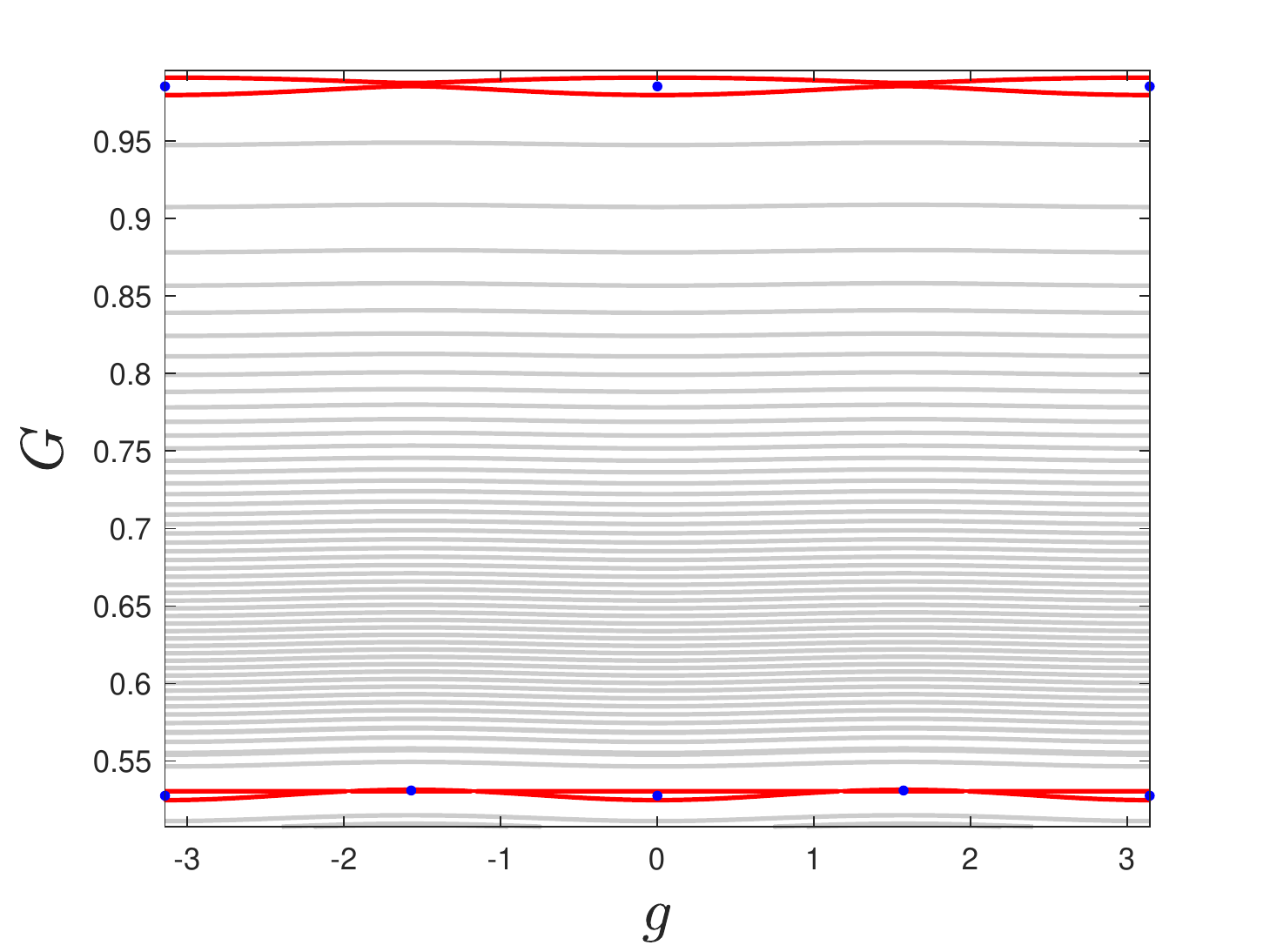}
			\caption{$\vert\rho\vert=0.21$}
			\label{fig:jcEx_34}
		\end{subfigure}
		\begin{subfigure}{1\textwidth}
			\centering
			\includegraphics[width=0.4\textwidth]{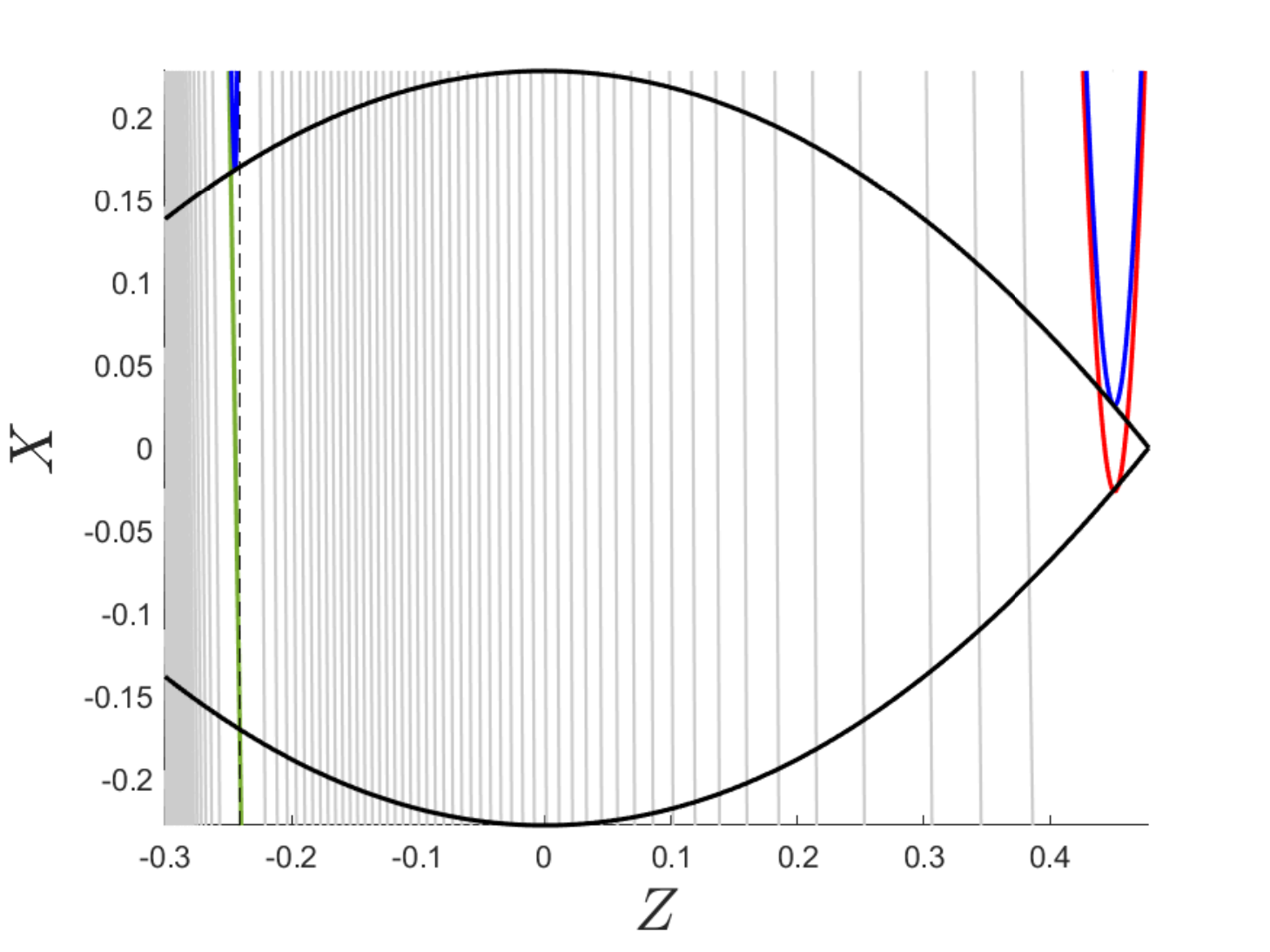}
			\includegraphics[width=0.4\textwidth]{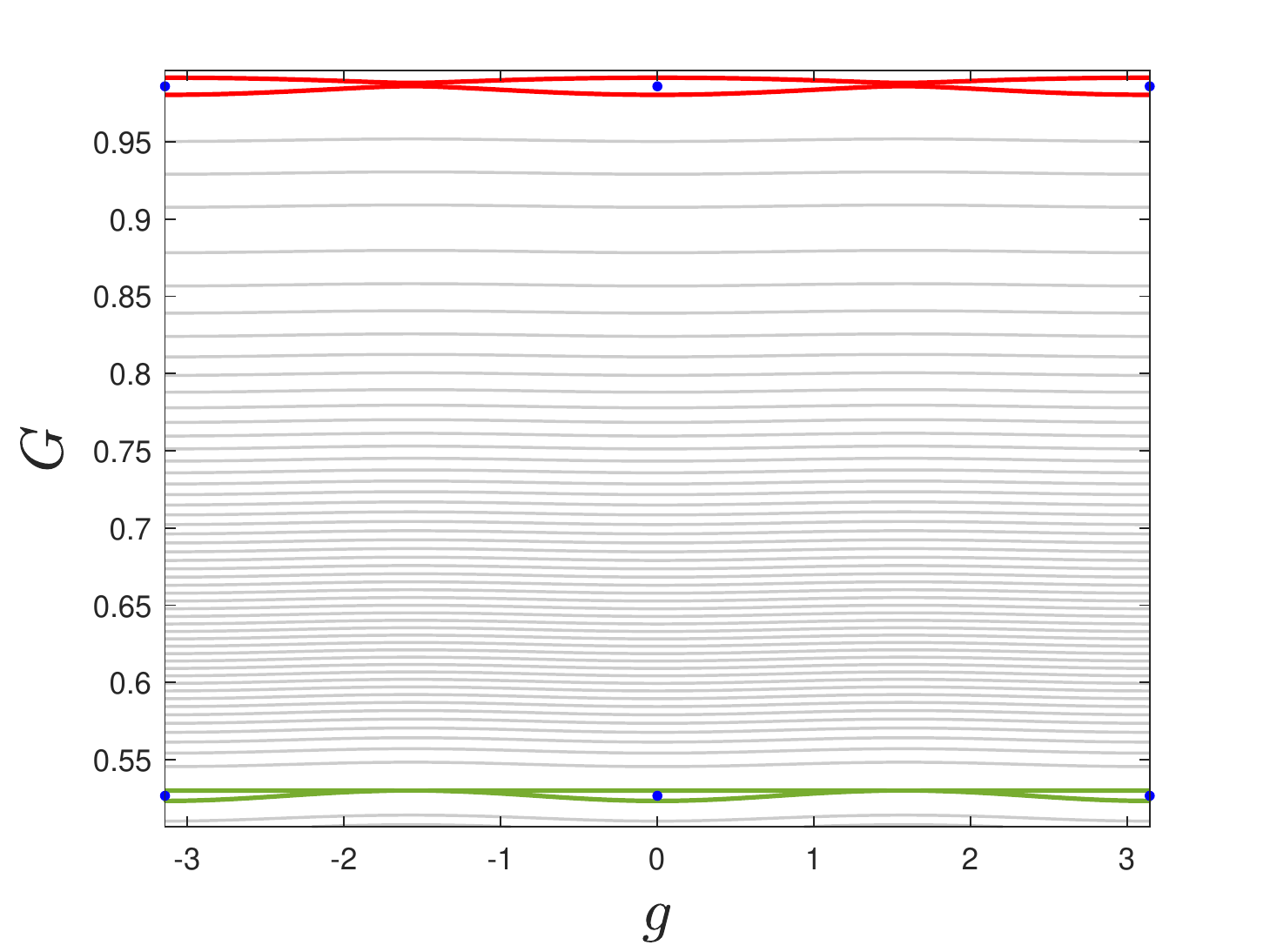}
			\caption{$\vert\rho\vert=\rho_{\square}$, $\rho_{\square}\sim 0.2098$}
			\label{fig:jcEx_4}
		\end{subfigure}
		\begin{subfigure}{1\textwidth}
			\centering
			\includegraphics[width=0.4\textwidth]{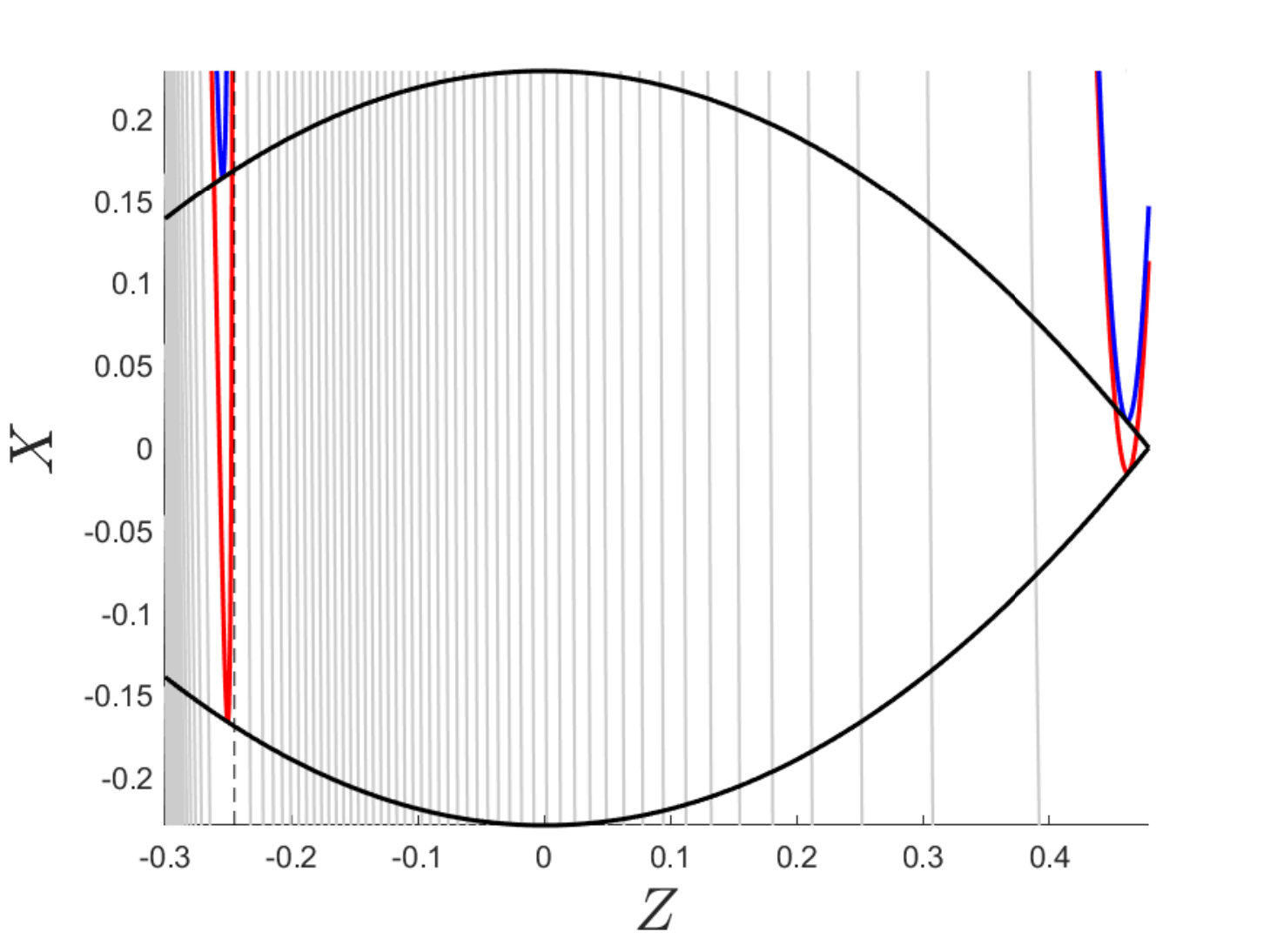}
			\includegraphics[width=0.4\textwidth]{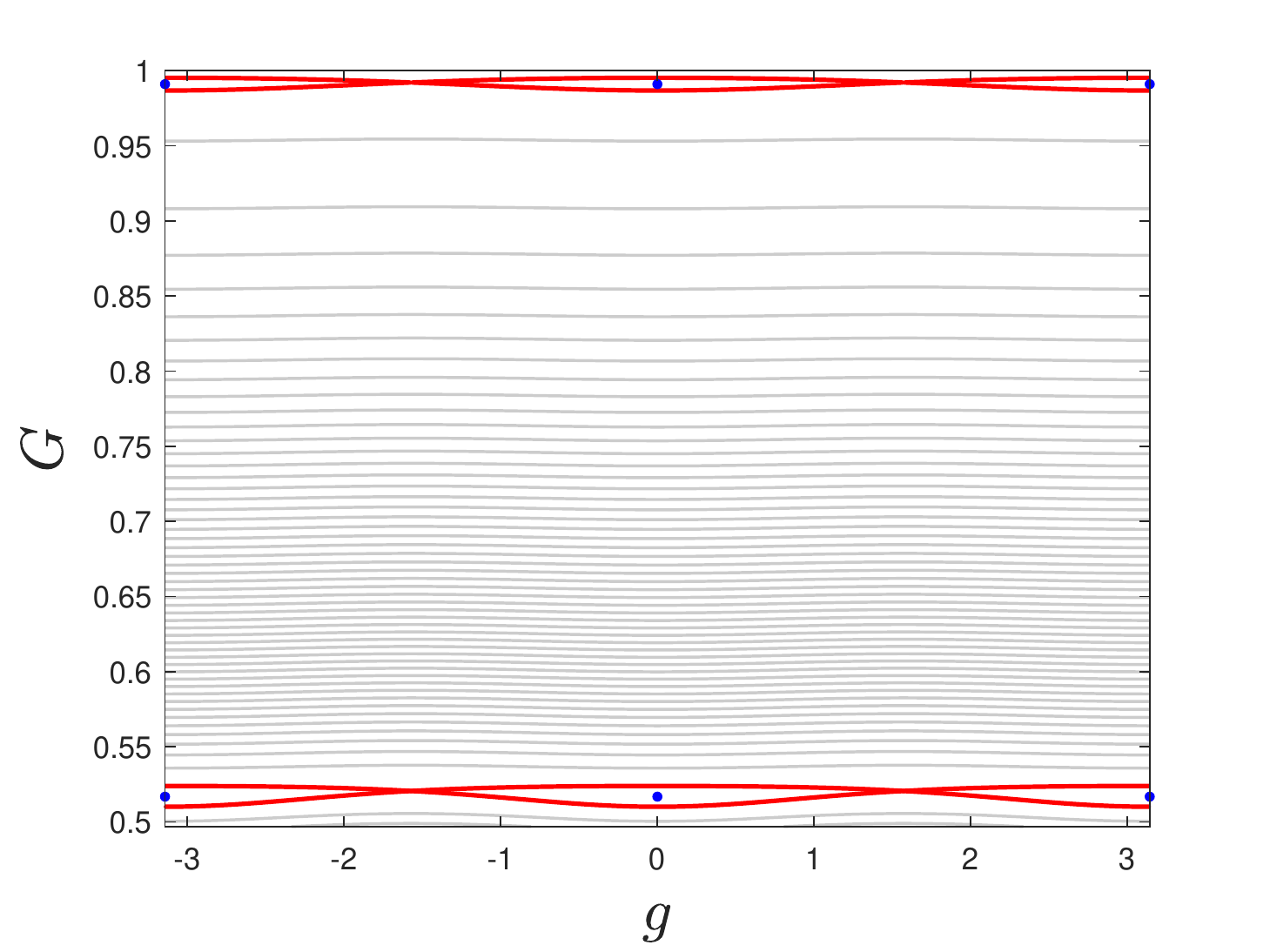}
			\caption{$\vert\rho\vert=0.207$}
			\label{fig:jcEx_45}
		\end{subfigure}
		\caption{Level curves for the $\J2$-problem with relativistic term with $\jjC=0.2$ and $\lambda=0.001$, for four different values of $\vert\rho\vert\in(\tilde{\rho}_-,\rho_{\diamond}]$. On the left, the levels curve are represented on the $(Z,X)$ plane. Enlargements of the  regions containing the equilibrium points are performed. On the right, the level curves are shown on corresponding enlargements in the $(g,G)$ plane. The same colour code used in Fig.\ref{fig:j413_bif1} is employed.}
	\end{figure}
	\begin{figure}
		\centering
		\begin{subfigure}{0.4\textwidth}
			\centering
			\includegraphics[width=1\textwidth]{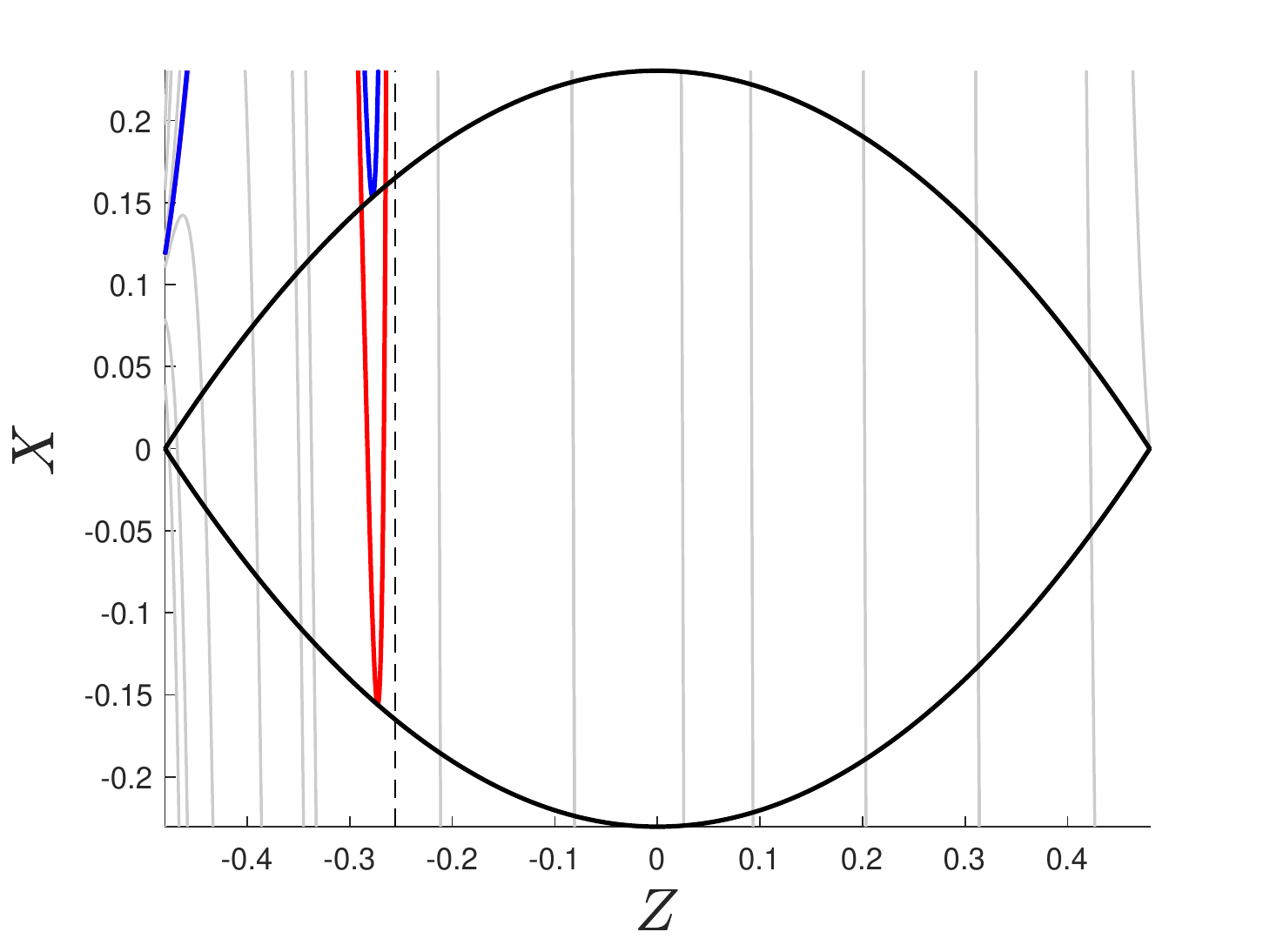}
		\end{subfigure}
		\begin{subfigure}{0.4\textwidth}
			\centering
			\includegraphics[width=1\textwidth]{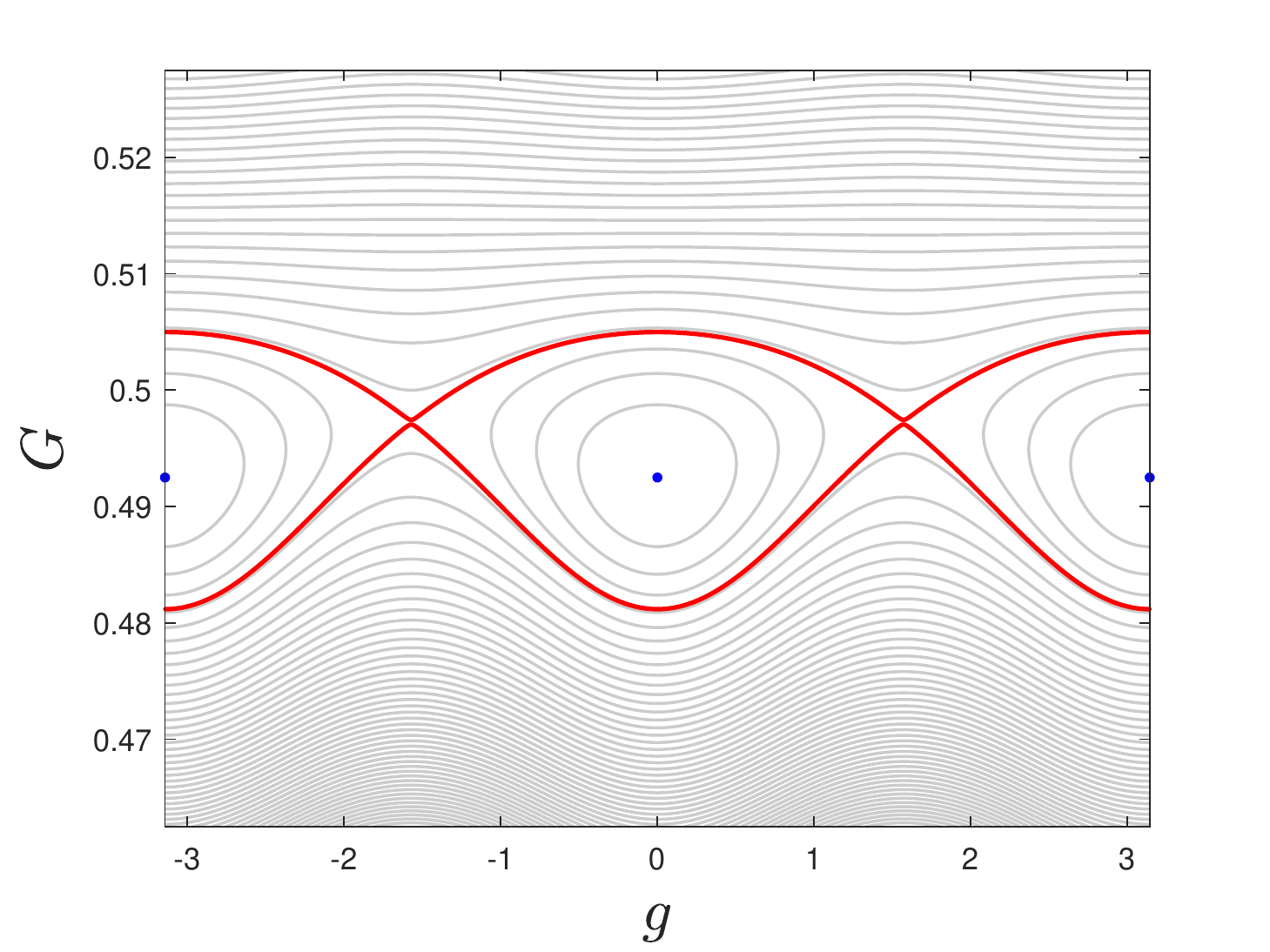}
		\end{subfigure}
		\caption{Level curves for the $\J2$-problem with relativistic term with $\jjC=0.2$ and $\lambda=0.001$ and for a value of $\vert\rho\vert<\tilde{\rho}_+$.Enlargements of the  regions containing the equilibrium points are performed. On the right, the level curves are shown on corresponding enlargements in the $(g,G)$ plane. The same colour code used in Fig.\ref{fig:j413_bif1} is employed.}
		\label{fig:jcEx_60}
	\end{figure}
	In conclusion, we have that
	\begin{itemize}
		\item  $\vert\rho\vert=\rho_{\blacksquare}$ and $\vert\rho\vert=\rho_{\blacklozenge}$ are saddle-node bifurcations, affecting the existence of the equilibrium points $E_{15}$, $E_{17}$, $E_{16}$ and $E_{18}$; for $\vert\rho\vert>\max(\rho_{\blacksquare},\rho_{\blacklozenge} )$ no equilibrium solution exist; 
		\item $\vert\rho\vert=\rho_{\diamond}$ and $\vert\rho\vert=\rho_{\square}$ are pitchfork bifurcation affecting the stability of the equilibrium points $E_{15}$ and $E_{16}$ and the existence of $\bar{E}_1$ and $\bar{E}_2$; 
		\item if existing, $\vert\rho\vert=\tilde{\rho}_+$ and $\vert\rho\vert=\tilde{\rho}_-$ are pitchfork bifurcations affecting the stability of $E_2$ and the existence of $E_{17}$ and $E_{18}$.
	\end{itemize}
	
	We give an example of the dynamical evolution setting $\lambda=0.001$ and $\jjC=0.2$. This last value is not realistic, but allows us to clearly illustrate the phenomenology just described. It holds $\rho_{\blacksquare}>\rho_{\blacklozenge}>\rho_{\diamond}>\rho_{\square}>\tilde{\rho}_->\tilde{\rho}_+$. After the saddle-node bifurcation at $\vert\rho\vert=\rho_{\blacksquare}$ (Fig.\ref{fig:jcEx_1}),  for $\rho_{\blacklozenge}<\vert\rho\vert<\rho_{\blacksquare}$ there exist the unstable equilibrium point $E_{18}$ and the stable $E_{16}$ (Fig.\ref{fig:jcEx_12}). After the second bifurcation (Fig.\ref{fig:jcEx_2}),  for $\rho_{\diamond}<\vert\rho\vert<\rho_{\blacklozenge}$ there exist also $E_{17}$, which is stable, and $E_{15}$ which is unstable (Fig.\ref{fig:jcEx_23}). At $\vert\rho\vert=\rho_{\diamond}$ $E_{15}$ coincide with $\bar{E}_1$ and $\bar{E}_2$ and it is degenerate (Fig.\ref{fig:jcEx_3}). For $=\rho_{\square}<\vert\rho\vert<\rho_{\diamond}$, $E_{15}$ is stable and $\bar{E}_1$ and $\bar{E}_2$ exist and are unstable (Fig.\ref{fig:jcEx_34}). At $\vert\rho\vert=\rho_{\square}$, $E_{17}$ coincides with $\bar{E}_1$ and $\bar{E}_2$ and it is degenerate (Fig.\ref{fig:jcEx_4}). After this last bifurcation, for $\tilde{\rho}_-<\vert\rho\vert<\rho_{\square}$, $\bar{E}_1$ and $\bar{E}_2$ do not exist, $E_{15}$ and $E_{17}$ are stable, and $E_{16}$ and $E_{18}$ are unstable (Fig.\ref{fig:jcEx_45}). After the last two bifurcations at $\vert\rho\vert=\tilde{\rho}_-$ and  $\vert\rho\vert=\tilde{\rho}_+$, there only exist the equilibrium point $E_{15}$, which is stable, and $E_{16}$ which is unstable (Fig.\ref{fig:jcEx_60}). 
	
	If $\jjC\ll1$, such that only the equilibrium points $E_{16}$ and $E_{15}$ exist, the dynamical evolution has no significant variation in comparison to the one of the $\J2$-problem.  It is the case of the Earth problem, since the values of $\jjC$ are typically very small (of the order of $10^{-6}$). Considering that $\rho_{\blacksquare},\rho_{\blacklozenge}\sim\frac{1}{\sqrt{80\jjC}}$ and $\rho_{\diamond},\rho_{\square}\sim\sqrt{\frac{7}{810\jjC}}$ in first approximation, our results are consistent with the outcomes of \cite{Jupp1991}.

	\section{Conclusions}
	\label{EndSect}
	We have described the existence and stability of frozen orbits in a gravity field expanded in even zonal terms. 
	The main focus has been given on the power of the geometric analysis of the reduced dynamics to highlight 
	the main features of these systems as they are determined by the presence of stable and unstable families. 
	In this respect, the study has been limited to the $J_2$ and $J_4$ problems and to the relativistic corrections, 
	showing the ability of the geometric invariant method to easily reproduce known results and predict new 
	features of higher-order terms. 
	The atlas of possible perturbations is wide and several other terms could be added. For many of them, this 
	approach requires very few changes and immediate results. For example, low-order tesseral terms, averaged 
	in order to preserve Brouwer structure, can be 
	easily analysed \citep{Palacian2007} without qualitative new results. Additional efforts are required for more 
	complex perturbations. Higher-degree zonal terms ($J_{2k}$ with $k \ge 3$) are the most promising since the 
	symmetry of the problem is preserved. Preliminary results like those presented in \cite{Coffey1994} can be 
	extended with a little effort. More general cases (odd zonal terms, higher-order tesserals, third-body effects, etc.) require a stronger commitment. However, in these cases, it is quite probable that difficulties arise more from the implementation 
	of the closed-form normalisation \citep{Palacian2002,IreneChristos2022} than from the use of the reduction 
	method.

	\section*{Acknowledgements}

	This work has been accomplished during the internship of I.C. 
	at the Department of Mathematics of the University of Rome Tor
	Vergata in the framework of the EU H2020 MSCA ETN Stardust-R (Grant Agreement 813644).
	G.P. acknowledges the support of MIUR-PRIN 20178 CJA2B
	``New Frontiers of Celestial Mechanics: theory and Applications''
	and the partial support of INFN and GNFM/INdAM.
	
	\section*{Compliance with ethical standards}
	
	{  Conflict of interest:} The authors declare that they have no conflict of interest.
	
	\section*{Appendix: Proof of inequalities \eqref{j2prob_derivativeP} and \eqref{j2prob_derivativeM}}
	We start by proving relation \eqref{j2prob_derivativeP}. 
	We have
	\[
	\frac{d {\rhoUJ2^2}_{2}}{dG}=\frac{1}{5\lambda}\frac{G}{C_+^2\sqrt{B_+}}\Big(-D_+\sqrt{B_+}+\lambda E_+\sqrt{B_+}+F_+\Big),
	\]
	where
	\begin{align*}
	D_+ = & -14400G^6+28800G^5+68640G^4,\\
	E_+ = & 4410G^4-14904G^3-14204G^2+48312G+28314, \\
	F_+ = & 20G^2D_+ + \lambda\big(-21600G^8+154080G^7-285920G^6-237600G^5+1006720G^4\big) \\
	& 
	+ \lambda^2(-129960G^6+470664G^5+679088G^4-2114832G^3-2381760G^2+1287528G+1774344),
	\end{align*}
	and where $B_+$ is defined in \eqref{rhoE312_j2}. It holds ${d {\rhoUJ2^2}_{2}}/{dG}>0$ if
	\[
	-D_+\sqrt{B_+}+\lambda E_+\sqrt{B_+}+F_+>0.
	\] 
	It is straightforward that $D_+>0$ and $E_+>0$. Moreover
	\[
	\begin{split}
	F_+ > & \Big( -288000{G}^{10}+576000{G}^{9}+1351200{G}^{8}+154080{G}^{7}-
	415880\,{G}^{6}+233064\,{G}^{5}\\
	& +1685808{G}^{4}-2114832\,{G}^{3}-
	2381760\,{G}^{2}+1287528\,G+1774344
	\Big)\lambda^2>0.
	\end{split}
	\]
	Thus, we need to verify whether
	\[
	\left(\lambda E_+\sqrt{B_+}+F_+\right)^2-D_+^2B_+>0.
	\]
	Since $B_+\ge(400G^8+240G^4\lambda+376\lambda^2)>(20G^4+6\lambda)^2$, we have
	\[
	\left(\lambda E_+\sqrt{B_+}+F_+\right)^2-D_+^2B_+> M_+,
	\]
	with
	\[
	\begin{split}
	M_+=\lambda^2E_+^2B_++F_+^2+2\lambda E_+F_+(20G^4+6\lambda)-D_+^2B_+ = M_+^{(4)}\lambda^4+ M_+^{(3)}\lambda^3+ M_+^{(2)}\lambda^2+ M_+^{(1)}\lambda,
	\end{split}
	\]
	where
	\begin{align*}
	M_+^{(4)} = &  23910365700\,{G}^{12}-181225103760\,{G}^{11}+120523638672\,{G}^{10}+
	1700153452368\,{G}^{9}\\
	& -1988897297356\,{G}^{8}-7508533086240\,{G}^{7}+
	6377312862144\,{G}^{6}+19130049303840\,{G}^{5}\\
	& -4594353603924\,{G}^{4}-
	23486016392784\,{G}^{3}-4409791599312\,{G}^{2}\\
	& +10749794943888\,G+
	4994571648924,\\
	M_+^{(3)} = &  -160G^4(103329675G^10-563720310G^9-911841039G^8+7739401536G^7\\
	& +2782008506G^6-39130317372G^5-12624306282G^4+84930202896G^3\\
	& +49983903915G^2-50940897150G-39230486295),\\
	M_+^{(2)}= &
	1600G^8(2772225G^8-11651040G^7-19496340G^6+86266224G^5+106893822G^4\\
	& -176211648G^3-404205252G^2-24689808G+315589417),\\
	M_+^{(1)}= &
	768000G^{12}(-30G^2+60G+143)(45G^2-72G-143)^2.
	\end{align*}
	We have $M_+^{(1)}>0$, $M_+^{(1)}+M_+^{(2)}>0$, $M_+^{(1)}+M_+^{(2)}+M_+^{(3)}>0$ and  $M_+^{(4)}>0$ $\forall G$; thus, using $\lambda<1$, it holds
	\[
	M_+> (M_+^{(4)}+M_+^{(3)}+M_+^{(2)}+M_+^{(1)})\lambda^4>0.
	\]
	Now, we prove relation \eqref{j2prob_derivativeM}. We have
	\[
	\frac{d {\rhoLJ2^2}_{2}}{dG}=\frac{1}{5\lambda}\frac{G}{C_-^2\sqrt{B_-}}\Big(-D_-\sqrt{B_-}+\lambda E_-\sqrt{B_-}+F_-\Big),
	\]
	where
	\begin{align*}
	D_- = &  20160G^6+28800G^5+5280G^4,\\
	E_- = & 22050G^4+43848G^3+21524G^2-10440G-4158 \\
	F_- = & 20G^2D_-(G) + \lambda\big(-846720\,{G}^{8}-1441440\,{G}^{7}+519680\,{G}^{6}+1680480\,{G}^{5}+
	352000\,{G}^{4}
	\big) \\ 
	& +\lambda^2(617400\,{G}^{6}+1375920\,{G}^{5}+392\,{G}^{4}-1902192\,{G}^{3}-968664
	\,{G}^{2}+554208\,G+211288
	),
	\end{align*}
	and $B_-$ is defined in \eqref{rhoE412_j2}.
	It holds ${d {\rhoLJ2^2}_{2}}/{dG}>0$ if
	\[
	-D_-\sqrt{B_-}+\lambda E_-\sqrt{B_-}+F_->0.
	\]
	It is straightforward that $D_->0$ and 
	\[
	\begin{split}
	F_- > & \Big( -1693440\,{G}^{8}-2882880\,{G}^{7}+1656760\,{G}^{6}+4736880\,{G}^{5}\\ &
	+704392\,{G}^{4}-1902192\,{G}^{3}-968664\,{G}^{2}+554208\,G+211288\Big)\lambda^2>0.
	\end{split}
	\]
	Instead $E_->0$ $\forall G\ge0.5$, while its sign changes if $G<0.5$.
	Let us consider $G\in(0,0.5]$; we need to verify whether
	\[
	F_-^2-(-D_-+\lambda E_-)^2B_->0.
	\]
	It holds
	\[
	F_-^2-(-D_-+\lambda E_-)^2B_- = P_-^{(4)}\lambda^4+ P_-^{(3)}\lambda^3+ P_-^{(2)}\lambda^2+ P_-^{(1)}\lambda,
	\]
	with
	\begin{align*}
	P_-^{(4)} = &  205663657500\,{G}^{12}+1286808541200\,{G}^{11}+2740720809456\,{G}^{10}
	+1509967832688\,{G}^{9}\\
	& -2841942124116\,{G}^{8}-4558328657760\,{G}^{7}-
	1319583449472\,{G}^{6}+1104354441696\,{G}^{5}\\
	&+231876171316\,{G}^{4}-
	430214668464\,{G}^{3}-18111422832\,{G}^{2}+89372757936\,G+17827435780,\\
	P_-^{(3)} = &  -160G^4(4650179625G^10+21853893474G^9+31457092779G^8-4333755960G^7\\
	& -49586843494G^6-32290996788G^5+11675441446G^4+12857914248G^3\\
	&-2955589987G^2-3112286430G-456484721),\\
	P_-^{(2)} = &  1600G^8(568229823G^8+2010142008G^7+1564140924G^6-2245174056G^5\\
	& -3853671558G^4-789759288G^3+1195807132G^2+548335656G+61908319),\\
	P_-^{(1)} = &  -768000G^{12}(42G^2+60G+11)(10647G^4+11808G^3-22398G^2-29664G-5269).
	\end{align*}
	For $G<0.5$, we have $P_-^{(1)}>0$, $P_-^{(2)}>0$, $P_-^{(3)}>0$ and $P_-^{(3)}+P_-^{(4)}>0$; thus
	\[
	F_-^2-(-D_-+\lambda E_-)^2B_->\lambda^4(P_-^{(3)}+P_-^{(4)})+ P_-^{(2)}\lambda^2+ P_-^{(1)}\lambda>0.
	\]
	Let us now consider $G\in[0.5,1]$. In this case, we need to verify whether
	\[
	\left(\lambda E_-\sqrt{B_-}+F_-\right)^2-D_-^2B_->0.
	\]
	Since $B_->400G^8$, 
	\[
	\left(\lambda E_-\sqrt{B_-}+F_-\right)^2-D_-^2B_-> M_-,
	\]
	with
	\[
	\begin{split}
	M_-=\lambda^2E_-^2B_-+F_-^2+40\lambda E_-F_-G^4-D_-^2B_- = M_-^{(4)}\lambda^4+ M_-^{(3)}\lambda^3+ M_-^{(2)}\lambda^2+ M_-^{(1)}\lambda,
	\end{split}
	\]
	where
	\begin{align*}
	M_-^{(4)} = &  976780822500\,{G}^{12}+4476174696000\,{G}^{11}+5453865257400\,{G}^{10}
	-5086142681760\,{G}^{9}\\
	&-16601363592772\,{G}^{8}-7099182483456\,{G}^{7}
	+11816572141456\,{G}^{6}+11183433731136\,{G}^{5}\\
	&-1358260437988\,{G}^{4
	}-4012690481472\,{G}^{3}-434866063560\,{G}^{2}+421070887776\,G+
	86153421508,\\
	M_-^{(3)} = &  -320G^4(3267280800G^10+12799767015G^9+9651811113G^8-23608483416G^7\\
	& -41966083012G^6-5277158466G^5+28856284782G^4+14899437888G^3\\
	& -4161615732G^2-3404287293G-463072687),
	\\
	M_-^{(2)}= &
	1600G^8(102880449G^8+321838272G^7+208527732G^6-295032528G^5-233557962G^4\\
	& +563365728G^3+780379476G^2+285203952G+31226833),
	\\
	M_-^{(1)}= &
	768000G^12(42G^2+60G+11)(63G^2+72G+11)^2.
	\end{align*}
	For $0.5\le G \le 1$, we have $M_+^{(1)}>0$, $M_+^{(2)}>0$, $M_+^{(2)}+M_+^{(3)}>0$ and $M_+^{(2)}+M_+^{(3)}+M_+^{(4)}>0$; thus,
	\[
	M_-> (M_+^{(4)}+M_+^{(3)}+M_+^{(2)})\lambda^4+M_+^{(1)}\lambda>0.
	\]
	
	\begin{align*}
	\hat{A}_- = & 80G^4 + \lambda\left((595G^2-1035)j_4 -175G^2-96G+189\right), \\
	\hat{B}_- = & \frac{\hat{A}_-^2-5\lambda\hat{C}_-\hat{D}_-}{16},\\
	\hat{C}_- = & (315G^2-539)j_4-63G^2-72G-11, & \\
	\hat{D}_- = &  {32}G^4 + \lambda\left((95G^2-175)j_4-35G^2-24G+49\right).
	\end{align*}
	
	\[ \begin{split} \frac{d^2\tilde{X}}{dZ^2}\pm\frac{d^2\hat{X}}{dZ^2}\sim & 16\sqrt{-80 j_C\rho^2+1}\Big(144000 j_C^3\rho^6-28400 j_C^2\rho^4+880 j_C\rho^2-7+\\ &\sqrt{-80 j_C\rho^2+1}(10000 j_C^2\rho^4-600 j_C\rho^2+7)\Big); \end{split} \]
	
	\[ \begin{split} \frac{d^2\tilde{X}}{dZ^2}\pm\frac{d^2\hat{X}}{dZ^2}\sim & 16\sqrt{-80 j_C\rho^2+1}\Big(-144000 j_C^3\rho^6+28400\jjC^2\rho^4-880 j_C\rho^2+7+\\ &\sqrt{-80 j_C\rho^2+1}(10000 j_C^2\rho^4-600 j_C\rho^2+7)\Big). \end{split} \]
	
	\begin{align*} \hat{A}_- = & 80G^4 + \lambda\left((595G^2-1035)j_4 -175G^2-96G+189\right), \\ \hat{B}_- = & \frac{\hat{A}_-^2-5\lambda\hat{C}_-\hat{D}_-}{16},\\ \hat{C}_- = & (315G^2-539)j_4-63G^2-72G-11, & \\ \hat{D}_- = &  {32}G^4 + \lambda\left((95G^2-175)j_4-35G^2-24G+49\right). \end{align*}

	\begin{align*} \hat{A}_- = & 80G^4 + \lambda\left((595G^2-1035)j_4 -175G^2-96G+189\right), \\ \hat{B}_- = & \frac{\hat{A}_-^2-5\lambda\hat{C}_-\hat{D}_-}{16},\\ \hat{C}_- = & (315G^2-539)j_4-63G^2-72G-11, & \\ \hat{D}_- = &  {32}G^4 + \lambda\left((95G^2-175)j_4-35G^2-24G+49\right). \end{align*}

\bibliographystyle{apalike}
\bibliography{refCP}

\end{document}